\documentclass[twocolumn]{aastex62}

\usepackage{amsmath}
\usepackage{natbib}
\usepackage{graphicx}
\usepackage{latexsym}
\usepackage{amssymb}
\usepackage{hyperref}

\newcommand{\mach}{M_{s}}
\newcommand{\bzero}{B_{0}}
\newcommand{\bsq}{b^2}
\newcommand{\vtotsq}{v_{\textrm{tot}}^2}
\newcommand{\vsolsq}{v_{\textrm{sol}}^2}
\newcommand{\vcompsq}{v_{\textrm{comp}}^2}

\received{}
\revised{}
\accepted{\today}

\begin{document}

\title{Generation of Solenoidal Modes and Magnetic Fields in Turbulence Driven by Compressive Driving}

\correspondingauthor{Jungyeon Cho}
\email{jcho@cnu.ac.kr}

\author{Jeonghoon Lim}
\affiliation{Department of Astronomy and Space Science, Chungnam National University, 99, Daehak-ro, Yuseong-gu, Daejeon, 34134, Republic of Korea; jhLim0918@o.cnu.ac.kr, jcho@cnu.ac.kr, hsyoon@kasi.re.kr}
\author{Jungyeon Cho}
\affiliation{Department of Astronomy and Space Science, Chungnam National University, 99, Daehak-ro, Yuseong-gu, Daejeon, 34134, Republic of Korea; jhLim0918@o.cnu.ac.kr, jcho@cnu.ac.kr, hsyoon@kasi.re.kr}
\affiliation{Korea Astronomy and Space Science Institute, 776, Daedeokdae-ro, Yuseong-gu, Daejeon, 34055, Republic of Korea}
\author{Heesun Yoon}
\affiliation{Korea Astronomy and Space Science Institute, 776, Daedeokdae-ro, Yuseong-gu, Daejeon, 34055, Republic of Korea}
\affiliation{Department of Astronomy and Space Science, Chungnam National University, 99, Daehak-ro, Yuseong-gu, Daejeon, 34134, Republic of Korea; jhLim0918@o.cnu.ac.kr, jcho@cnu.ac.kr, hsyoon@kasi.re.kr}

\begin{abstract}
We perform numerical simulations of hydrodynamic (HD) and magnetohydrodynamic (MHD) turbulence driven by compressive driving to study generation of solenoidal velocity component and small-scale magnetic field. We mainly focus on the effects of mean magnetic field ($B_0$) and the sonic Mach number ($M_s$). We also consider two different driving schemes in terms of correlation timescale of forcing vectors: a finite-correlated driving and a delta-correlated driving. The former has a longer correlation timescale of forcing vectors, which is comparable to large-eddy turnover time, than the latter. Our findings are as follows. First, when we fix the value of $\bzero$, the level of solenoidal velocity component after saturation increases as $\mach$ increases. A similar trend is observed for generation of magnetic field when $\bzero$ is small. Second, when we fix the value of $\mach$, HD and MHD simulations result in similar level of the solenoidal component when $\bzero$ $\lesssim$ 0.2 (or Alfven Mach number of $\sim$ 5). However, the level increases when $B_0$ $\gtrsim$ 0.2. Roughly speaking, the magnetic energy density after saturation is a linearly increasing function of $\bzero$ irrespective of $M_s$. Third, generation of solenoidal velocity component is not sensitive to numerical resolution, but that of magnetic energy density is mildly sensitive. Lastly, when initial conditions are same, the finite-correlated driving always produces more solenoidal velocity and small-scale magnetic field components than the delta-correlated driving. We additionally analyze the vorticity equation to understand why higher $\mach$ and $\bzero$ yield larger quantity of the solenoidal velocity component.

\end{abstract}


\section{Introduction} \label{sec:sec1}

Turbulence is ubiquitous in almost all astrophysical media ranging from the interstellar medium (ISM) (e.g., \citealt{L1981,ES04,MK04,MO07}) to the intracluster medium (ICM) (e.g., \citealt{Kulsrud97,Sc04,Ryu08,Hitomi16,Vazza17}). Turbulence in cold (10K) and dense interstellar molecular clouds has Mach numbers ($M_s$) of $\sim$ a few or even larger than 10 (see, e.g., \citealt{L1981}). Such supersonic turbulence plays essential roles in star formation processes in the ISM (see \citealt{MK04} for a review). Unlike this, ICM turbulence is usually subsonic with $M_s$ $\lesssim$ 1/2 (see, e.g., \citealt{Ryu08,BJ14}). It also has crucial impacts on astrophysical phenomena in the ICM, including amplification of weak seed magnetic fields (e.g., \citealt{Schekochihin04,Ryu08,Cho14}).

Magnetic fields permeated in turbulence appear in a variety of astrophysical objects and have huge impacts on them. For instance, they significantly affect evolution of molecular clouds and fragmentation process of cores in the ISM (see \citealt{HI19} for a review). Moreover, magnetic fields and turbulence in the ICM can accelerate cosmic ray electrons and protons (see \citealt{BJ14} and references therein).

The strength of magnetic fields varies from the ISM to the ICM. Observations of dust polarization in interstellar molecular cloud cores suggest that the strength is typically $\sim$ mG (see \citealt{Cr2012} and references therein). On the other hand, observations of synchrotron emission from galaxy clusters and Faraday rotation reveal that magnetic fields of the order of $\mu$G exist in the ICM (see, e.g., \citealt{CT02, GF04, Ryu12}). 

Turbulence in astrophysical media has a mixture of both solenoidal ($\nabla \cdot \mathbf{v}$ = 0) and compressive ($\nabla \times \mathbf{v}$ = 0) velocity components, where $\mathbf{v}$ is velocity. Similarly, turbulence driving force ($\mathbf{f}$) can also have solenoidal ($\nabla \cdot \mathbf{f}$ = 0) and compressive ($\nabla \times \mathbf{f}$ = 0) components. In turbulence studies, solenoidal driving has been predominantly used. However, there also have been multiple literatures that make use of compressive driving. Earlier studies have shown that driving mechanism affects characteristics of turbulence or related physical phenomena. For example, compressively driven turbulence has a wider probability density function (PDF) of density \citep{F08, F10} and has more intermittent structures \citep{Fe09,YHS19} than solenoidally driven one. 

 When solenoidal driving forces turbulence, it is evident that solenoidal energy dominates over compressive one. This is independent of both its Mach number (see, e.g., \citealt{Berto2001,F11}) and of its degree of magnetization (for hydrodynamic turbulence, see \citealt{Kritsuk07,F10,F13}; for magnetohydrodyanmic turbulence, see \citealt{Boldrev2002,Cho03,Kritsuk10,F11,Porter15}).

When compressive driving forces turbulence, previous numerical studies have shown that solenoidal velocity component in such turbulence can be generated at shocks and amplified by vortex stretching \citep{F11,Porter15}. In particular, \citet{F11} found that solenoidal energy accounts for up to $\sim$ 40\% of total kinetic energy when turbulence is supersonic and a weak mean magnetic field is present. However, since they considered only a single mean magnetic field strength, the dependence of the solenoidal ratio (i.e., the ratio of solenoidal to total kinetic energies) on the mean magnetic field strength in compressively driven turbulence is not fully determined yet. In this regard, we mainly aim at determining the solenoidal ratio in compressively driven turbulence by taking various strengths of the mean magnetic field into account. We also investigate the role of the magnetic fields in the generation of solenoidal modes.  

Amplification of magnetic field on scales comparable to or smaller than the driving scale by turbulent motions is known as small-scale turbulence dynamo. In this process, turbulent motions stretch, twist, and fold magnetic field lines, which in turn results in conversion of kinetic energy of turbulence to magnetic energy (see, e.g., \citealt{1950RSPSA.201..405B,CV00,Haugen04,Schekochihin04,2005PhR...417....1B,Schekochihin07,Cho09} for details; see also Appendix in \citealt{Cho14}). Since the dynamo action is mainly achieved by solenoidal motions of turbulence, it is apparent for types of driving to have influence on the process. For solenoidal driving, comprehensive studies exist related to the turbulence dynamo. Those studies have numerically shown that solenoidal driving efficiently amplifies small-scale magnetic field via field line stretching, and the resulting magnetic energy becomes comparable to kinetic energy at saturation \citep{CV00,Haugen03,Schekochihin04,Schekochihin07,Ryu08,Cho09,F11,Cho12,Porter15}. On the other hand, compressive driving cannot efficiently excite small-scale magnetic field because it does not produce enough solenoidal velocity components to amplify the magnetic field. As a result, the fully excited field is dynamically insignificant \citep{F11, Porter15}. Apart from its inefficiency, earlier studies have revealed relatively fewer facts for turbulence dynamo in compressively driven turbulence. In this paper, we provide more comprehensive study on this topic. To be specific, we consider a wide range of the mean magnetic field strengths and try to estimate an upper limit of magnetic saturation level in compressively driven turbulence when numerical resolution is very high and the mean magnetic field is very weak. 

The paper is organized as follows. We explain our numerical method in Section \ref{sec:sec2}. We present results related to generation of solenoidal modes and small-scale magnetic fields in Sections \ref{sec:sec3} and \ref{sec:sec4}, respectively. We discuss our findings in Sections \ref{sec:sec5} and \ref{sec:sec6}, and give summary in Section \ref{sec:sec7}.

\begin{deluxetable*}{ccccccccc}
\tablenum{1}
\tabletypesize{\scriptsize}
\tablecaption{Results of simulations\label{tab:tab1}}
       \tablewidth{0.5\textwidth}
\tablehead{
\colhead{Run} & 
\colhead{Driving} & 
\colhead{Resolution} &     
\colhead{$\mach$ \tablenotemark{a}} &
\colhead{$\bzero$ \tablenotemark{b}} &
\colhead{$\vsolsq/\vtotsq$ \tablenotemark{c}} &
\colhead{$\bsq/\vtotsq$ \tablenotemark{d}} & 
\colhead{($t_1$,$t_2$) \tablenotemark{e}}  &
}
\startdata
F1024MS1-$\bzero$0.1 & Finite-correlated compressive & $1024^3$ & $\sim$ 1  & 0.1 & 0.262 & 0.194 & (3,5.5) \\
\hline
F512MS0.5-$\bzero$0.05 &  & $512^3$  & $\sim$ 0.5  & 0.05 & 0.099 & 0.059 & (5,10) \\
F512MS0.5-$\bzero$0.1 &  & $512^3$  & $\sim$ 0.5  & 0.1 & 0.099 & 0.091 & (5,10) \\
F512MS1-$\bzero$0.001 &  & $512^3$  & $\sim$ 1  & 0.001 & 0.249 & 0.033 & (40,70) \\
F512MS1-$\bzero$0.01 &  & $512^3$  & $\sim$ 1  & 0.01 & 0.246 & 0.044 & (20,30) \\
F512MS1-$\bzero$0.05 &  & $512^3$  & $\sim$ 1  & 0.05 & 0.280 & 0.091 & (3,5.5)  \\
F512MS1-$\bzero$0.1 &  & $512^3$  & $\sim$ 1  & 0.1 & 0.275 & 0.162 & (3,5.5) \\
F512MS1-$\bzero$0.2 &  & $512^3$  & $\sim$ 1  & 0.2 & 0.307 & 0.250 & (3,5.5) \\
F512MS1-$\bzero$0.6 &  & $512^3$  & $\sim$ 1  & 0.6 & 0.491 & 0.441 & (3,5.5) \\
F512MS1-$\bzero$1  &  & $512^3$  & $\sim$ 1  & 1 & 0.654 & 0.526 & (3,6)  \\
F512MS1-Hydro &  & $512^3$ & $\sim$ 1  & - & 0.317 & - & (3,6)  \\
\hline
F256MS0.5-$\bzero$0.001 & & $256^3$  & $\sim$ 0.5  & 0.001 & 0.147 & 0.004 & (140,180)  \\
F256MS0.5-$\bzero$0.01 & & $256^3$  & $\sim$ 0.5  & 0.01 & 0.130 & 0.011 & (30,75)  \\
F256MS0.5-$\bzero$0.05 & & $256^3$ & $\sim$ 0.5  & 0.05 & 0.096 & 0.042 & (8,12)  \\
F256MS0.5-$\bzero$0.1  &  & $256^3$  & $\sim$ 0.5  & 0.1 & 0.093 & 0.072 & (8,12) \\
F256MS0.5-$\bzero$0.2 &  & $256^3$  & $\sim$ 0.5  & 0.2 & 0.101 & 0.102 & (8,12) \\
F256MS0.5-$\bzero$1 &  & $256^3$  & $\sim$ 0.5  & 1 & 0.447 & 0.475 & (3,10)  \\
F256MS1-$\bzero$0.001 &  & $256^3$  & $\sim$ 1  & 0.001 & 0.245 & 0.015 & (80,160)  \\
F256MS1-$\bzero$0.01 &  & $256^3$  & $\sim$ 1  & 0.01 & 0.237 & 0.026 & (30,160) \\
F256MS1-$\bzero$0.05 &  & $256^3$  & $\sim$ 1  & 0.05 & 0.272 & 0.065 & (3.5.5) \\
F256MS1-$\bzero$0.1 &  & $256^3$  & $\sim$ 1  & 0.1 & 0.262 & 0.128 & (3,5.5) \\
F256MS1-$\bzero$0.2 &  & $256^3$  & $\sim$ 1  & 0.2 & 0.284 & 0.209 & (3,5.5) \\
F256MS1-$\bzero$0.6 &  & $256^3$  & $\sim$ 1  & 0.6 & 0.459 & 0.415 & (3,5.5) \\
F256MS1-$\bzero$1  &  & $256^3$  & $\sim$ 1  & 1 & 0.615 & 0.504 & (3,6)  \\
F256MS3-$\bzero$0.001 &  & $256^3$  & $\sim$ 3  & 0.001 & 0.332 & 0.018 & (80,160)  \\
F256MS3-$\bzero$0.01 &  & $256^3$  & $\sim$ 3  & 0.01 & 0.336 & 0.024 & (30,50)  \\
F256MS3-$\bzero$0.05 &  & $256^3$  & $\sim$ 3  & 0.05 & 0.367 & 0.084 & (5,12)  \\
F256MS3-$\bzero$0.1 &  & $256^3$  & $\sim$ 3  & 0.1 & 0.452 & 0.157 & (5,12)  \\
F256MS3-$\bzero$0.2 &  & $256^3$  & $\sim$ 3  & 0.2 & 0.560 & 0.275 & (5,12)  \\
F256MS3-$\bzero$1  &  & $256^3$ & $\sim$ 3  & 1 & 0.761 & 0.423 & (3,10)  \\
F256MS10-$\bzero$1  &  & $256^3$ & $\sim$ 10  & 1 & 0.761 & 0.355 & (3,10) \\
F256MS0.5-Hydro &  & $256^3$ & $\sim$ 0.5  & - & 0.132 & - & (3,10)  \\
F256MS1-Hydro &  & $256^3$ & $\sim$ 1  & - & 0.300 & - & (3,6)  \\
F256MS3-Hydro &  & $256^3$ & $\sim$ 3  & - & 0.350 & - & (3,10)  \\
\hline
\hline
D512MS1-$\bzero$0.01 & Delta-correlated compressive & $512^3$  & $\sim$ 1  & 0.01 & 0.161 & 0.023 & (20,30) \\
D512MS1-$\bzero$1  &  & $512^3$  & $\sim$ 1  & 1 & 0.471 & 0.511 & (3,6) \\
\hline
D256MS0.5-$\bzero$0.01 &  & $256^3$ & $\sim$ 0.5  & 0.01 & 0.063 & 0.006 & (70,95)  \\
D256MS1-$\bzero$0.01 &  & $256^3$ & $\sim$ 1  & 0.01 & 0.130 & 0.012 & (40,100)  \\
D256MS3-$\bzero$0.01 &  & $256^3$ & $\sim$ 3  & 0.01 & 0.259 & 0.018 & (40,80) \\
D256MS0.5-$\bzero$1  &  & $256^3$ & $\sim$ 0.5  & 1 & 0.309 & 0.376 & (3,10) \\
D256MS1-$\bzero$1  &  & $256^3$ & $\sim$ 1  & 1 & 0.417 & 0.467 & (3,6) \\
D256MS3-$\bzero$1  &  & $256^3$ & $\sim$ 3  & 1 & 0.590 & 0.384 & (3,10) \\
D256MS10-$\bzero$1  & & $256^3$ & $\sim$ 10  & 1 & 0.641 & 0.306 & (3,10)  \\
D256MS0.5-Hydro &  & $256^3$ & $\sim$ 0.5  & - & 0.066 & - & (3,10)   \\
D256MS1-Hydro &  & $256^3$ & $\sim$ 1  & - & 0.165 & - & (3,10)  \\
D256MS3-Hydro &  & $256^3$ & $\sim$ 3  & - & 0.255 & - & (3,10)  \\
\hline
\hline
Sol-F256MS1-$\bzero$0.001 & Finite-correlated solenoidal & $256^3$ & $\sim$ 1  & 0.001 & 0.958 & 0.262 & (30,80) \\
\hline
Sol-D256MS1-$\bzero$0.001 & Delta-correlated solenoidal & $256^3$ & $\sim$ 1  & 0.001 & 0.886 & 0.091 & (30,80)  
\enddata
\tablenotetext{a}{The sonic Mach number.}
\tablenotetext{b}{The strength of the mean magnetic field, which in our units is actually the Alfven speed. Note that in our simulations, the rms velocity is $\sim$ 1.}
\tablenotetext{c}{The solenoidal ratio after saturation. Here $\mathbf{v_{\textrm{sol}}}$ is solenoidal velocity component and $\mathbf{v_{\textrm{tot}}}$ is total velocity.}
\tablenotetext{d}{The magnetic saturation level after saturation. Here $b^2$ is small-scale magnetic energy density.}
\tablenotetext{e}{The time interval in the unit of large-eddy turnover time ($t_{\textrm{ed}}$) for averaging the physical quantities.}
\end{deluxetable*}

\section{Numerical Method}\label{sec:sec2}


\subsection{Numerical Code\label{sec:sec2.1}}

We use an Essentially Non-Oscillatory (ENO) scheme \citep[see][]{CL02} to solve the ideal magnetohydrodynamic (MHD) equations in a periodic box of size 2$\pi$:

\begin{equation}
\label{eq:eq1}
\frac{\partial{\rho}}{\partial{t}}+\nabla \cdot \left(\rho\mathbf{v}\right)=0,
\end{equation}

\begin{equation}\label{eq:eq2}
\rho\left(\frac{\partial{\mathbf{v}}}{\partial{t}}+\mathbf{v} \cdot \nabla \mathbf{v}\right) +c_s^2 \nabla \rho -\left(\nabla \times \mathbf{B}\right) \times \mathbf{B}  =  \rho \mathbf{f},
\end{equation}

\begin{equation}\label{eq:eq3}
\frac{\partial{\mathbf{B}}}{\partial{t}}-\nabla \times \left(\mathbf{v} \times \mathbf{B}\right) = 0,
\end{equation}

\begin{equation}\label{eq:eq4}
\nabla \cdot \mathbf{B} = 0,  
\end{equation}
with an isothermal equation of state $p$ = $c_s^2\rho$, where $c_s$ is the sound speed, $\rho$ is the density, $p$ is the gas pressure, $\mathbf{f}$ is a driving force (see Section \ref{sec:sec2.2} for details),  $\mathbf{v}$ is the velocity, and $\mathbf{B}$ is the magnetic field divided by $\sqrt{4\pi}$. For hydrodynamic (HD) simulations, we only solve Equations (\ref{eq:eq1}) and (\ref{eq:eq2}) with $\mathbf{B}$ = 0. The magnetic field consists of two components: a uniform mean field ($\mathbf{\bzero}$) and a fluctuating random field ($\mathbf{b}$), so that $\mathbf{B}$ = $\mathbf{\bzero}$ + $\mathbf{b}$. At t = 0, the density and velocity are set to 1 and zero, respectively, to assume a static medium with a constant density. In cases of MHD, only a mean magnetic field exists at the beginning.

\subsection{Forcing Schemes\label{sec:sec2.2}}

In our simulations, we drive turbulence in Fourier space using either solenoidal ($\nabla \cdot \mathbf{f}$ = 0) or compressive ($\nabla \times \mathbf{f}$ = 0) driving. We consider two different types of driving in terms of different correlation timescale of forcing vectors: a finite-correlated driving and a delta-correlated driving. In the former, forcing vectors continuously change with a correlation timescale comparable to the large-eddy turnover time. In the latter, both the direction and amplitude of driving change in a very short timescale $\Delta t = 0.001$ in code units, which roughly corresponds to a few thousandths of the large-eddy turnover time (see \citealt{YHS16} for details). In summary, we consider the following forcing schemes:
\begin{enumerate}
	\item Finite-correlated compressive driving.
	\item Finite-correlated solenoidal driving.
	\item Delta-correlated compressive driving.
	\item Delta-correlated solenoidal driving.
\end{enumerate}

\begin{figure*}[ht!]
\centering
\includegraphics[scale=0.15]{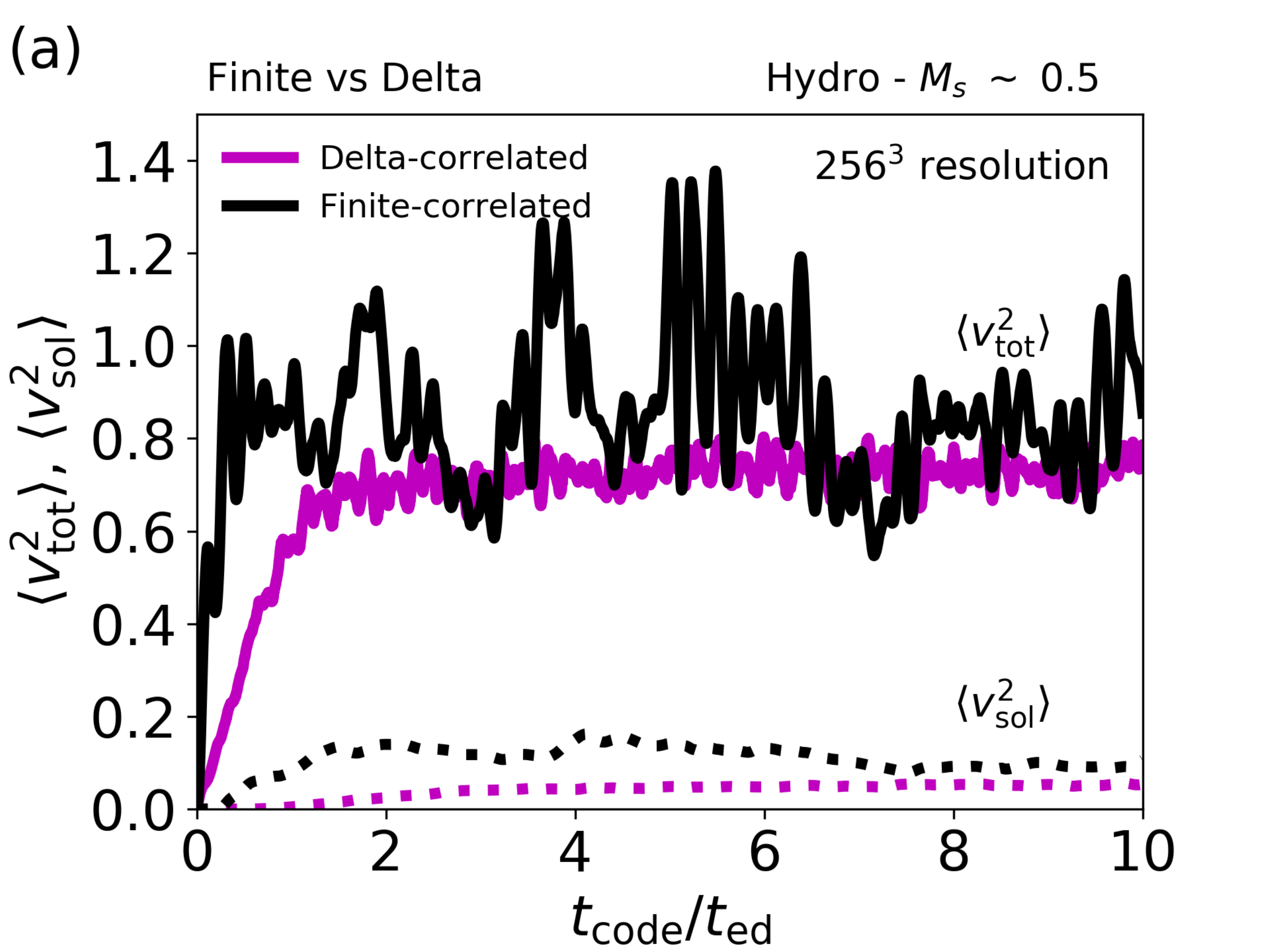}
\includegraphics[scale=0.15]{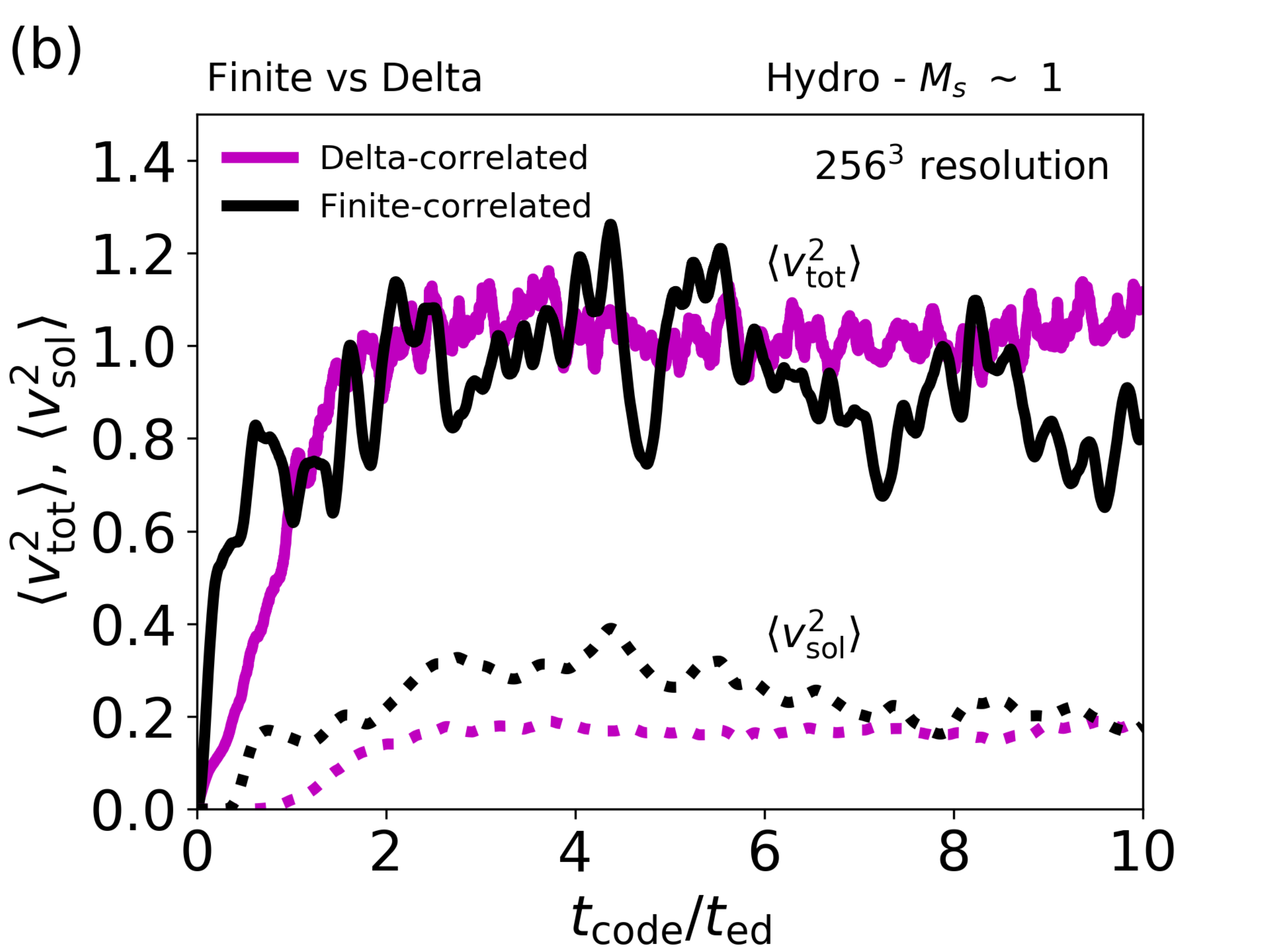} 
\includegraphics[scale=0.15]{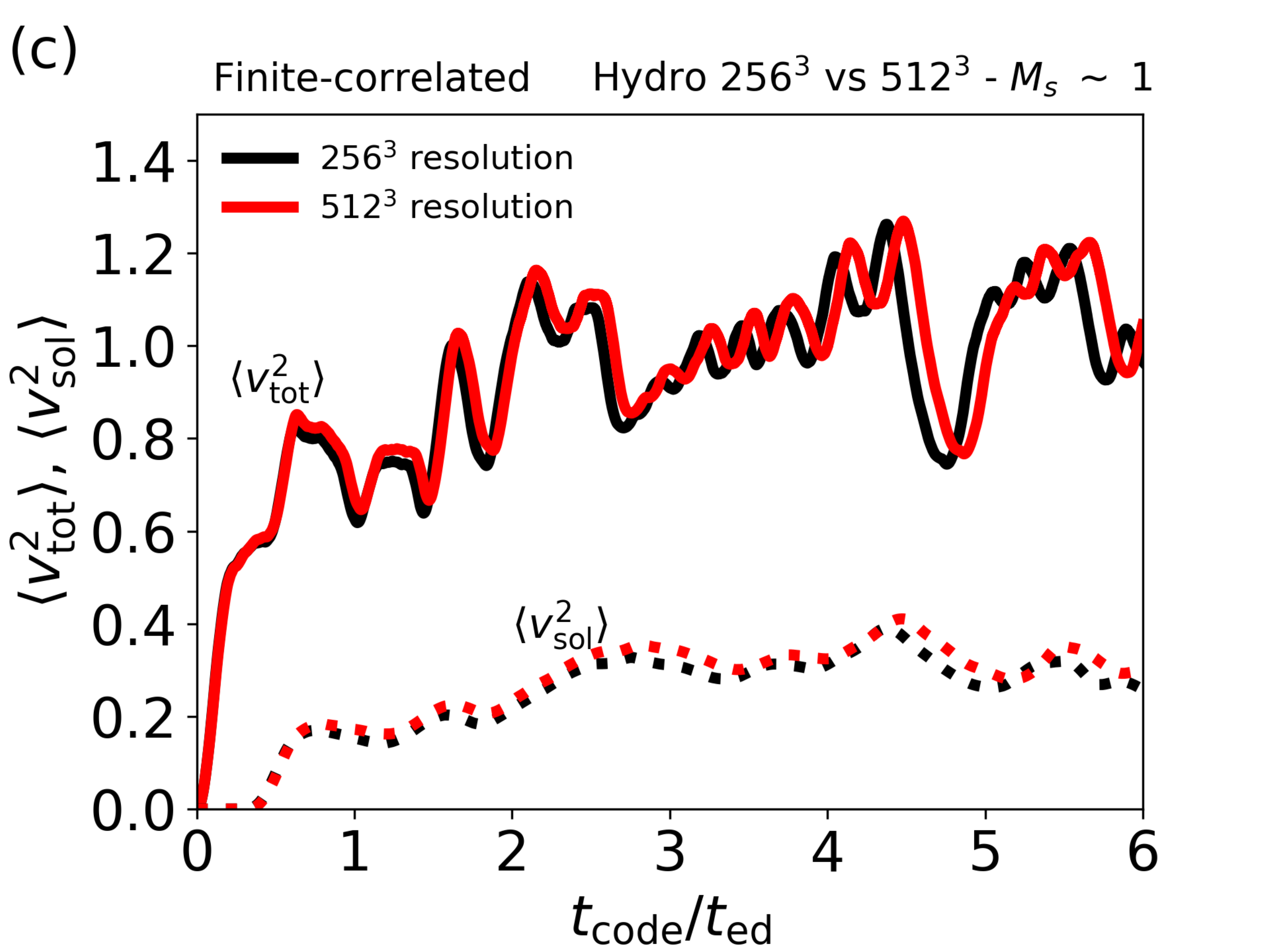}
\caption{Time evolution of $\langle \vtotsq \rangle$ (solid curves) and $\langle \vsolsq \rangle$ (dotted curves) for HD turbulence driven by compressive driving, where $\langle \cdots \rangle$ denotes spatial average. (a) $\mach$ $\sim$ 0.5 and $256^3$ resolution. (b) $\mach$ $\sim$ 1 and $256^3$ resolution. (c) Resolution study only for the finite-correlated compressive driving and $\mach$ $\sim$ 1. Black and magenta curves in the left and the middle panels represent the finite-correlated and the delta-correlated compressive drivings, respectively. Black and red curves in the right panel correspond to $256^3$ and $512^3$ resolutions, respectively.
\label{fig:fig1}}
\end{figure*}

\subsection{Simulations\label{sec:sec2.3}}
We use up to $1024^3$ grid points in our periodic computational box. In all simulations, energy injection peaks at k $\approx$ 2.5, where k is the wavenumber. The strength of the mean magnetic field ($B_0$) ranges from 0.001 to 1.0 in MHD turbulence simulations, which is actually the Alfven speed of the mean magnetic field. In our simulations, the rms velocity ($v_{\textrm{rms}}$) is roughly one when turbulence is fully developed. We change the isothermal sound speed $c_s$ to control the sonic Mach number $\mach$ $\equiv$ $v_{\textrm{rms}}/c_s$. The resulting sonic Mach number ranges from $\sim$ 0.5 to $\sim$ 10. When we present time evolutions of physical quantities, we use a normalized time: $t \ = t_{\textrm{code}}/t_{\textrm{ed}}$. Here $t_{\textrm{code}}$ is time in code units, and $t_{\textrm{ed}} = L_f/v_{\textrm{rms}}$ is large-eddy turnover time. The driving scale of turbulence $L_{f}$ is about 2.5 times smaller than the computational box.   

Table \ref{tab:tab1} lists our simulations. We use the notation $X_{1}X_{2}$MS$X_{3}$-$B_{0}X_{4}$, where $X_{1}$ = F or D refers to either the finite-correlated or the delta-correlated driving\footnote{We additionally consider two simulations with solenoidal driving. They are notated by Sol-D256MS1-$\bzero$0.001 and Sol-F256MS1-$\bzero$0.001, respectively. All other simulations, which do not start with Sol-, are for compressive driving (see Table \ref{tab:tab1}).}; $X_{2}$ =  256, 512, or 1024 refers to the number of grid points in each spatial direction; $X_{3}$ = 0.5, 1, 3, or 10 refers to the sonic Mach number; $X_{4}$ = 0.001, 0.01, 0.05, 0.1, 0.2, 0.6, or 1.0 refers to the strength of the mean magnetic field, and $B_0 X_{4}$ = Hydro refers to HD simulation.

We use the following notations in this paper:
\begin{enumerate}
	\item $\mathbf{v_{\textrm{tot}}}$ ($\equiv$ $\mathbf{v_{\textrm{sol}}}$ + $\mathbf{v_{\textrm{comp}}}$): total velocity.
	\item $\mathbf{v_{\textrm{sol}}}$, $\mathbf{v_{\textrm{comp}}}$: solenoidal and compressive velocity components, respectively.
	\item $\vsolsq/\vtotsq$ ($\equiv$ $\langle \vsolsq \rangle/ \langle\vtotsq \rangle$): we refer to this as \textit{solenoidal ratio}. Here $\langle \cdots \rangle$ denotes spatial average.
	\item $B$ ($\equiv$ $\sqrt{\bzero^2 + \bsq}$): total magnetic field strength.
	\item $\bzero$: mean magnetic field strength. Note that, in our units, $B_0$ is actually the Alfven speed of the mean field.
	\item $b$ ($\equiv$ $\sqrt{\langle b^2 \rangle}$): random magnetic field strength.
	\item $\bsq/\vtotsq$ ($\equiv$ $\langle \bsq\rangle/\langle \vtotsq \rangle$): we refer to this as \textit{magnetic saturation level}.
\end{enumerate}


\begin{figure}[]
\includegraphics[scale=0.25]{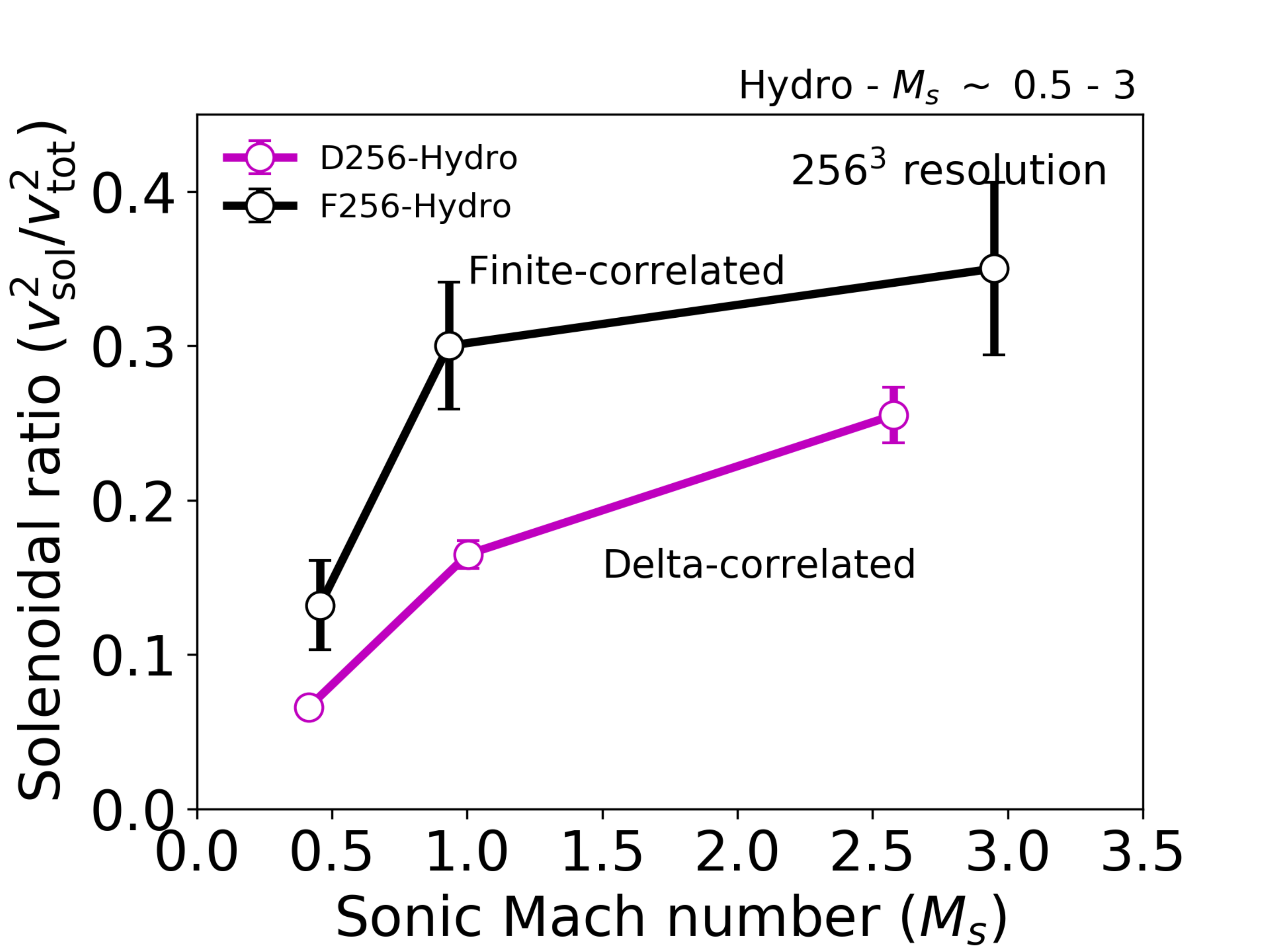}
\caption{Average values of the solenoidal ratio ($\vsolsq/\vtotsq$) as a function of the sonic Mach number ($\mach$) for the HD simulations. Black and magenta circles indicate the finite-correlated and the delta-correlated compressive drivings for $256^3$ resolution, respectively. The error bars represent standard deviations. In Table \ref{tab:tab1}, the average values and the time intervals for taking average are shown. 
\label{fig:fig2}}
\end{figure}

\section{Generation of solenoidal modes\label{sec:sec3}}
In this section, we investigate generation of solenoidal velocity component in compressively driven turbulence. In Section \ref{sec:sec3.1}, we consider HD turbulence. We describe effects of the sonic Mach number ($\mach$) and the mean magnetic field ($\bzero$) for MHD turbulence in Sections \ref{sec:sec3.2} and \ref{sec:sec3.3}, respectively.

\subsection{Effects of the Sonic Mach Number in HD Turbulence\label{sec:sec3.1}}
Figure \ref{fig:fig1} shows time evolution of $\langle \vtotsq \rangle$ and $\langle \vsolsq \rangle$ in HD turbulence driven by compressive driving. Figures \ref{fig:fig1}(a) and \ref{fig:fig1}(b) are for comparison of the finite-correlated compressive driving with the delta-correlated compressive driving at $256^3$ resolution, which are represented as black and magenta curves, respectively. Figure \ref{fig:fig1}(c) shows resolution study for the finite-correlated compressive driving, in which black and red curves denote $256^3$ and $512^3$ resolutions, respectively. The sonic Mach number is $\sim$ 0.5 in the left panel, and $\sim$ 1 in the middle and the right panels. 

 As we can see from the figure, turbulence seems to saturate before approximately $3t_{\textrm{ed}}$, and the level of $\langle \vsolsq \rangle$ at saturation stage  is much lower than that of $\langle \vtotsq \rangle$ in all simulations presented. In addition, we can clearly see from Figures \ref{fig:fig1}(a) and \ref{fig:fig1}(b) that the finite-correlated compressive driving (see black curves) yields nearly same levels of $\langle \vtotsq \rangle $ regardless of $\mach$, but $\langle \vsolsq \rangle$ for $M_s$ $\sim$ 1 is larger than that for $M_s$ $\sim$ 0.5. For the delta-correlated compressive driving (see magenta curves), the level of both $\langle \vtotsq\rangle $ and $\langle \vsolsq\rangle $ for $M_s$ $\sim$ 1 is larger than that for $M_s$ $\sim$ 0.5. Regarding numerical resolution effect, Figure \ref{fig:fig1}(c) clearly shows that the time evolutions in both resolutions are virtually identical, which means that the generation of solenoidal modes is very insensitive to the numerical resolution.
 
Figure \ref{fig:fig2} shows average values of the solenoidal ratio as a function of $M_s$. Black and magenta circles correspond to the finite-correlated and the delta-correlated compressive drivings, respectively. The figure apparently presents that solenoidal ratios increase as $M_s$ increases. When $M_s$ $\sim$ 3, the solenoidal ratio is $\sim$ 0.35. For comparison, \citet{F10}, who used a finite-correlated driving, obtained the ratio of $\sim$ 0.4 in compressively driven HD turbulence at $M_s$ $\sim$ 5. The higher ratio they obtained may stem from a higher value of $M_s$ than ours. On top of that, when $M_s$ is similar, solenoidal ratios of the finite-correlated compressive driving are always higher than those of the delta-correlated compressive driving. Therefore, in HD turbulence driven by compressive driving, generation of solenoidal motions is dependent on both $M_s$ and driving schemes. 

\begin{figure*}[t!]
\centering
\includegraphics[scale=0.15]{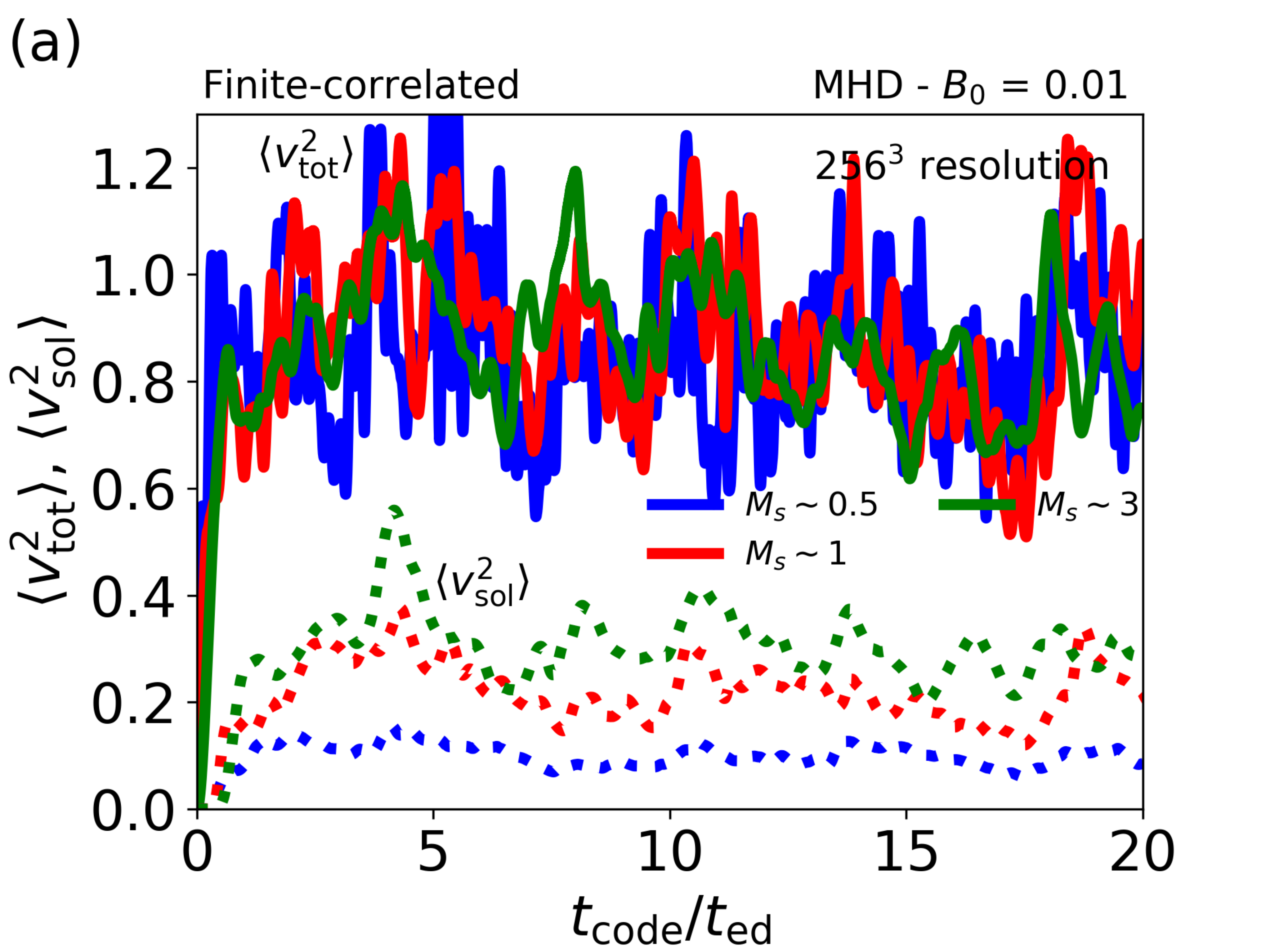}
\includegraphics[scale=0.15]{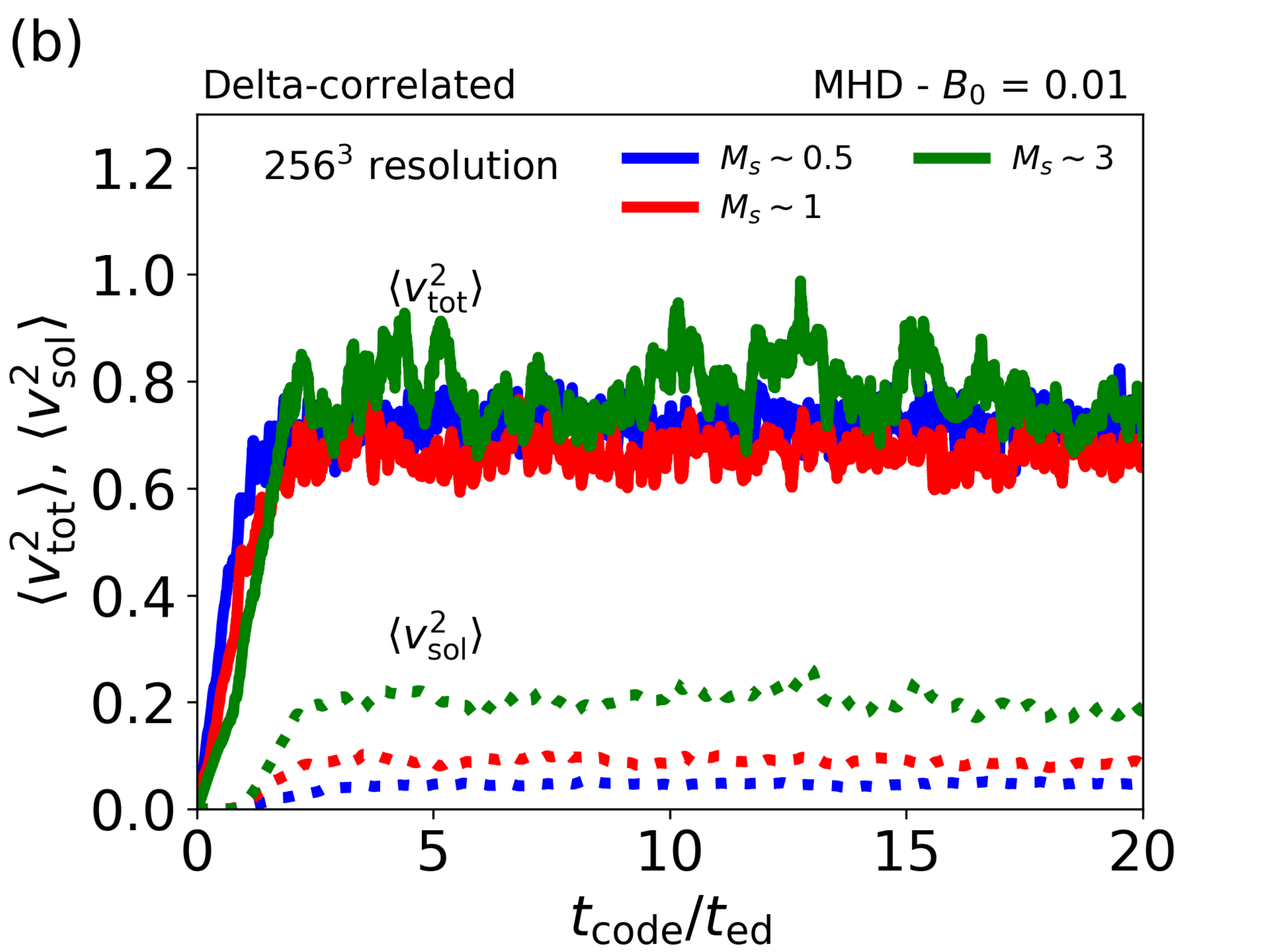} 
\includegraphics[scale=0.15]{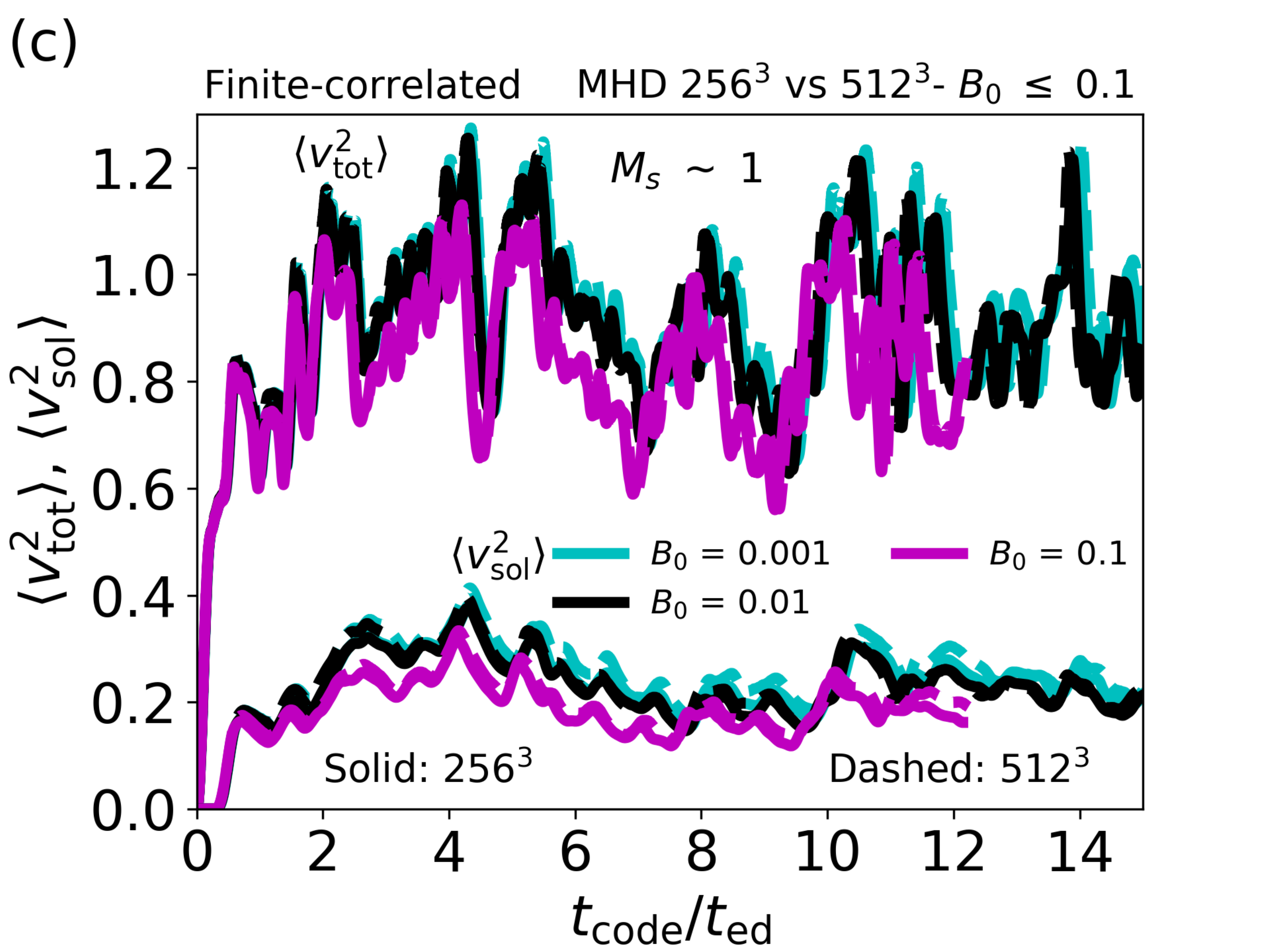} 
\caption{Time evolution of $\langle \vtotsq\rangle$ and $\langle \vsolsq\rangle$ for MHD turbulence simulations with $\bzero$ $\leq$ 0.1. (a) The finite-correlated compressive driving with $B_0$ = 0.01. (b) The delta-correlated compressive driving with $B_0$ = 0.01. (c) Resolution study for $B_0$ $\leq$ 0.1 and $\mach$ $\sim$ 1. Blue, red, and green solid (dotted) curves in the left and the middle panels represent $\langle \vtotsq\rangle$ ($\langle \vsolsq\rangle$) for $\mach$ $\sim$ 0.5, $\sim$ 1, and $\sim$ 3, respectively. In the right panel, cyan, black, and magenta solid (dashed) curves correspond to $B_0$ = 0.001, 0.01, and 0.1 for $256^3$ ($512^3$) resolution, respectively. Only the finite-correlated driving is considered in the right panel.
\label{fig:fig3}}
\end{figure*}

\begin{figure*}
\centering
\includegraphics[scale=0.15]{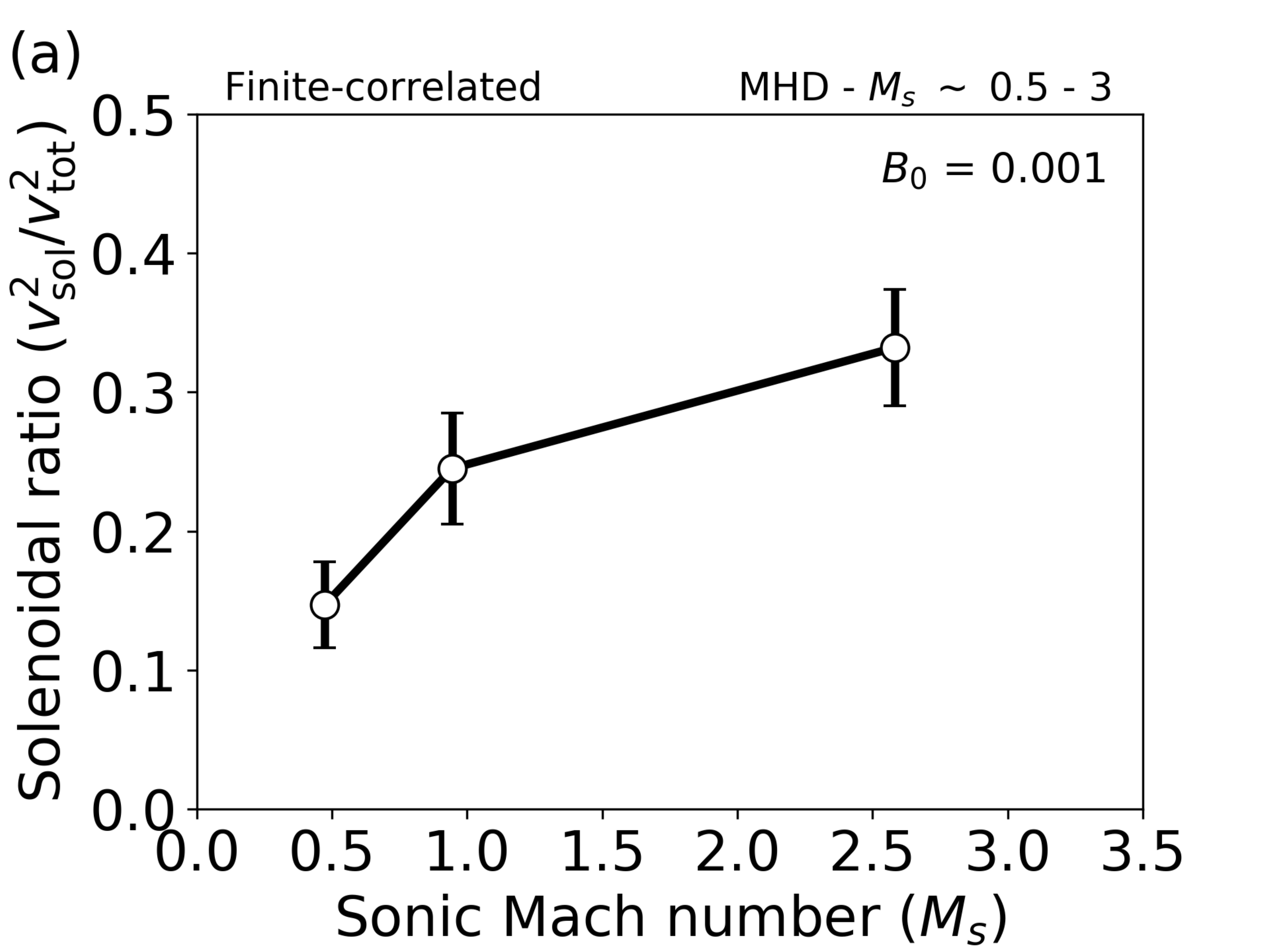}
\includegraphics[scale=0.15]{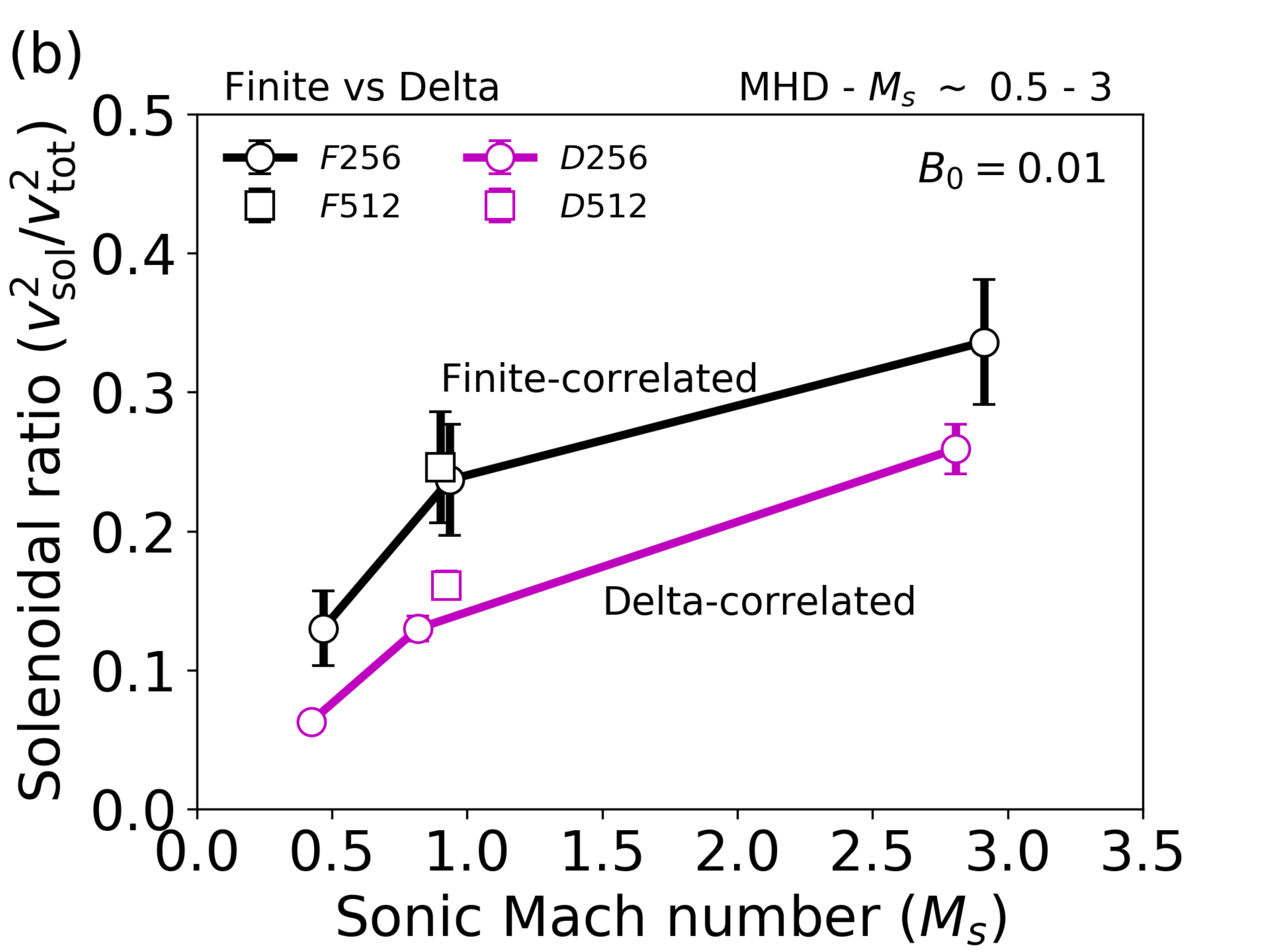} 
\includegraphics[scale=0.15]{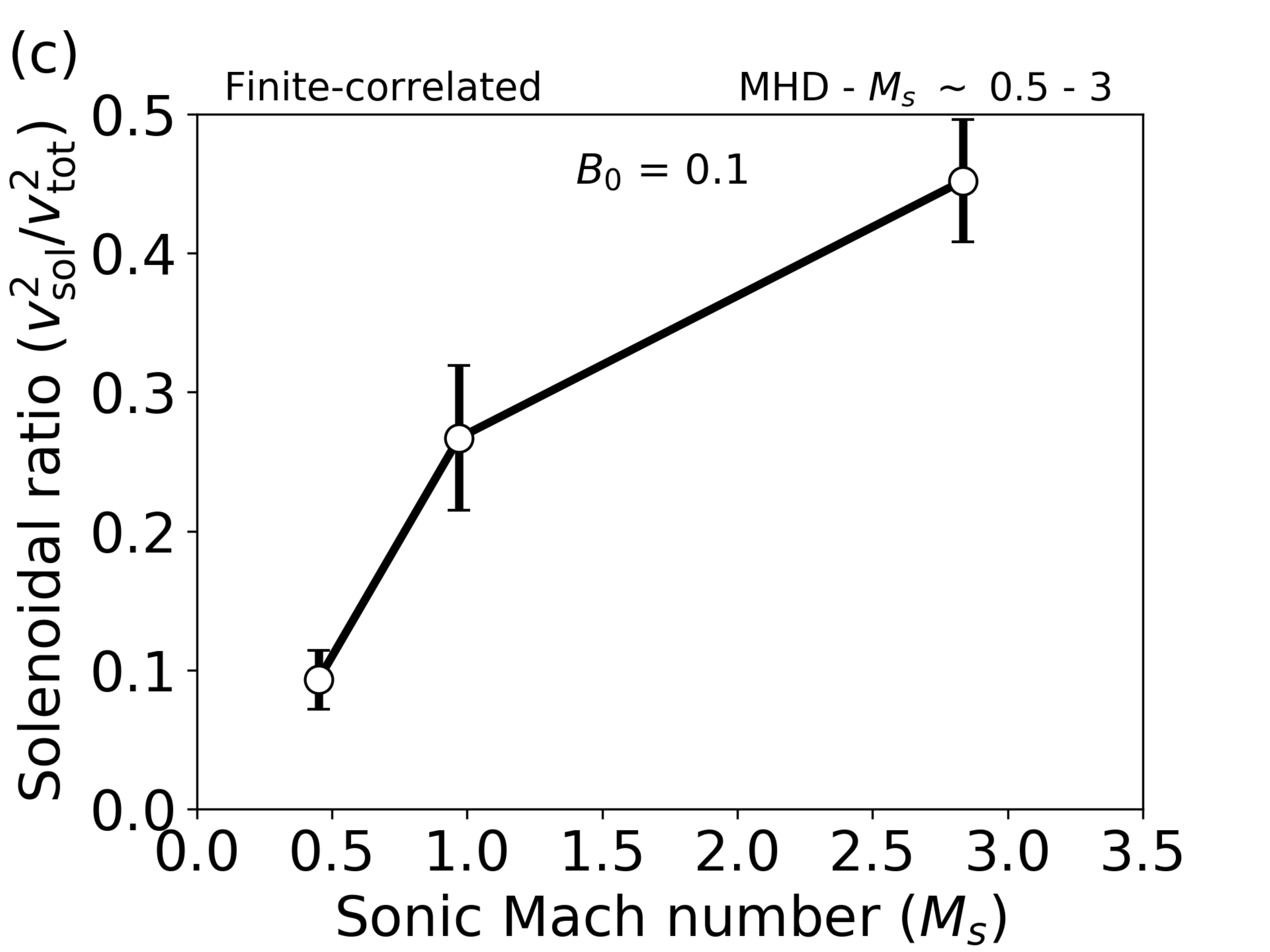} 
\caption{Average values of the solenoidal ratio ($\vsolsq/\vtotsq$) as a function of the sonic Mach number ($M_s$) for the MHD simulations with $B_{0}$ $\leq$ 0.1. (a) $B_0$ = 0.001. (b) $B_0$ = 0.01. (c) $B_0$ = 0.1. In the left ($B_0$ = 0.001) and the right panels ($B_0$ = 0.1), we present the ratio of the finite-correlated compressive driving only. In the middle panel ($B_0$ = 0.01), black and magenta circles (squares) indicate the finite-correlated and the delta-correlated compressive drivings for $256^3$ ($512^3$) resolution, respectively. The error bars in each panel denote standard deviations. In Table \ref{tab:tab1}, the average values and the time intervals for taking average are shown. 
\label{fig:fig4}}
\end{figure*}


\subsection{Effects of the Sonic Mach Number in MHD Turbulence\label{sec:sec3.2}}
Here we mainly study how the sonic Mach number $M_s$ influences generation of solenoidal motions in MHD turbulence driven by compressive driving. We study effects of driving schemes and numerical resolution as well. We consider weak mean magnetic field cases ($B_0$ $\leq$ 0.1) in Section \ref{sec:sec3.2.1} and a strong mean magnetic field case ($B_0$ = 1.0) in Section \ref{sec:sec3.2.2}.

\subsubsection{Weak $B_0$ Cases ($B_0$ $\leq$ 0.1)\label{sec:sec3.2.1}}
Figures \ref{fig:fig3} shows time evolution of $\langle \vtotsq\rangle $ and $\langle \vsolsq\rangle$ in MHD turbulence driven by compressive driving. Blue, red, and green curves in Figures \ref{fig:fig3}(a) and \ref{fig:fig3}(b) correspond to $M_s$ $\sim$ 0.5, $\sim$ 1, and $\sim$ 3, respectively. We separately consider the finite-correlated and the delta-correlated compressive drivings in Figures \ref{fig:fig3}(a) and \ref{fig:fig3}(b), respectively. According to those two figures, the evolution of $\langle \vtotsq\rangle $ is similar irrespective of $M_s$. However, the level of $\langle \vsolsq\rangle$ at saturation increases as $M_s$ increases in both driving schemes.

Figure \ref{fig:fig3}(c) shows the effects of numerical resolution and $B_0$ in the case of $M_s$ $\sim$ 1 and the finite-correlated compressive driving. We plot results of six simulations. In the figure, different colors of curves represent different values of $B_0$: cyan, black, and magenta colors correspond to $B_0$ = 0.001, 0.01, and 0.1, respectively. Each color has two different line styles: solid and dashed curves correspond to $256^3$ and $512^3$ resolutions, respectively. As in the case of HD turbulence, the resolution effect seems insignificant: in the figure, solid and dashed curves with the same color virtually coincide, which means that results are nearly resolution-independent. In addition, time evolution of all curves looks similar, which means that the effect of $B_0$ also seems insignificant for $M_s$ $\sim$ 1 and $B_0$ $\leq$ 0.1  (see Section \ref{sec:sec3.3} for $B_0$ higher than 0.1).  

\begin{figure*}[t!]
\centering
\includegraphics[scale=0.15]{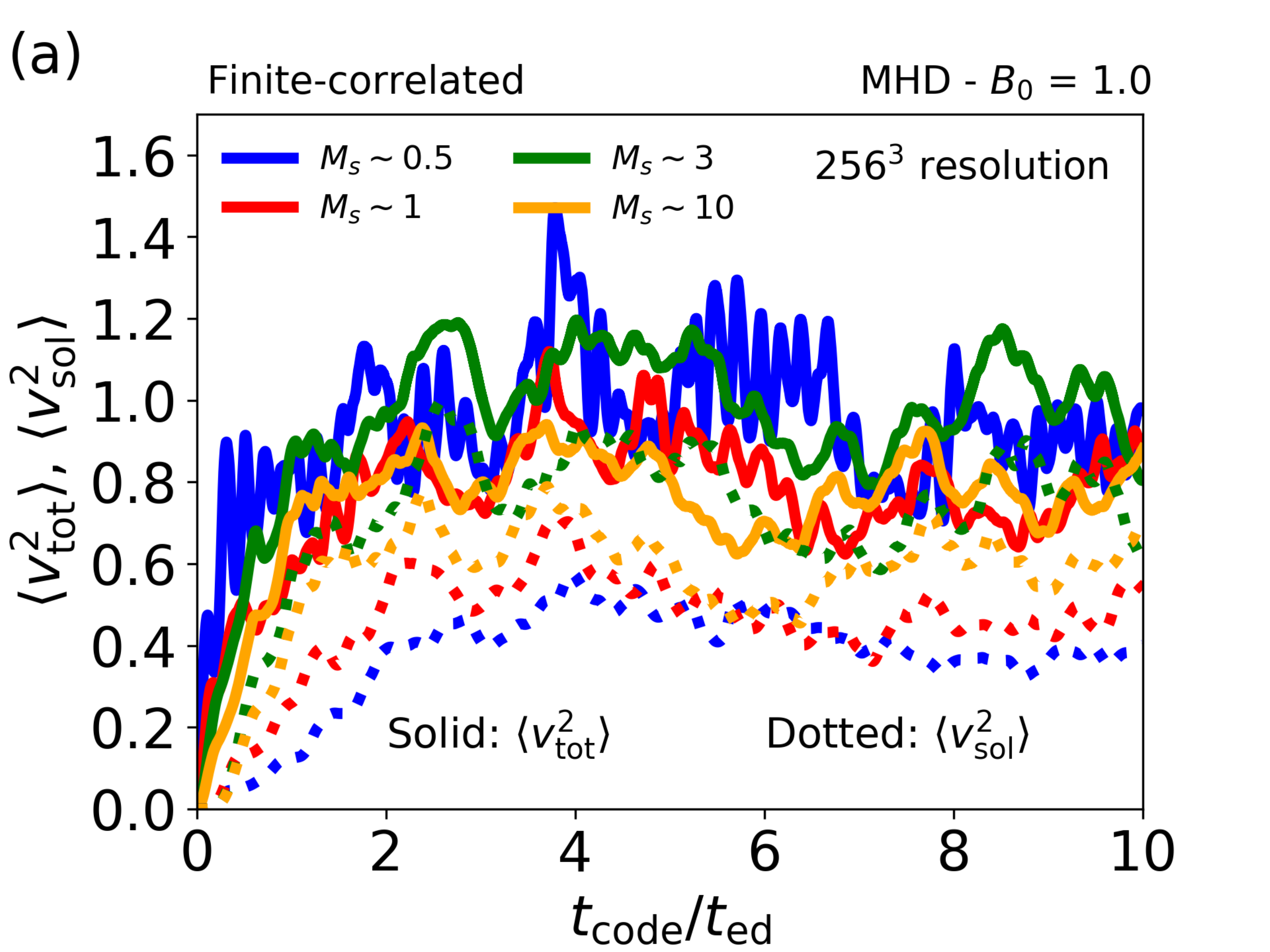}
\includegraphics[scale=0.15]{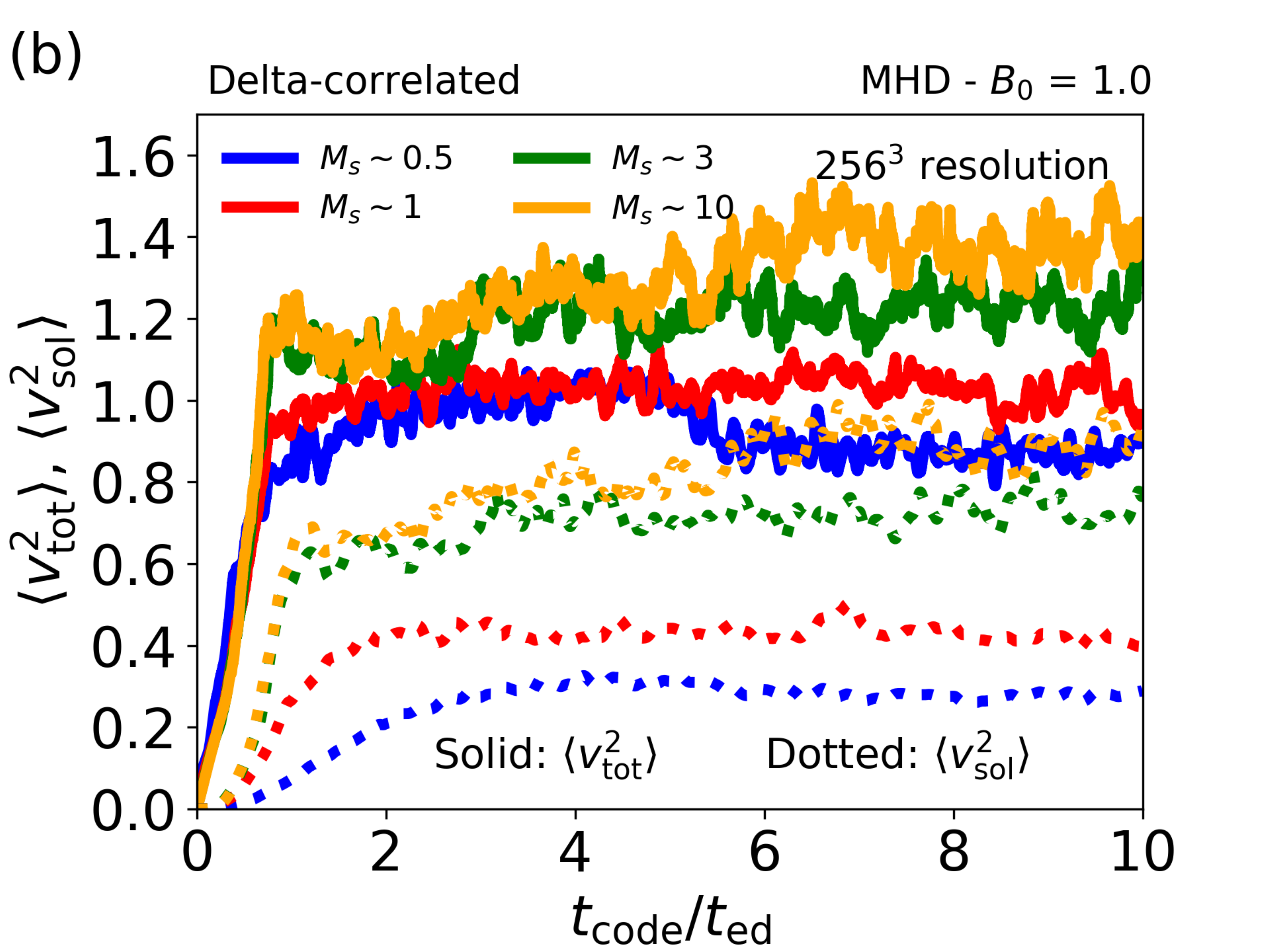} 
\includegraphics[scale=0.15]{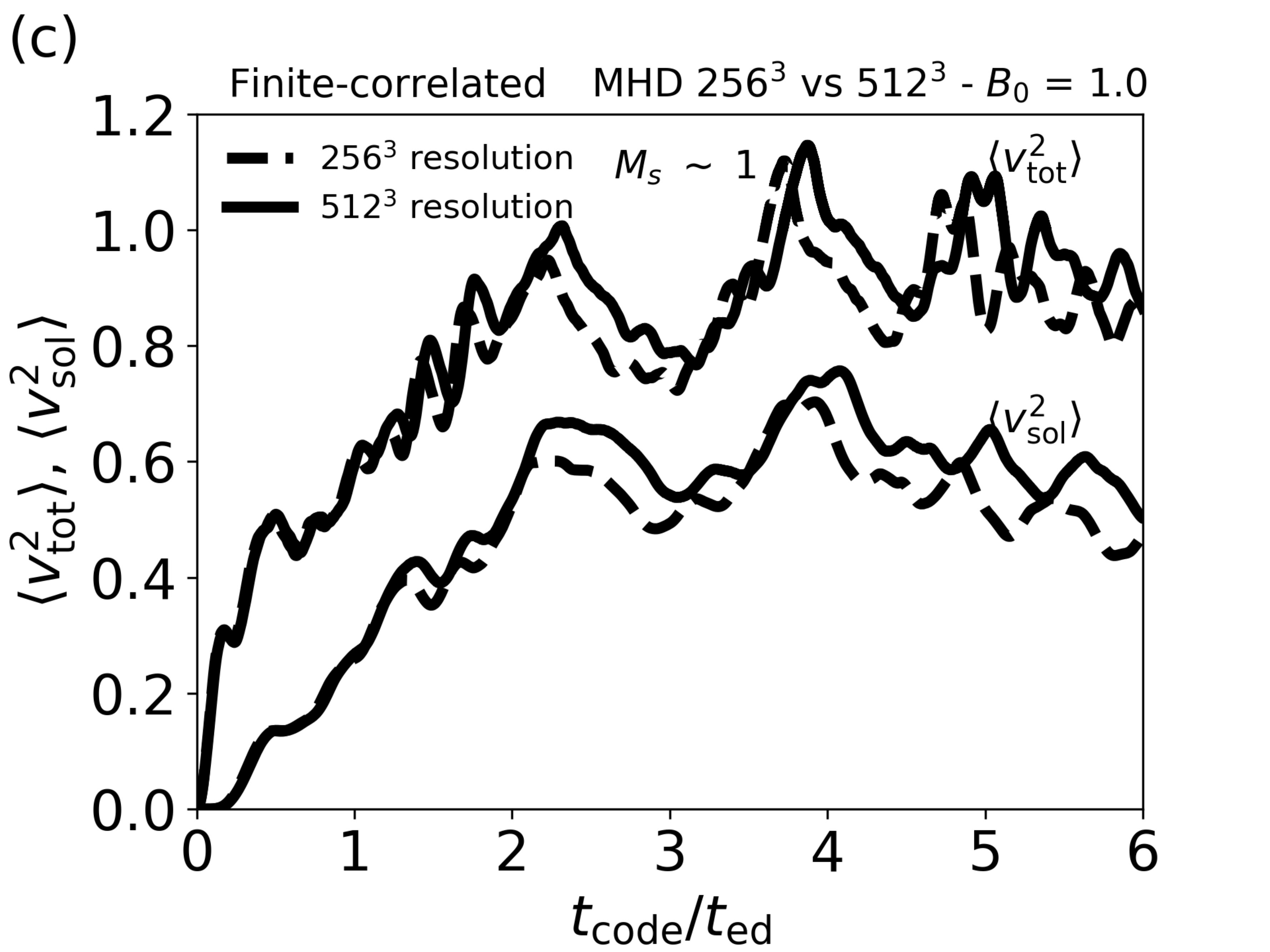} 
\caption{Similar to Figure \ref{fig:fig3}, but for $\bzero$ = 1.0. Blue, red, green, and orange curves in the left and the middle panels represent $M_s$ $\sim$ 0.5, $\sim$ 1, $\sim$ 3 and $\sim$ 10, respectively. In the right panel, dashed and solid curves correspond to $256^3$ and $512^3$ resolutions for the finite-correlated compressive driving, respectively.
\label{fig:fig5}}
\end{figure*}


\begin{figure}
\centering
\includegraphics[scale=0.25]{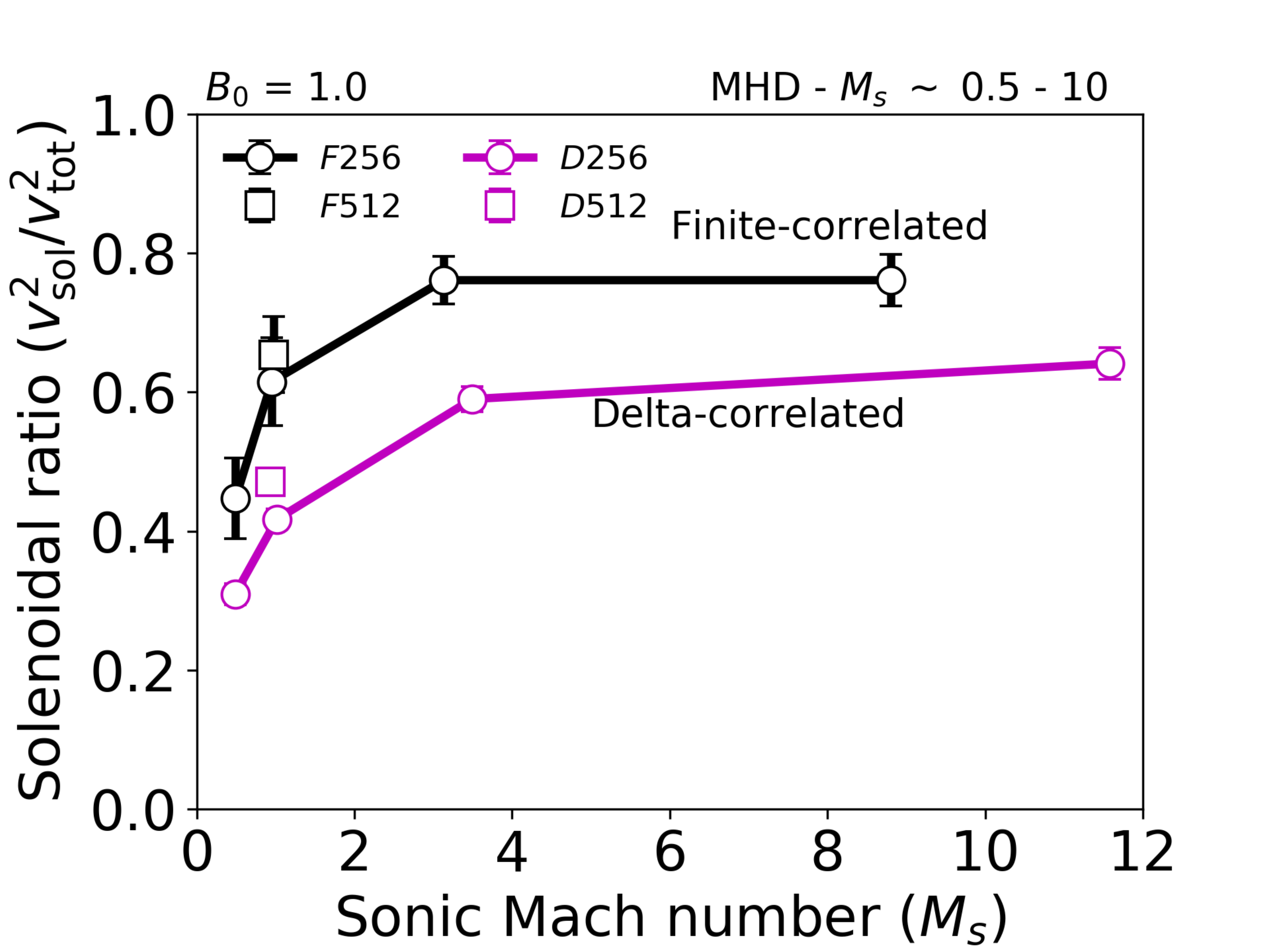} 
\caption{The same as Figure \ref{fig:fig4}(b), but for $\bzero$ = 1.0. We consider $M_s$ up to $\sim$ 10. 
\label{fig:fig6}}
\end{figure}

Figure \ref{fig:fig4} shows averaged solenoidal ratio as a function of $M_s$ for simulations with $M_s$ $\lesssim$ 3 and $B_0$ $\leq$ 0.1. Figures \ref{fig:fig4}(a)-(c) correspond to the ratio for $B_0$ = 0.001, 0.01, and 0.1, respectively. In Figures \ref{fig:fig4}(a) and \ref{fig:fig4}(c), we consider only the finite-correlated compressive driving. In Figure \ref{fig:fig4}(b) (see the middle panel), black and magenta circles (squares) correspond to the finite-correlated and the delta-correlated compressive drivings for $256^3$ ($512^3$) resolution, respectively. 

As in the case of HD turbulence, we can observe the following trends in Figure \ref{fig:fig4}. First, for $M_s$ effects, we can clearly see that as $M_s$ increases, the solenoidal ratio also increases regardless of $B_0$. When $M_s$ $\sim$ 3, the ratio is as large as $\sim$ 0.35 for $B_0$ = 0.001 and 0.01, and $\sim$ 0.45 for $B_0$ = 0.1. The trend we observe is consistent with the result from \citet{F11}, who considered a much lower mean magnetic field strength than ours and used a finite-correlated compressive driving. They obtained solenoidal ratios of $\sim$ 0.10, $\sim$ 0.27, and $\sim$ 0.32 for $M_s$ $\sim$ 0.5, $\sim$ 1.0, and $\sim$ 3.0, respectively. Those values are similar to ours: $\sim$ 0.13, $\sim$ 0.24, and $\sim$ 0.33 for similar $M_s$'s and $B_0$ = 0.01. This implies that the solenoidal ratio is not sensitive to $\bzero$ as long as $B_0$ is sufficiently small (see Section \ref{sec:sec3.3} for further discussions). Second, as we showed in Figure \ref{fig:fig4}(b), the solenoidal ratio does not seem sensitive to numerical resolution. This can be also supported by comparing the result of \citet{Porter15} with ours. They forced turbulence using a delta-correlated compressive driving to have $M_s$ $\sim$ 0.5 and considered a mean magnetic field strength of $\bzero$ $\sim$ 0.001 in our units. The solenoidal ratio is $\sim$ 0.07 in their simulation with $1024^3$ resolution. Taking into account the fact that the solenoidal ratio is insensitive to numerical resolution, we can conclude that their result is consistent with ours of $\sim$ 0.06 from $256^3$ resolution (Run D256MS0.5-$B_0$0.01). Third, we can see from Figure \ref{fig:fig4}(b) that, when $M_s$ is similar and numerical resolution is same, the finite-correlated compressive driving results in higher solenoidal ratio than the delta-correlated compressive driving.

\subsubsection{A Strong $B_0$ Case ($B_0$ = 1)\label{sec:sec3.2.2}} 
Figure \ref{fig:fig5} is similar to Figure \ref{fig:fig3}, but we consider $B_0$ = 1 and $M_s$ up to $\sim$ 10. In Figures \ref{fig:fig5}(a) and \ref{fig:fig5}(b), blue, red, cyan, and orange solid (dotted) curves represent $\langle \vtotsq\rangle$ $(\langle \vsolsq\rangle)$ for $M_s$ $\sim$ 0.5, $\sim$ 1, $\sim$ 3, and $\sim$ 10, respectively. Figures \ref{fig:fig5}(a) and \ref{fig:fig5}(b) are for the finite-correlated and the delta-correlated compressive drivings, respectively.

First of all, when we compare Figure \ref{fig:fig5} with Figure \ref{fig:fig3}, we can note that the levels of $\langle \vsolsq\rangle $ at saturation are much higher. Second, as we can see from Figure \ref{fig:fig5}(a), the finite-correlated compressive driving yields similar levels of $\langle \vtotsq\rangle $ irrespective of $M_s$. However, $\langle \vsolsq \rangle$ shows dependence on $M_s$. Roughly speaking, the level of $\langle \vsolsq\rangle$ increases as $M_s$ increases for $M_s$ $\lesssim$ 3 (see Figure \ref{fig:fig6} for more quantitative evaluation for this). Third, as we can see in Figure \ref{fig:fig5}(b), the level of $\langle \vsolsq\rangle $ increases with $M_s$ for the delta-correlated compressive driving.

Figure \ref{fig:fig5}(c) shows results of resolution study for the finite-correlated compressive driving in the case of $M_s$ $\sim$ 1. Dashed and solid curves are for $256^3$ and $512^3$ resolutions, respectively. Similar to the case of $\bzero$ = 0.01, the effect of resolution on $\langle \vsolsq\rangle$ does not seem very significant. 

Figure \ref{fig:fig6} shows the solenoidal ratio as a function of $M_s$ for $\bzero$ = 1.0. Black and magenta circles (squares) correspond to the finite-correlated and the delta-correlated compressive drivings for $256^3$ ($512^3$) resolution, respectively. In the figure, we can observe the same trend of the solenoidal ratio as in Figures \ref{fig:fig2} and \ref{fig:fig4}; the ratio increases with $M_s$ and is larger for the finite-correlated compressive driving than for the delta-correlated compressive driving at a similar $M_s$. It is worth noting that the solenoidal ratio exceeds 0.5 for both driving schemes in supersonic regime. This means that although turbulence is driven by compressive driving, solenoidal modes eventually dominate over compressive ones in the presence of a strong mean magnetic field. This result may imply that Alfven modes become more important than compressive modes in the regime.

\begin{figure*}[ht!]
\centering
\includegraphics[scale=0.15]{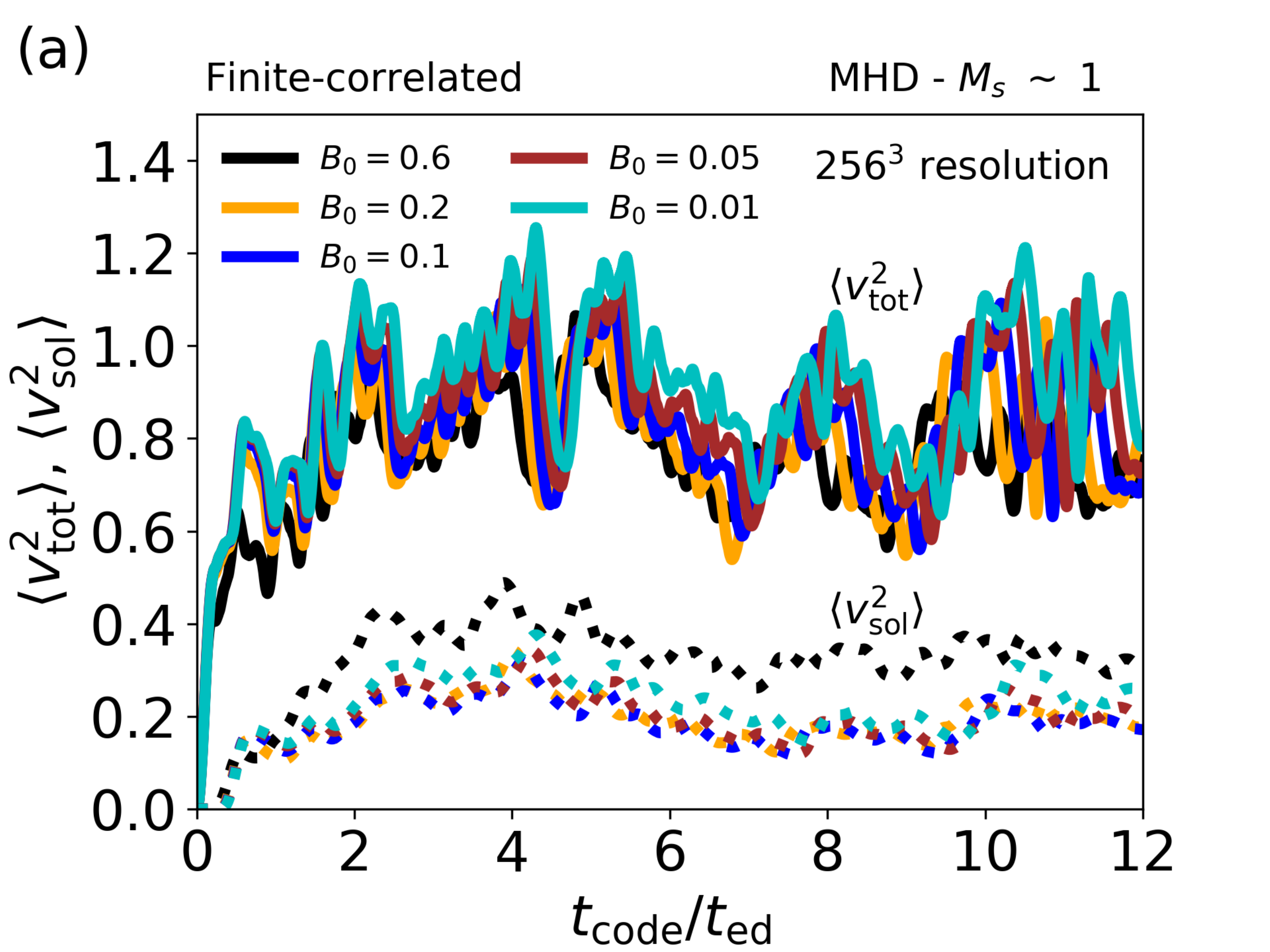}
\includegraphics[scale=0.15]{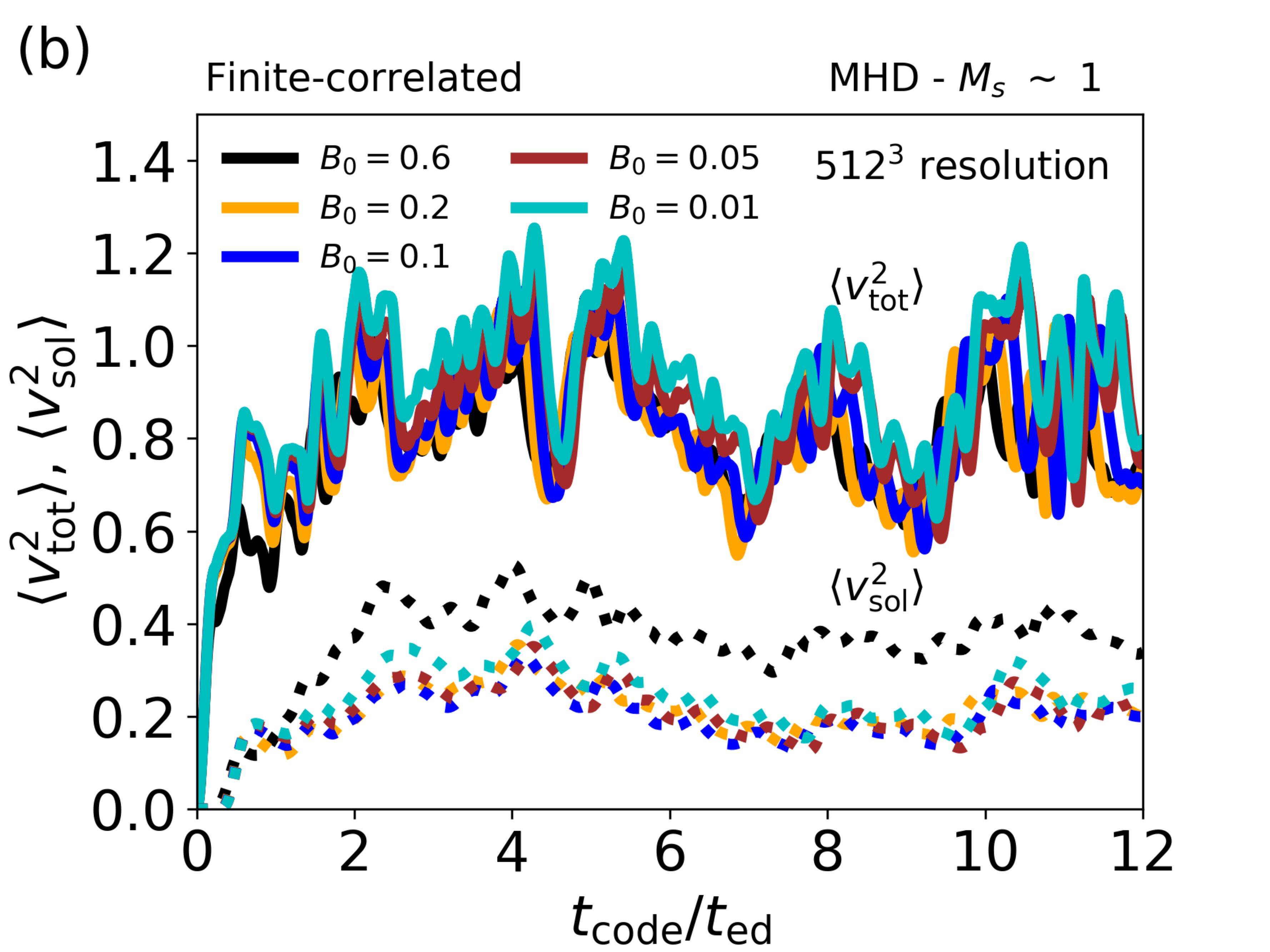} 
\includegraphics[scale=0.15]{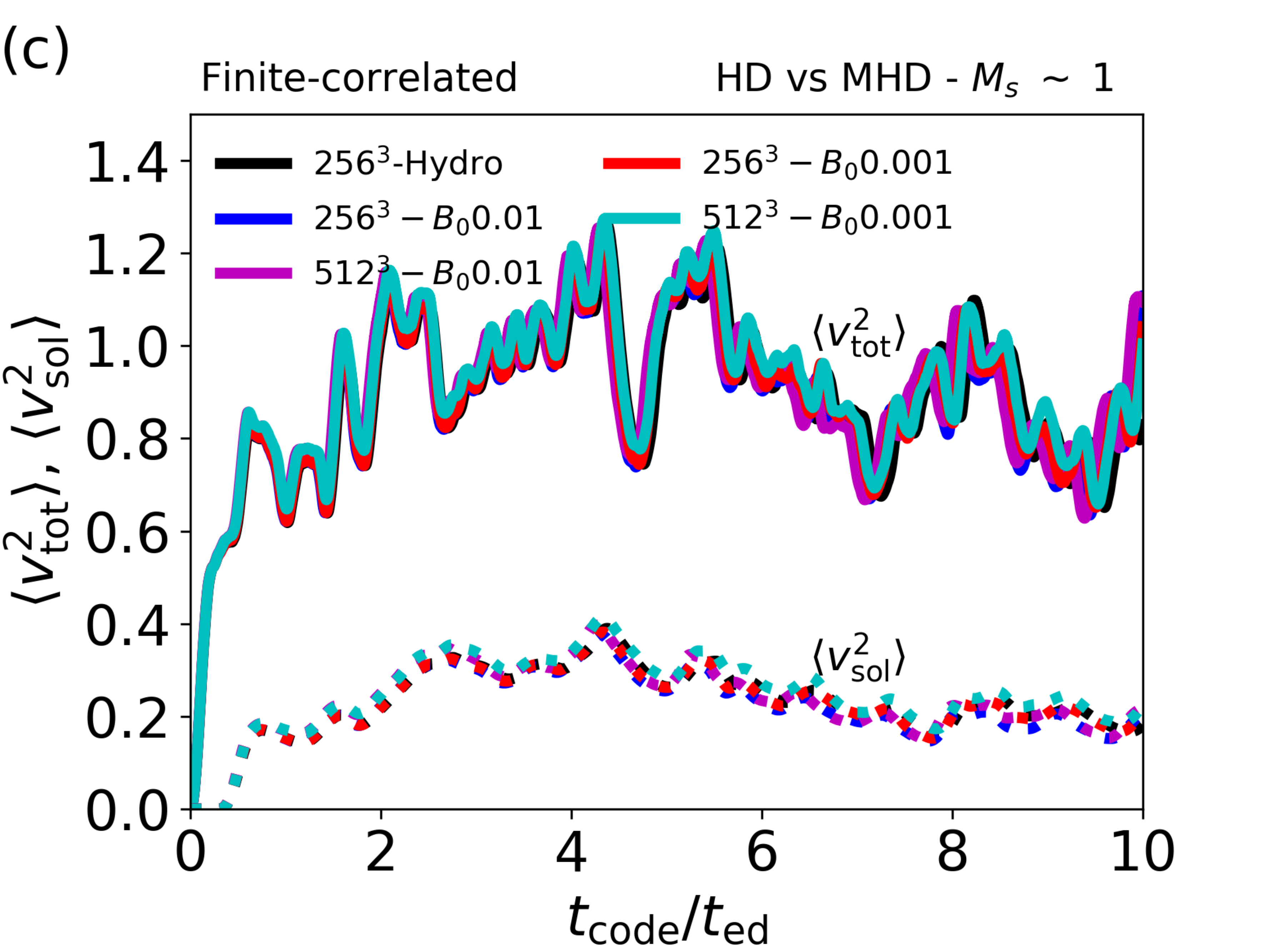} 
\caption{Time evolution of $\langle \vtotsq\rangle$ (solid curves) and $\langle \vsolsq\rangle$ (dotted curves) for MHD turbulence simulations with $M_s$ $\sim$ 1. (a) The finite-correlated compressive driving with $256^3$ resolution. (b) The finite-correlated compressive driving with $512^3$ resolution. (c) Comparison of HD and MHD simulations. Cyan, brown, blue, orange, and black curves in the left and the middle panels represent $\bzero$ = 0.01, 0.05, 0.1, 0.2, and 0.6, respectively. In the right panel, black curves represent Run F256MS1-Hydro simulation. Blue and magenta curves correspond to Run F256MS1-$\bzero$0.01 and Run F512MS1-$\bzero$0.01, respectively. Red and cyan curves correspond to Run F256MS1-$\bzero$0.001 and Run F512MS1-$\bzero$0.001, respectively.
\label{fig:fig7}}
\end{figure*}


\begin{figure*}
\centering
\includegraphics[scale=0.15]{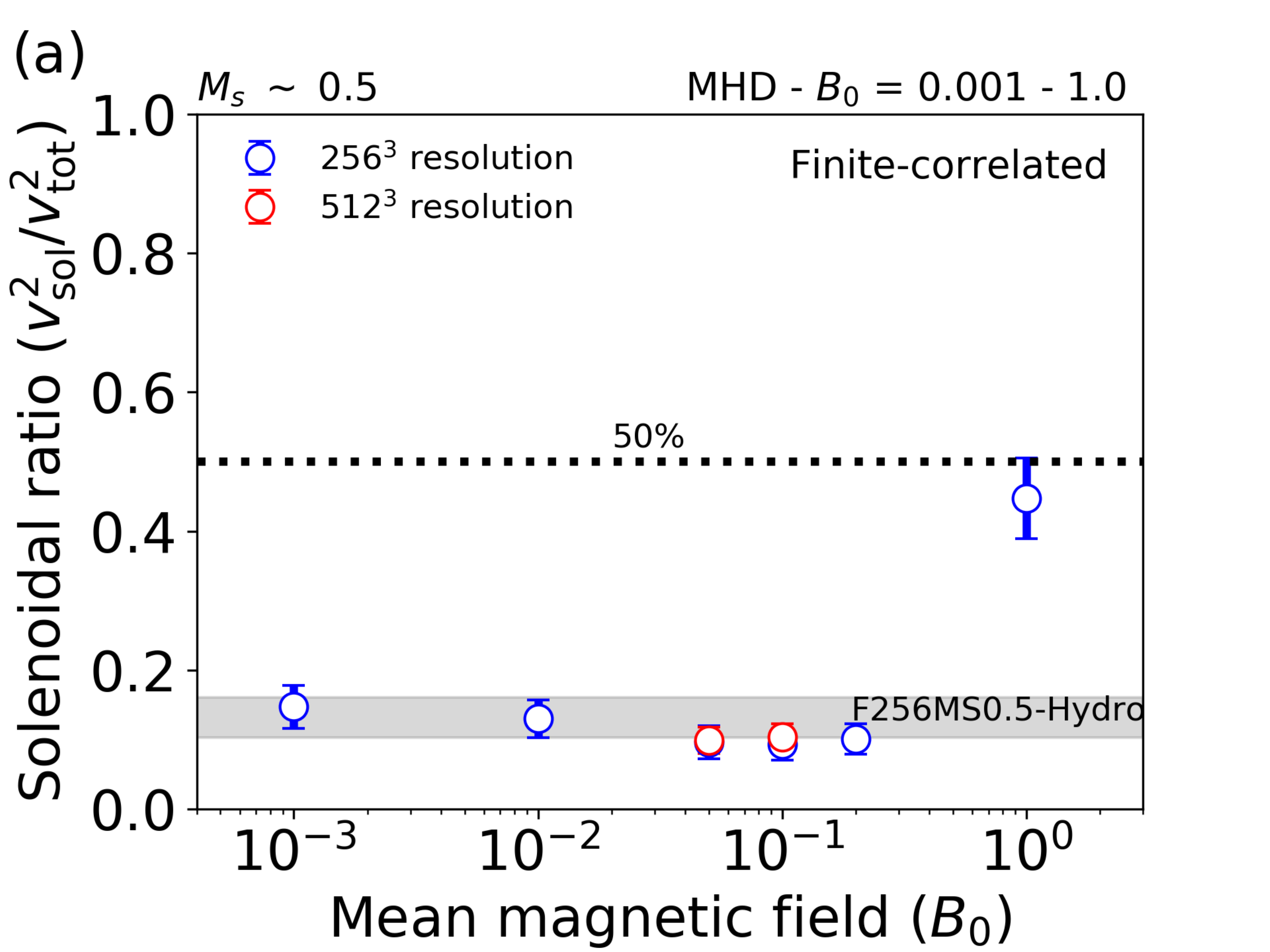}
\includegraphics[scale=0.15]{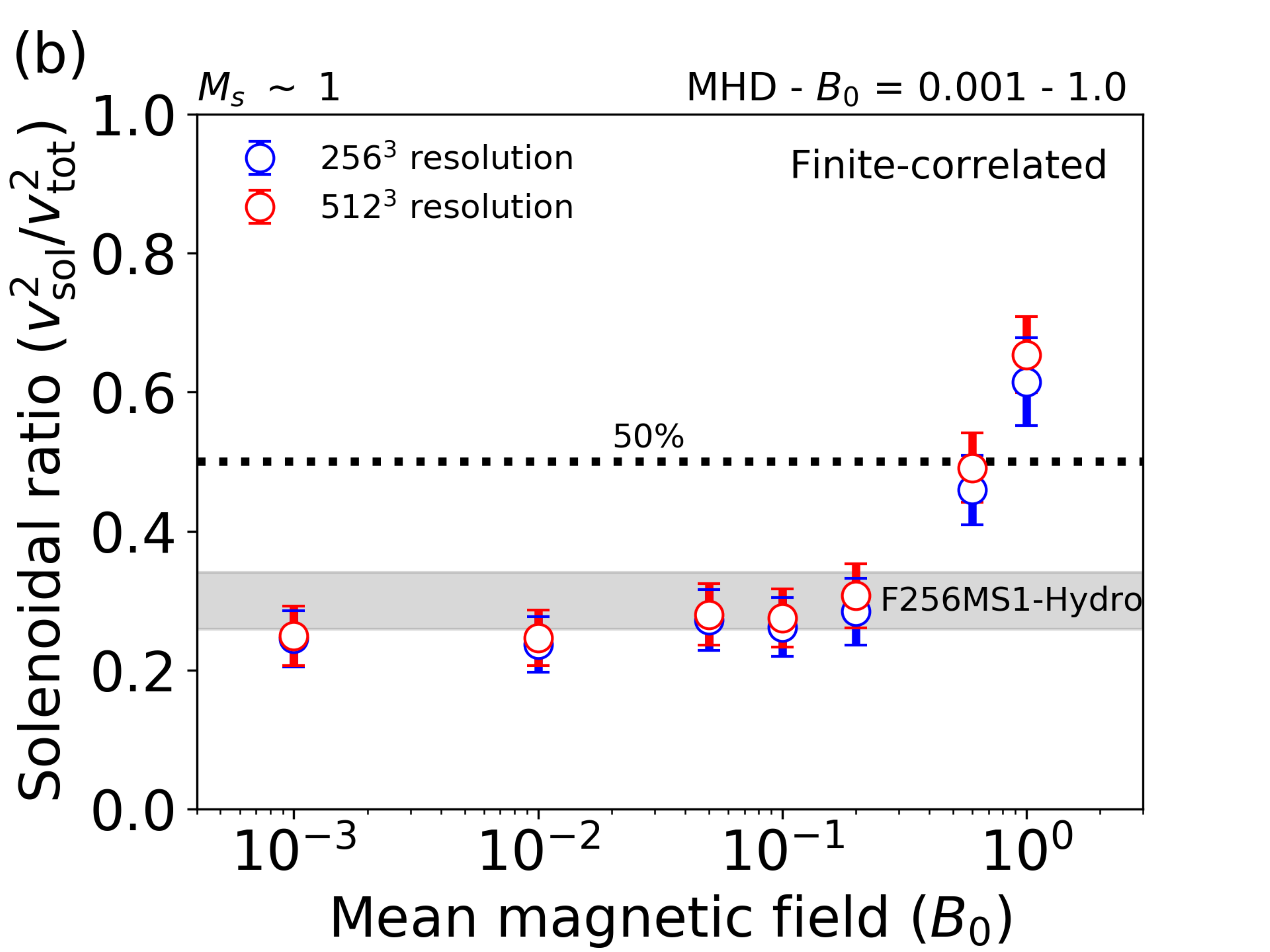} 
\includegraphics[scale=0.15]{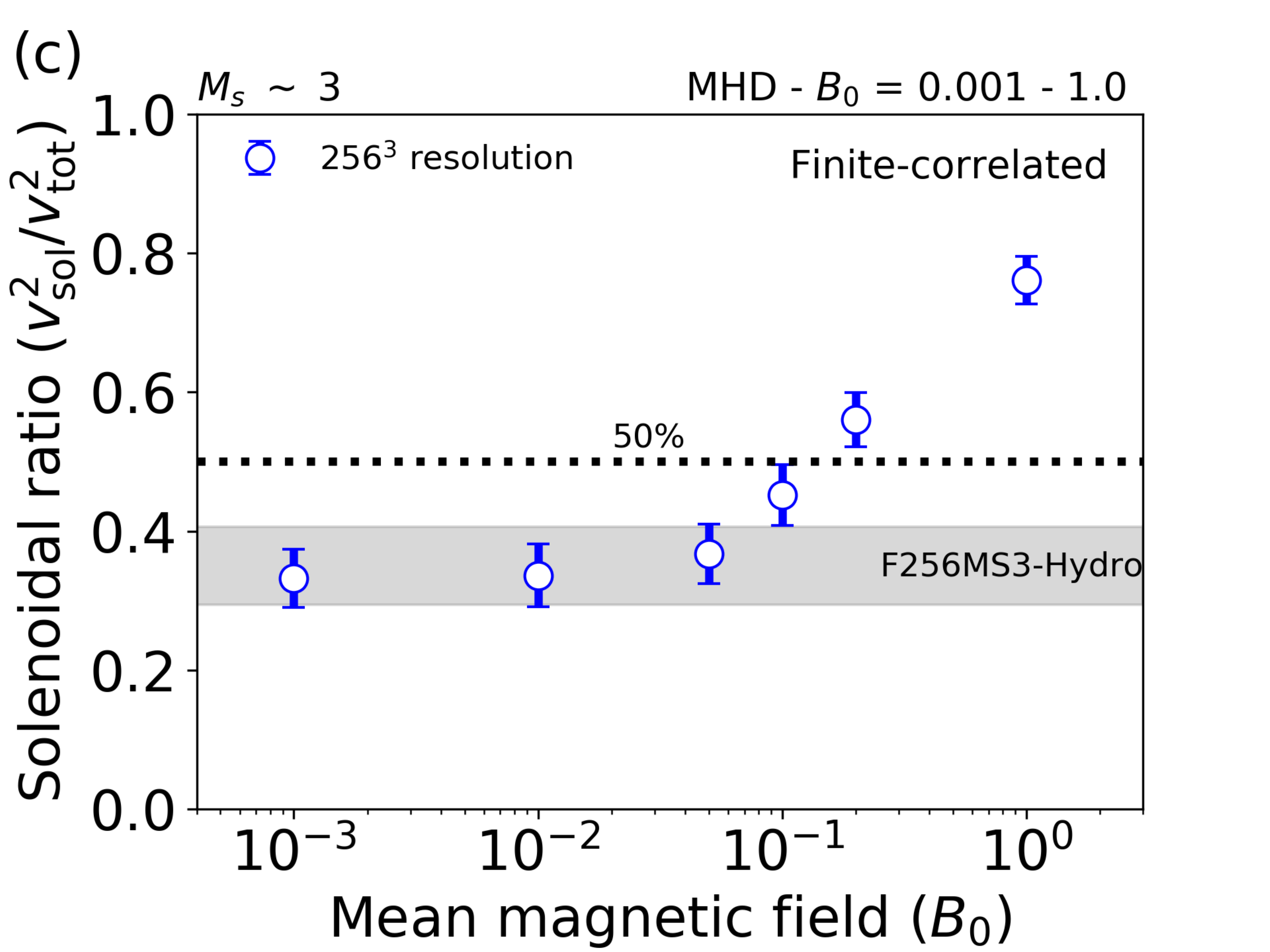} 
\caption{Average values of the solenoidal ratio ($\vsolsq/\vtotsq$) as a function of the mean magnetic field ($\bzero$) for the MHD simulations with $M_{s}$ $\lesssim$ 3. (a) $M_s$ $\sim$ 0.5. (b) $M_s$ $\sim$ 1. (c) $M_s$ $\sim$ 3. In the right panel ($M_s$ $\sim$ 3), we present results from $256^3$ resolution simulations only. In the left and the middle panels, blue and red circles denote solenoidal ratios of the simulations with $256^3$ and $512^3$ resolutions, respectively. The grey shaded region in each panel shows the 1$\sigma$-dispersion about the average ratio from HD simulations. The error bars show standard deviations. In Table \ref{tab:tab1}, the average values and the time intervals for taking average are shown. 
\label{fig:fig8}}
\end{figure*}

\subsection{Effects of the Mean Magnetic Field\label{sec:sec3.3}}
In this subsection, we deal with effects of the mean magnetic field ($\bzero$) on the generation of solenoidal modes in turbulence driven by compressive driving. For this purpose, we consider $M_s$ $\lesssim$ 3 and $\bzero$ ranging from 0.001 (very weak mean field case) to 0.6 (marginally strong mean field case).  We do not study effects of driving schemes here; we consider only the finite-correlated compressive driving. 

Figure \ref{fig:fig7} shows time evolution of $\langle \vtotsq\rangle$ and $\langle \vsolsq\rangle$ in MHD turbulence driven by compressive driving. Cyan, brown, blue, orange, and black curves in Figures \ref{fig:fig7}(a) and \ref{fig:fig7}(b) correspond to $\bzero$ = 0.01, 0.05, 0.1, 0.2, and 0.6, respectively. The numerical resolutions in Figures \ref{fig:fig7}(a) and \ref{fig:fig7}(b) are $256^3$ and $512^3$, respectively. Figure \ref{fig:fig7}(c) shows simulation results for $B_0$ $\leq$ 0.01, in which magenta, blue, cyan, red, and black curves correspond to Run F512MS1-$\bzero$0.01, Run F256MS1-$\bzero$0.01, Run F512MS1-$\bzero$0.001, Run F256MS1-$\bzero$0.001, and Run F256MS1-Hydro, respectively. Note that all solid curves virtually coincide and so do all dotted curves.

\begin{figure*}[ht!]
\centering
\includegraphics[scale=0.15]{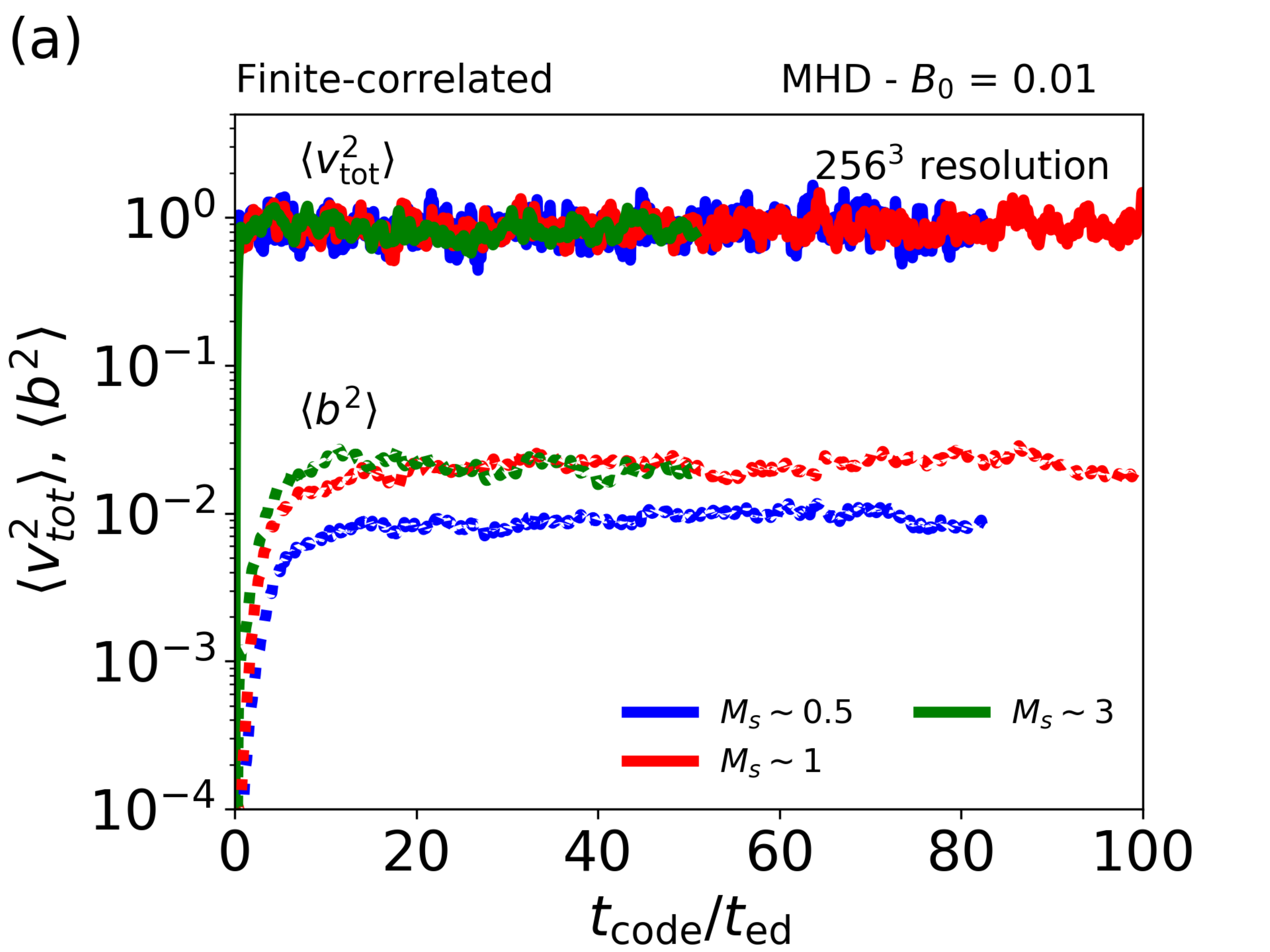}
\includegraphics[scale=0.15]{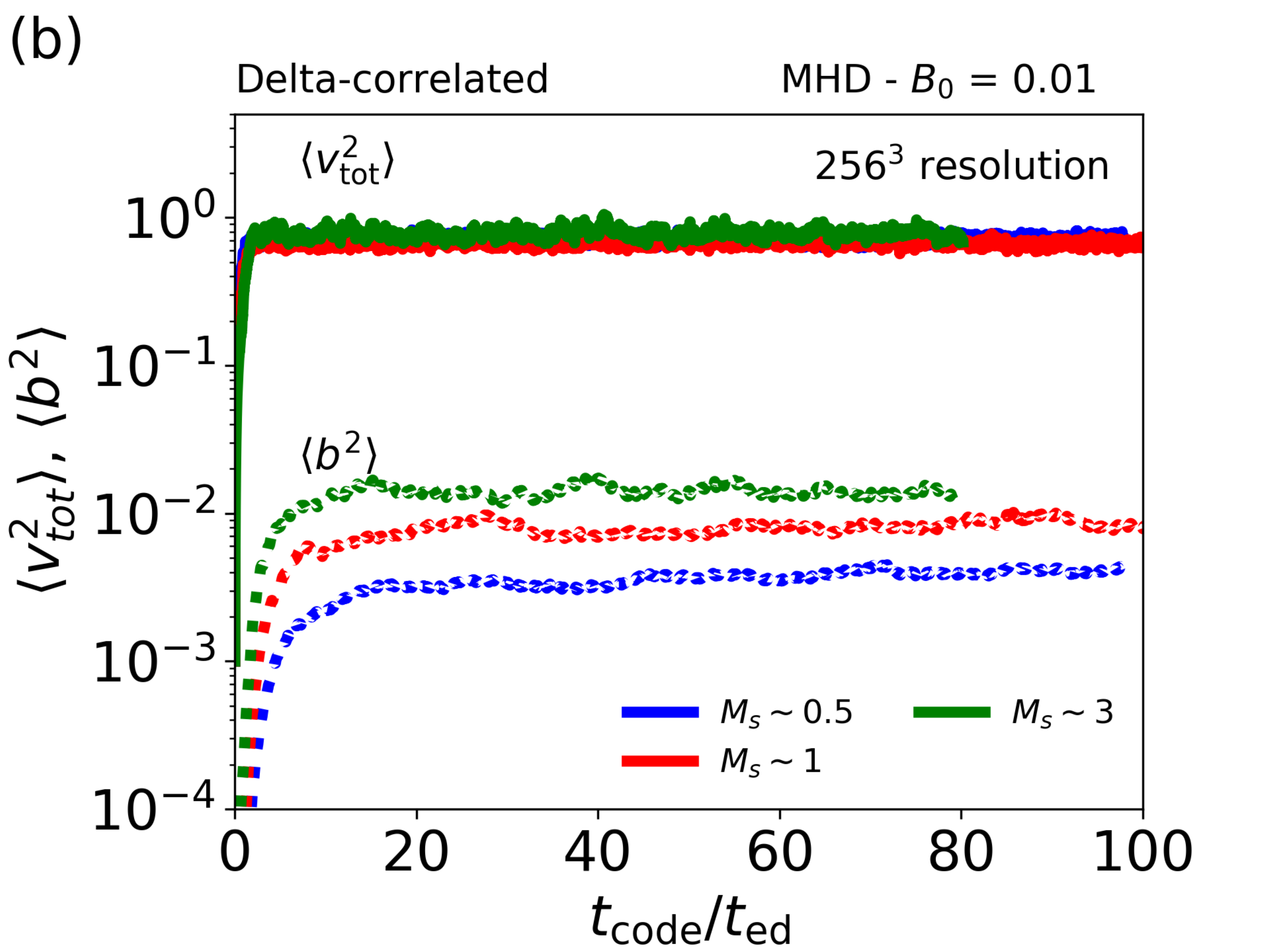} 
\includegraphics[scale=0.15]{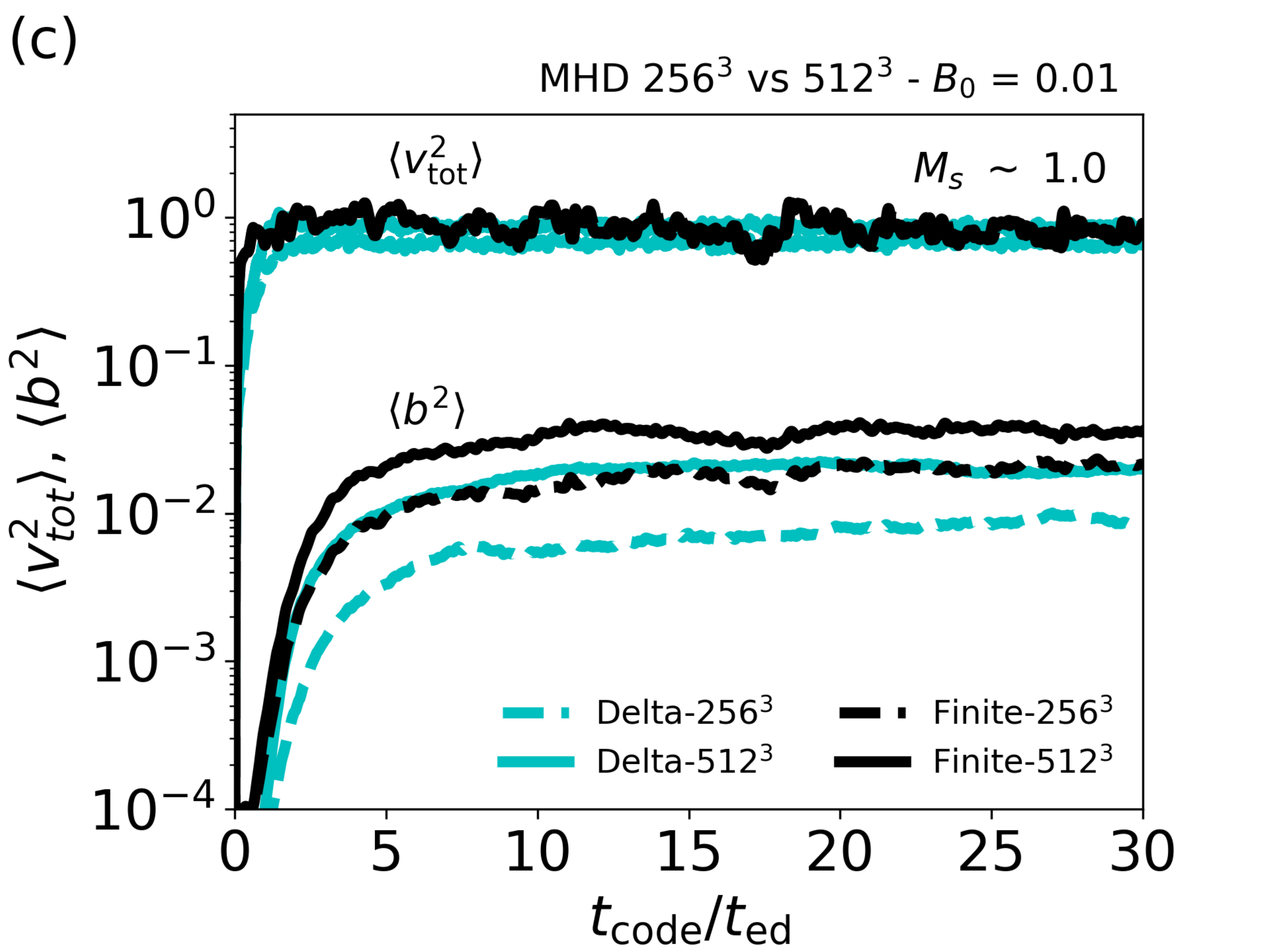} 
\caption{Time evolution of $\langle \vtotsq\rangle$ and $\langle \bsq\rangle$ for MHD turbulence simulations with $\bzero$ = 0.01. (a) The finite-correlated compressive driving. (b) The delta-correlated compressive driving. (c) Resolution study for both the finite-correlated and the delta-correlated compressive drivings in the case of $\mach$ $\sim$ 1. Blue, red, and green solid (dotted) curves in the left and the middle panels represent $\langle \vtotsq\rangle$ ($\langle \bsq\rangle$) for $\mach$ $\sim$ 0.5, $\sim$ 1, and $\sim$ 3, respectively. In the right panel, black dashed and solid curves correspond to $256^3$ and $512^3$ resolutions for the finite-correlated compressive driving, respectively. Cyan dashed and solid curves correspond to $256^3$ and $512^3$ resolutions for the delta-correlated compressive driving, respectively. For $\langle b^2 \rangle$, the red dotted curves in the left and the middle panels are the same as the black dashed and cyan dashed curves in the right panel, respectively. 
\label{fig:fig9}}
\end{figure*}


\begin{figure}
\centering
\includegraphics[scale=0.25]{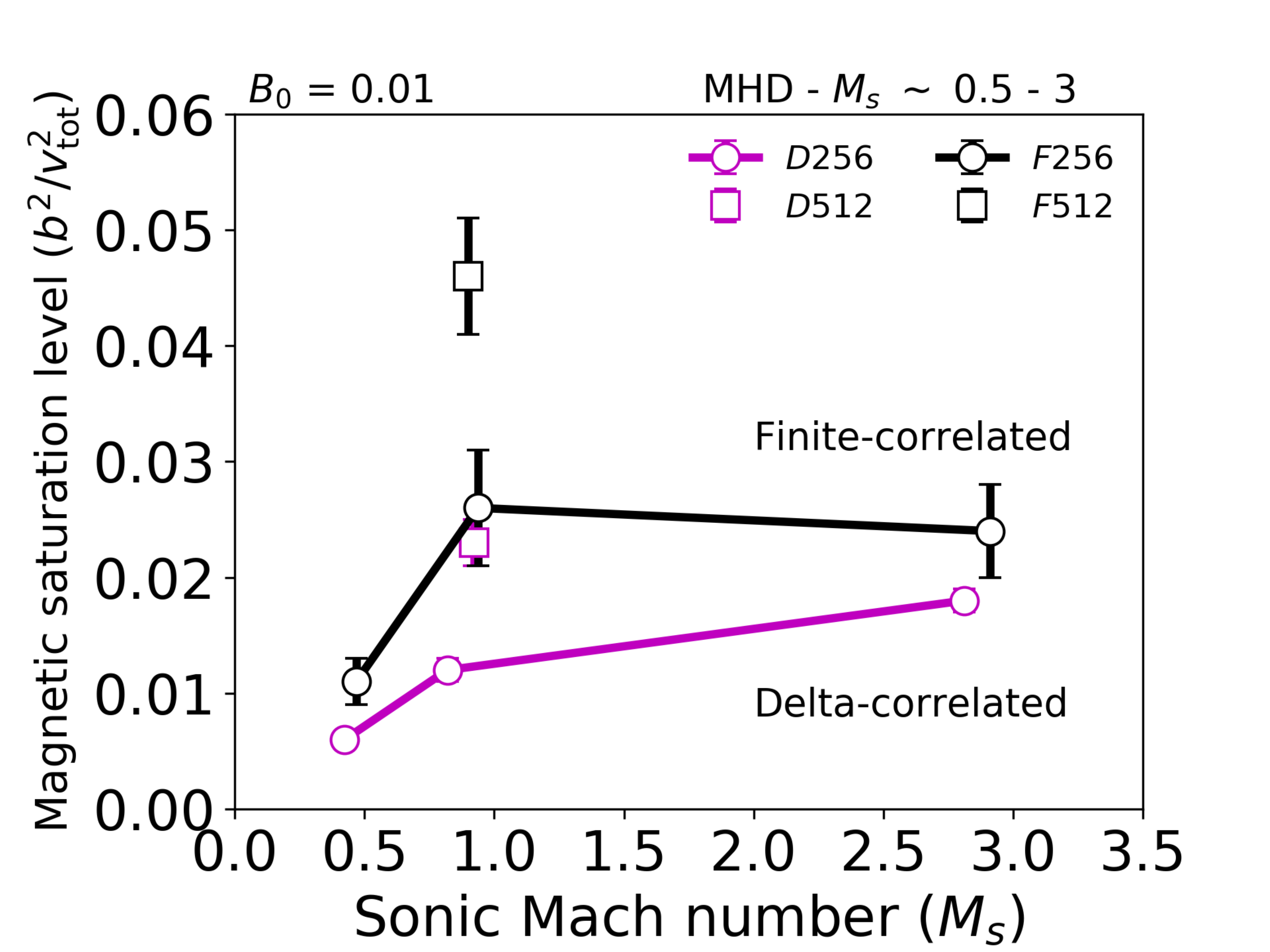}
\caption{Averaged values of the magnetic saturation level ($\bsq/\vtotsq$) as a function of the sonic Mach number ($\mach$) for $B_{0}$ = 0.01. Black and magenta circles (squares) indicate the finite-correlated and the delta-correlated compressive drivings for $256^3$ ($512^3$) resolution, respectively. The error bars represent standard deviations. In Table \ref{tab:tab1}, the average values and the time intervals for taking average are shown. 
\label{fig:fig10}}
\end{figure}


First, $\langle \vtotsq\rangle$ and $\langle \vsolsq\rangle$ for all simulations in Figure \ref{fig:fig7} seem to saturate roughly after 3$t_{\textrm{ed}}$. Second, in Figures \ref{fig:fig7}(a) and \ref{fig:fig7}(b), the evolution of $\langle \vtotsq\rangle $ is almost same irrespective of both $\bzero$ and the numerical resolution. Furthermore, when $\bzero$ $\leq$ 0.2, the evolution of $\langle \vsolsq\rangle$ at saturation is nearly indistinguishable in both numerical resolutions (i.e., in both Figures \ref{fig:fig7}(a) and \ref{fig:fig7}(b)). On the other hand, the level of $\langle \vsolsq\rangle$ for $B_0$ = 0.6 at saturation is notably higher than those for $B_0$ $\leq$ 0.2 in both numerical resolutions. Third, according to Figure \ref{fig:fig7}(c), it is noticeable that both numerical resolution and the degree of magnetization do not strongly affect the evolutions of $\langle \vtotsq\rangle $ and $\langle \vsolsq\rangle$ when $B_0$ is weak or zero.

Figure \ref{fig:fig8} shows average values of the solenoidal ratio as a function of $\bzero$. Figures \ref{fig:fig8}(a)-(c) correspond to the ratio for $M_s$ $\sim$ 0.5, $\sim$ 1, and $\sim$ 3, respectively. In Figure \ref{fig:fig8}(c), we present the results of simulations with $256^3$ resolution only. In Figures \ref{fig:fig8}(a) and \ref{fig:fig8}(b), blue and red circles represent $256^3$ and $512^3$ resolutions, respectively. The grey shaded region in each panel indicates the 1$\sigma$-dispersion about the average solenoidal ratio from the HD simulations\footnote{Note that, contrary to time evolutions of HD and MHD simulations presented in Figure \ref{fig:fig7}(c), which virtually coincide, Figure \ref{fig:fig8}(b) shows that the solenoidal ratios for $B_0$ = 0.001 and 0.01 are slightly lower than that for Run F256MS1-Hydro. This originates from different time intervals for averaging the ratios; we use (3,6) for Run F256MS1-Hydro and much later times for runs with $B_0$ = 0.001 and 0.01 (see Table \ref{tab:tab1}). As we can see from Figure \ref{fig:fig7}(c), the level of $\langle \vsolsq\rangle$ between $t_{\textrm{code}}/t_{\textrm{ed}}$ = 3 and 6 is larger than that after 6$t_{\textrm{ed}}$. If we average the ratio for Run F256MS1-Hydro between $t_{\textrm{code}}/t_{\textrm{ed}}$ = 6 and 10, we obtain $\sim$ 0.247, which is close to the ratios of the simulations with $B_0$ = 0.001 and 0.01.}.

First, Figure \ref{fig:fig8} clearly shows that the solenoidal ratio is lower when $M_s$ is lower at the same $B_0$, as we discussed earlier in Sections \ref{sec:sec3.1} and \ref{sec:sec3.2}. Second, when $B_0$ is small, the solenoidal ratio is virtually the same as the HD value. However, as $B_0$ exceeds a certain strength, the ratio begins to deviate from the HD value. The solenoidal ratio seems to exceed 0.5 when $B_0$ $\geq$ 0.6 for $M_s$ $\sim$ 1 and $B_0$ $\geq$ 0.2 for $M_s$ $\sim$ 3. Third, we note that deviation of the ratio from the HD values occurs roughly at $B_0$ = 0.1. However, it shows a weak dependence on $M_s$.  When $M_s$ $\lesssim$ 1, the solenoidal ratios for $\bzero$ $\leq$ 0.2 are not considerably different from the ratio of the HD simulations. On the contrary, in the case of $M_s$ $\sim$ 3, this happens at a smaller $B_0$ ($\sim$ 0.05). Lastly, in Figures \ref{fig:fig8}(a) and \ref{fig:fig8}(b), the numerical resolution does not seem to have substantial impacts on the ratio especially for $\bzero$ $\leq$ 0.1. Although the ratio slightly depends on numerical resolution, the difference is within the error bar.

\begin{figure*}[ht!]
\centering
\includegraphics[scale=0.15]{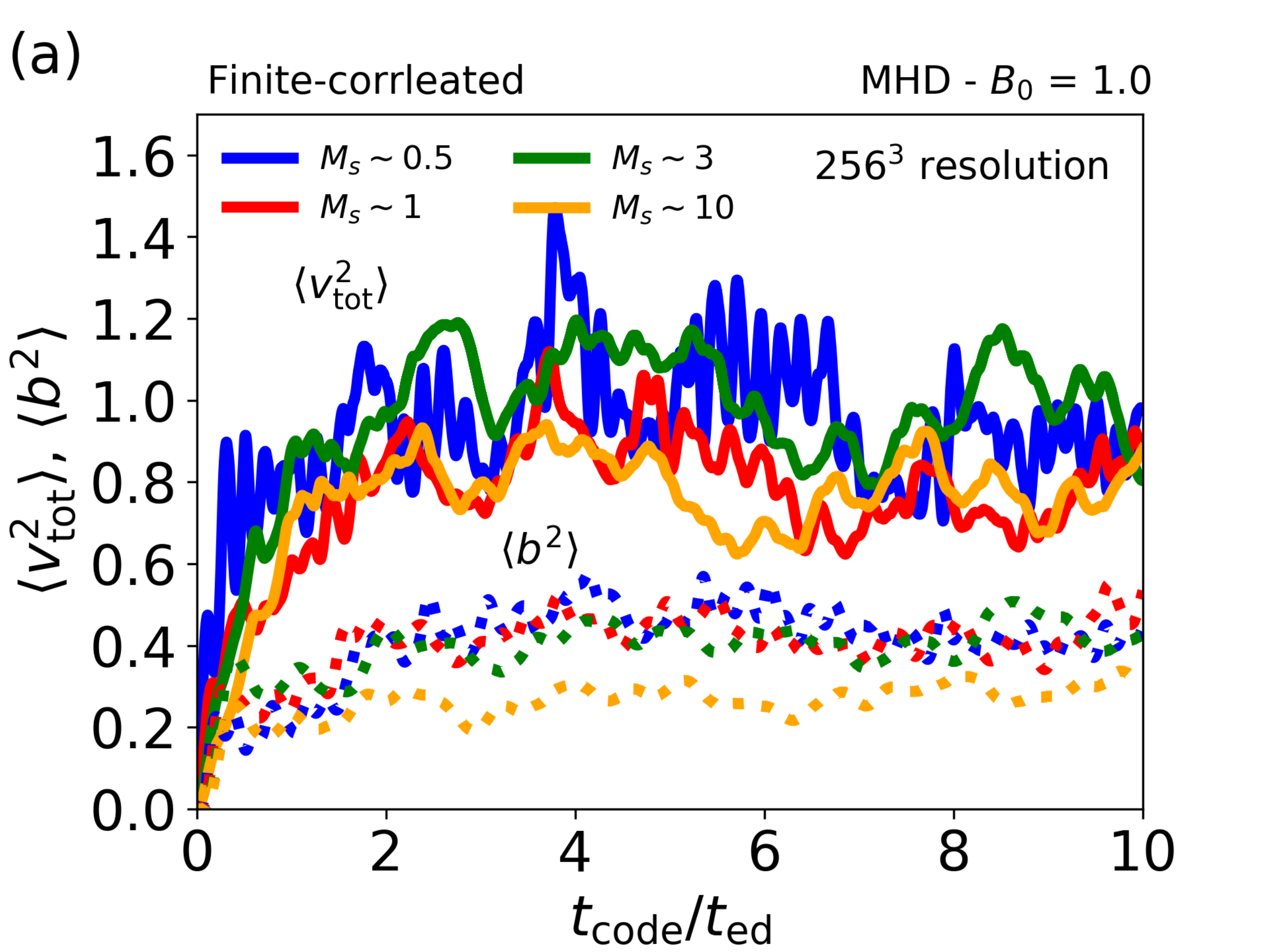}
\includegraphics[scale=0.15]{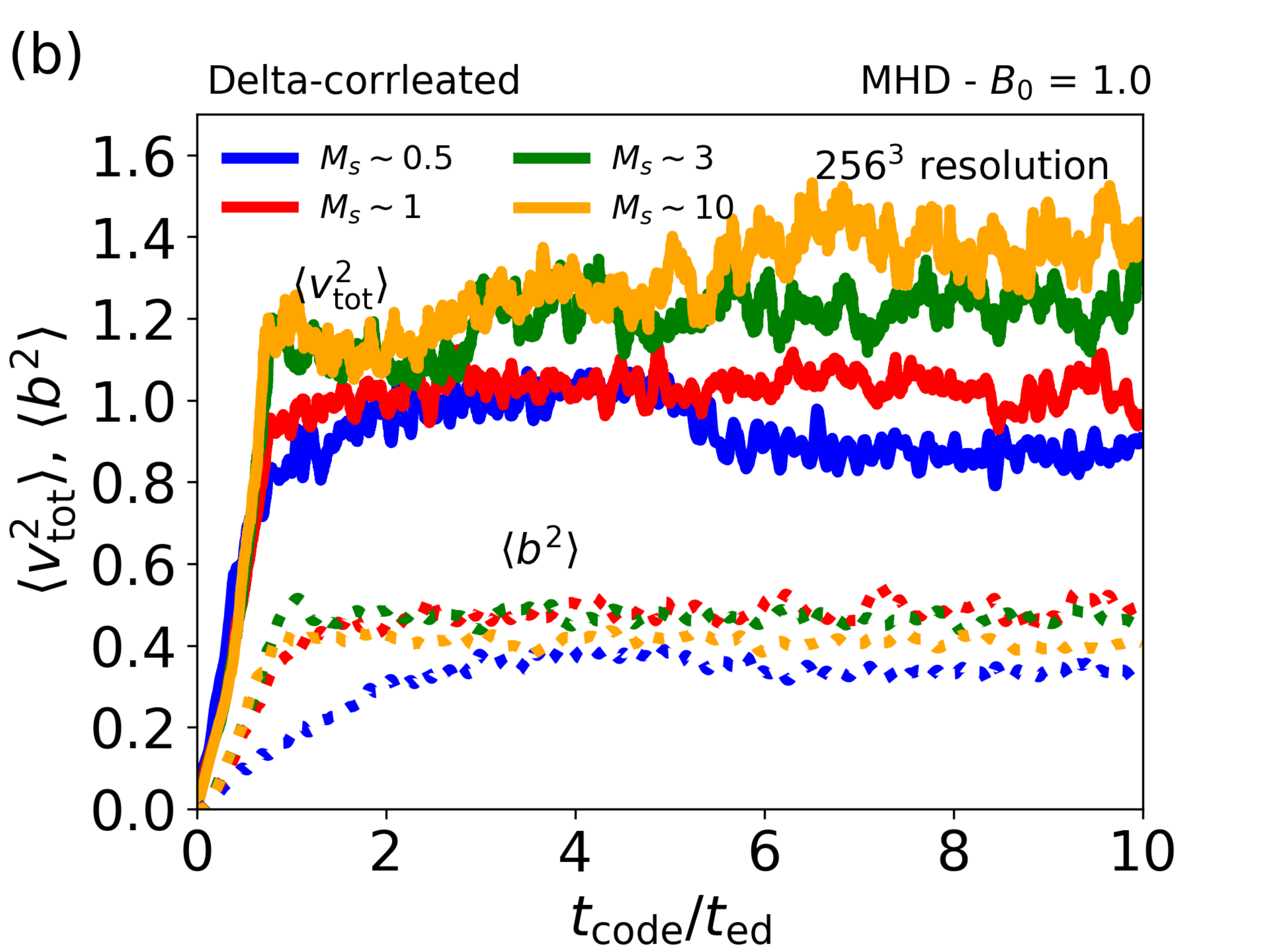} 
\includegraphics[scale=0.15]{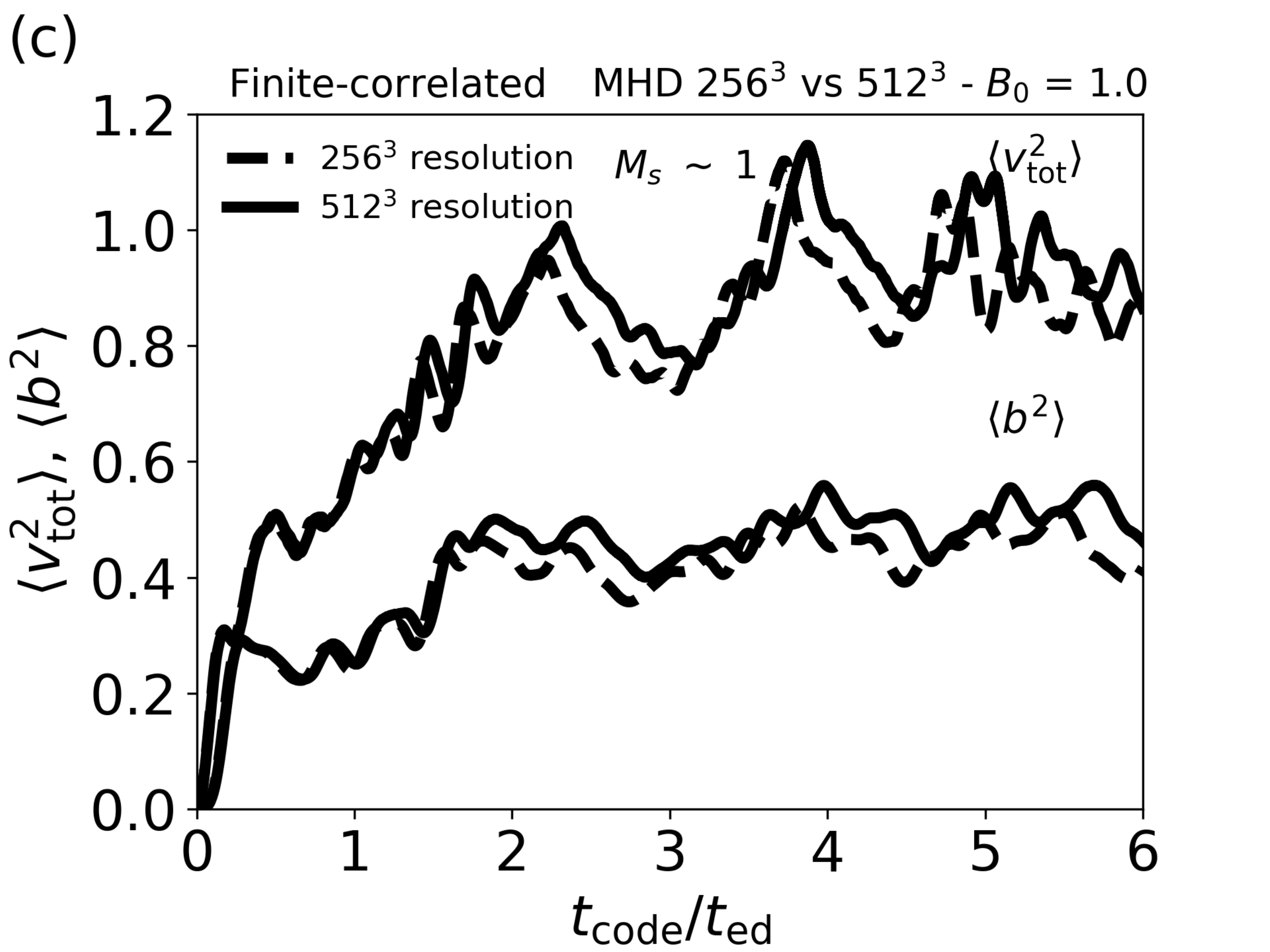} 
\caption{Similar to Figure \ref{fig:fig9}, but for $\bzero$ = 1.0. Blue, red, green, and orange curves in the left and the middle panels represent $\mach$ $\sim$ 0.5, $\sim$ 1, $\sim$ 3, and $\sim$ 10, respectively. In the right panel, dashed and solid curves correspond to $256^3$ and $512^3$ resolutions for the finite-correlated compressive driving, respectively.\label{fig:fig11}}
\end{figure*}

\begin{figure}
\centering
\includegraphics[scale=0.25]{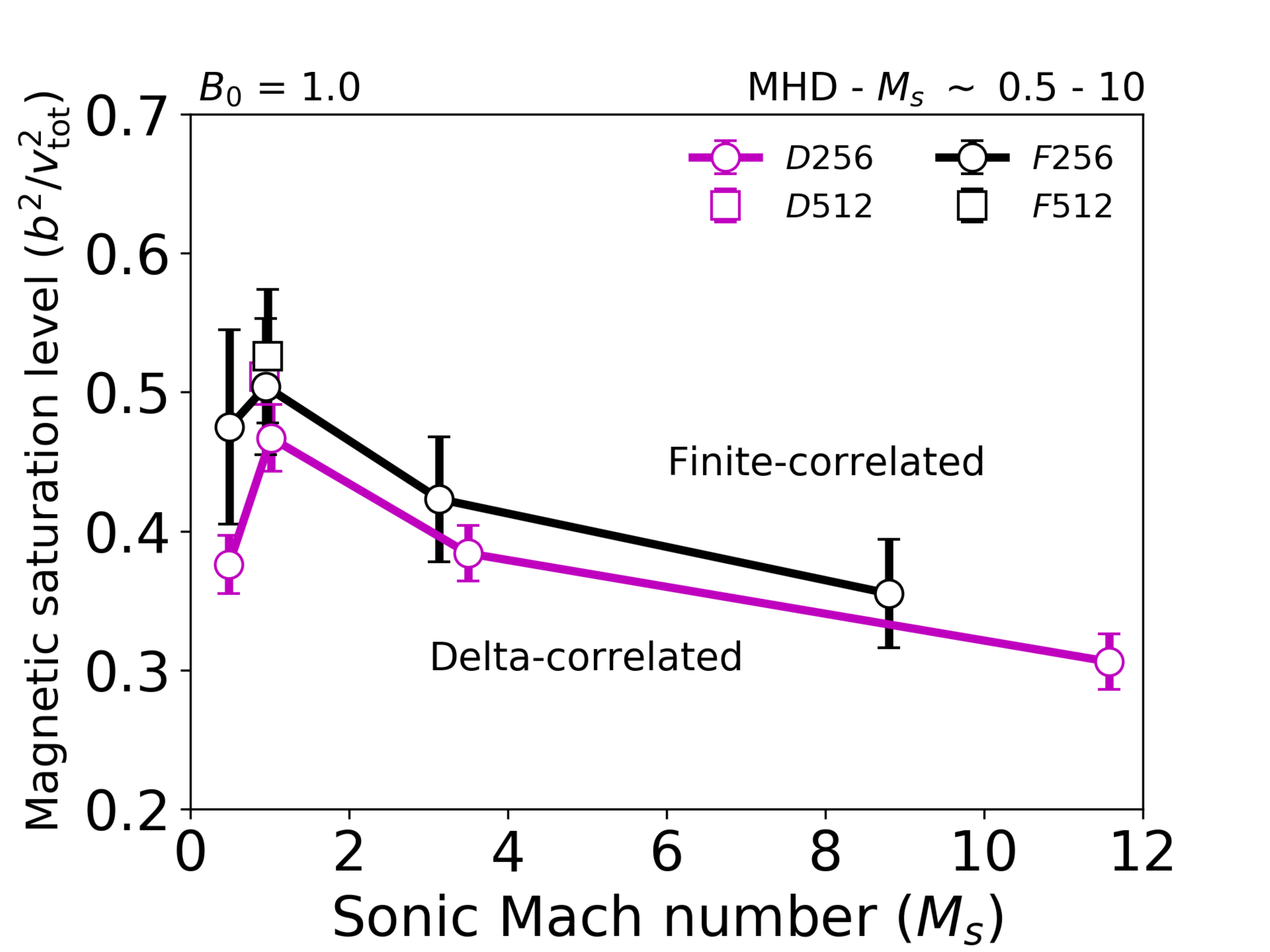}
\caption{The same as Figure \ref{fig:fig10}, but for $\bzero$ = 1.0. We consider $M_s$ up to $\sim$ 10. 
\label{fig:fig12}}
\end{figure}


\section{Generation of small-scale random magnetic fields\label{sec:sec4}}
In this section, we investigate generation of small-scale magnetic field in compressively driven turbulence. The strength of the small-scale magnetic field is defined by $b$ = $\sqrt{B^2 - \bzero^2}$ as we described in Section \ref{sec:sec2}. At the beginning of simulations, no random magnetic field exists. Then, as simulations go on, turbulence develops and amplifies small-scale magnetic field via stretching effect, which is mainly provided by solenoidal velocity component. In Sections \ref{sec:sec4.1} and \ref{sec:sec4.2}, we deal with effects of the sonic Mach number ($M_s$) and the mean magnetic field ($\bzero$), respectively.

\subsection{Effects of the Sonic Mach Number\label{sec:sec4.1}}

\subsubsection{Weak $B_0$ Cases ($B_0$ = 0.01)\label{sec:sec4.1.1}}
Figure \ref{fig:fig9} shows time evolution of $\langle \vtotsq\rangle$ and $\langle \bsq\rangle$ in MHD turbulence driven by compressive driving. Blue, red, and green curves in Figures \ref{fig:fig9}(a) and \ref{fig:fig9}(b) correspond to $\mach$ $\sim$ 0.5, $\sim$ 1, and $\sim$ 3, respectively. Figures \ref{fig:fig9}(a) and \ref{fig:fig9}(b) are for the finite-correlated and the delta-correlated compressive drivings, respectively. Figure \ref{fig:fig9}(c) shows results of resolution study. In the figure, different colors of curves represent different driving schemes and different line styles different numerical resolutions. Black dashed and black solid curves indicate the finite-correlated compressive driving for $256^3$ and $512^3$ resolutions, respectively. Cyan dashed and cyan solid curves indicate the delta-correlated compressive driving for $256^3$ and $512^3$ resolutions, respectively. The vertical axis is in logarithmic scale in all panels.


\begin{figure*}[ht!]
\centering
\includegraphics[scale=0.15]{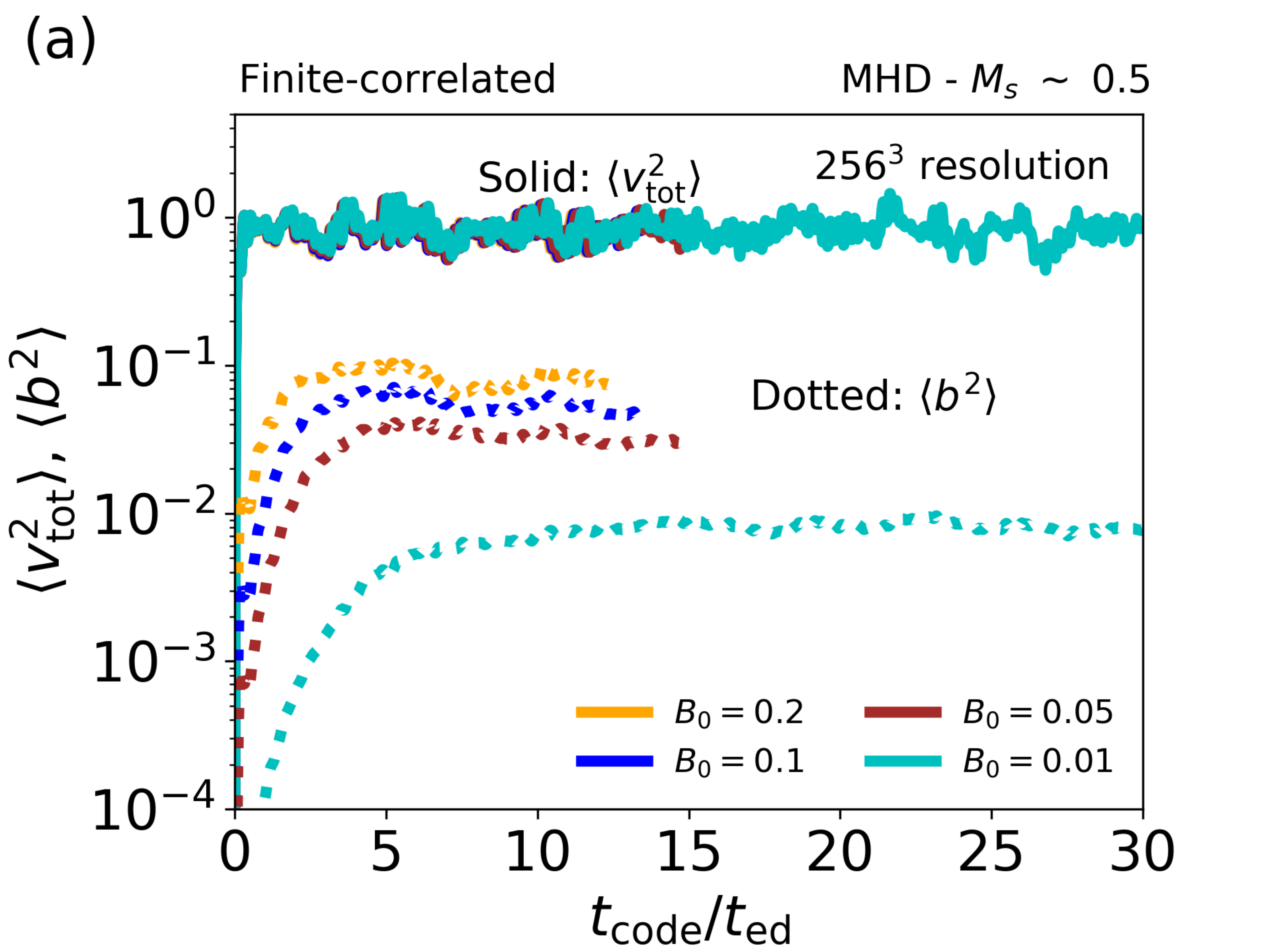}
\includegraphics[scale=0.15]{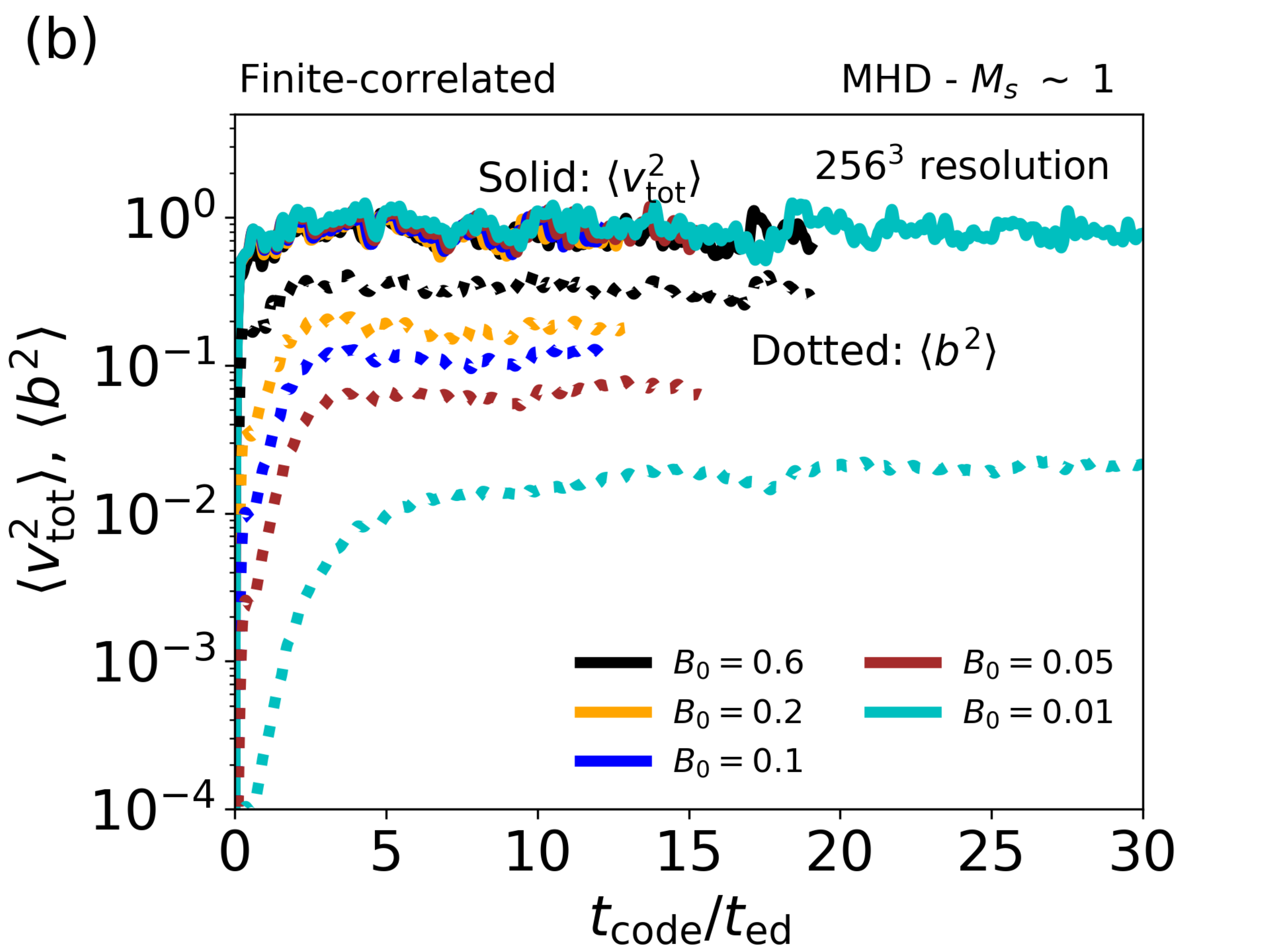} 
\includegraphics[scale=0.15]{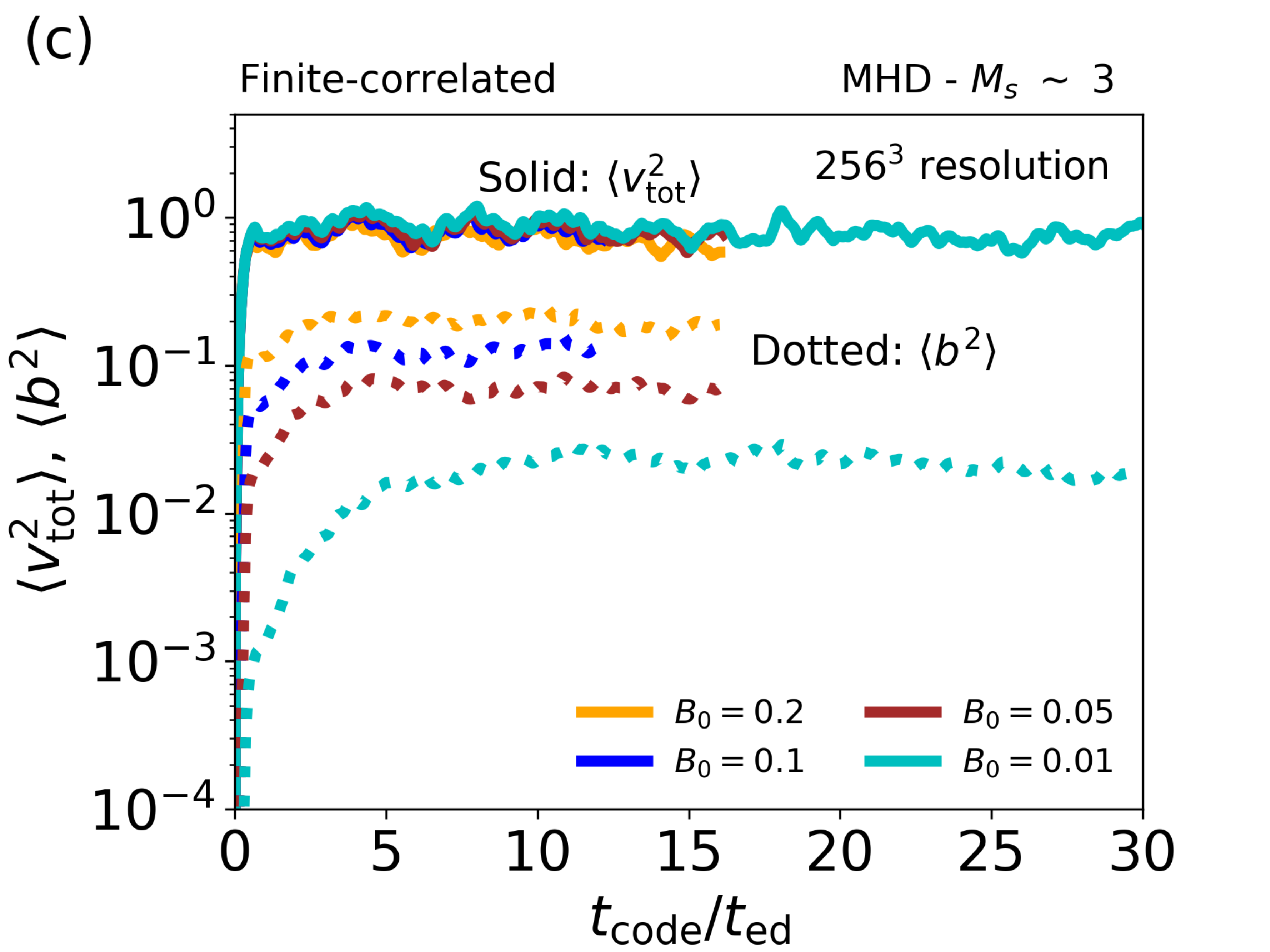} 
\caption{Time evolution of $\langle \vtotsq\rangle $ (solid curves) and $\langle \bsq\rangle$ (dotted curves) for MHD turbulence simulations with $256^3$ resolution and $\mach$ $\lesssim$ 3. (a) $M_s$ $\sim$ 0.5. (b) $M_s$ $\sim$ 1. (c) $M_s$ $\sim$ 3. In each panel, cyan, brown, blue, orange, and black curves represent $\bzero$ = 0.01, 0.05, 0.1, 0.2, and 0.6, respectively. 
\label{fig:fig13}}
\end{figure*}


\begin{figure*}[ht!]
\centering
\includegraphics[scale=0.15]{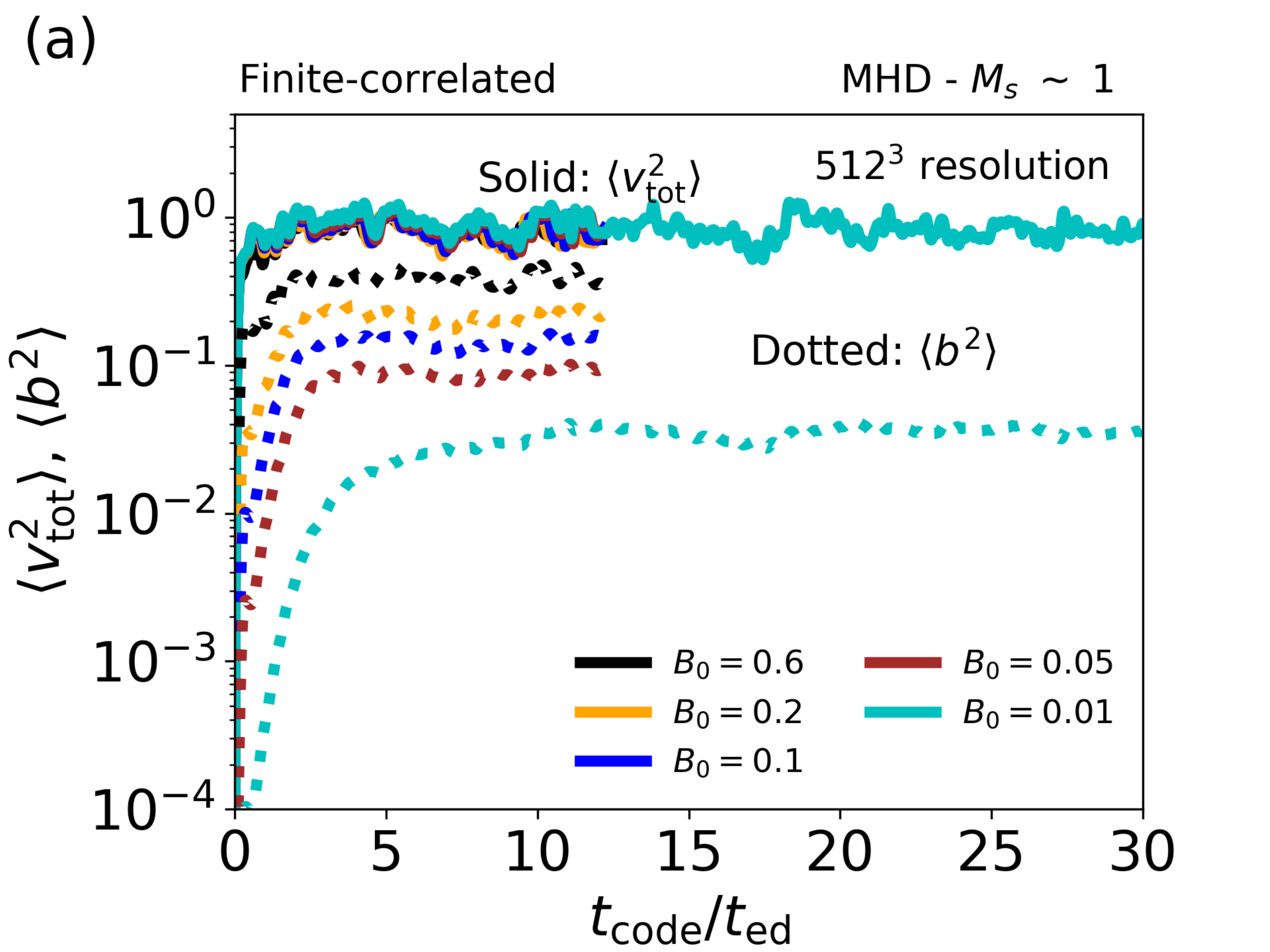}
\includegraphics[scale=0.15]{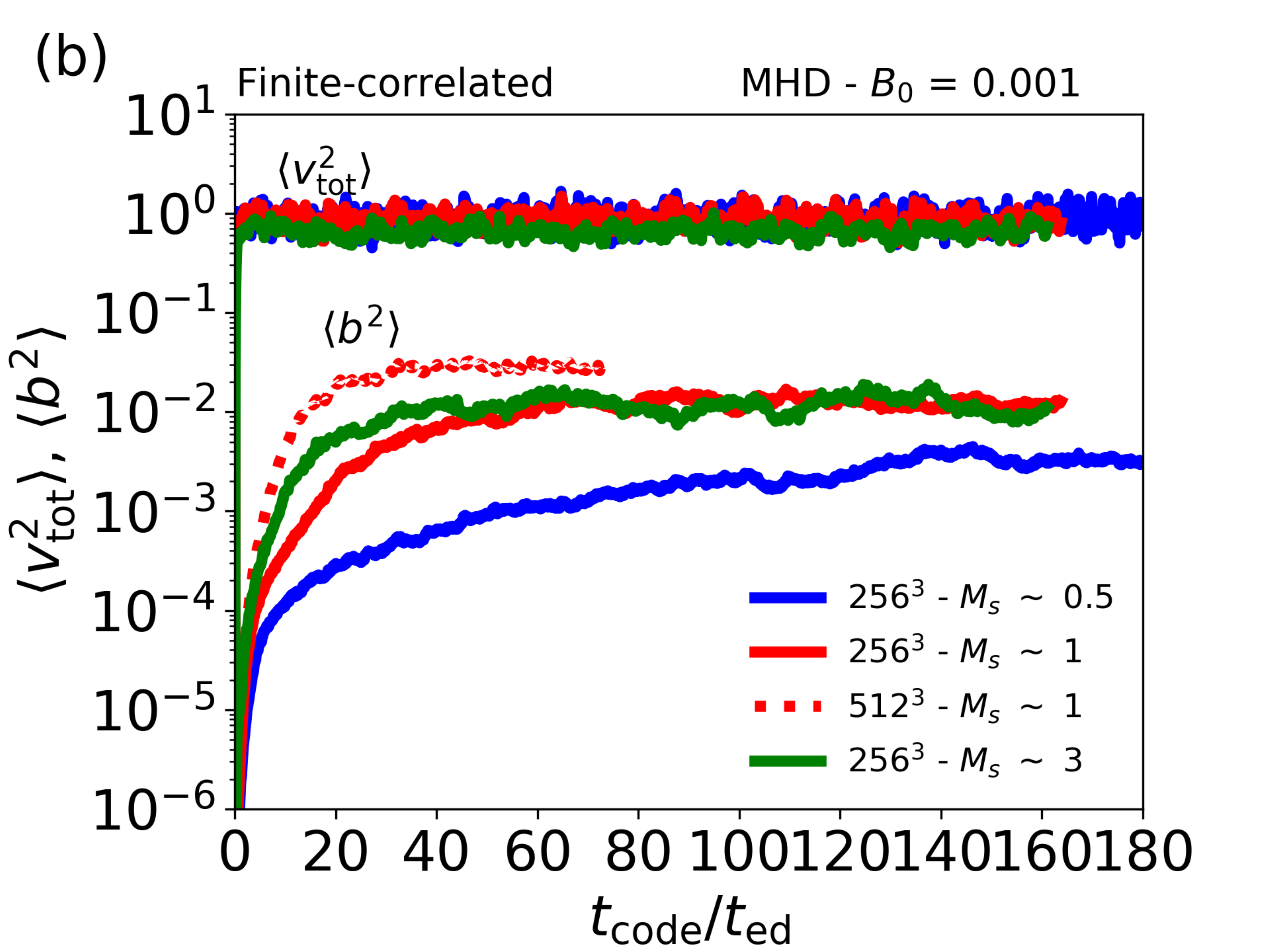} 
\includegraphics[scale=0.15]{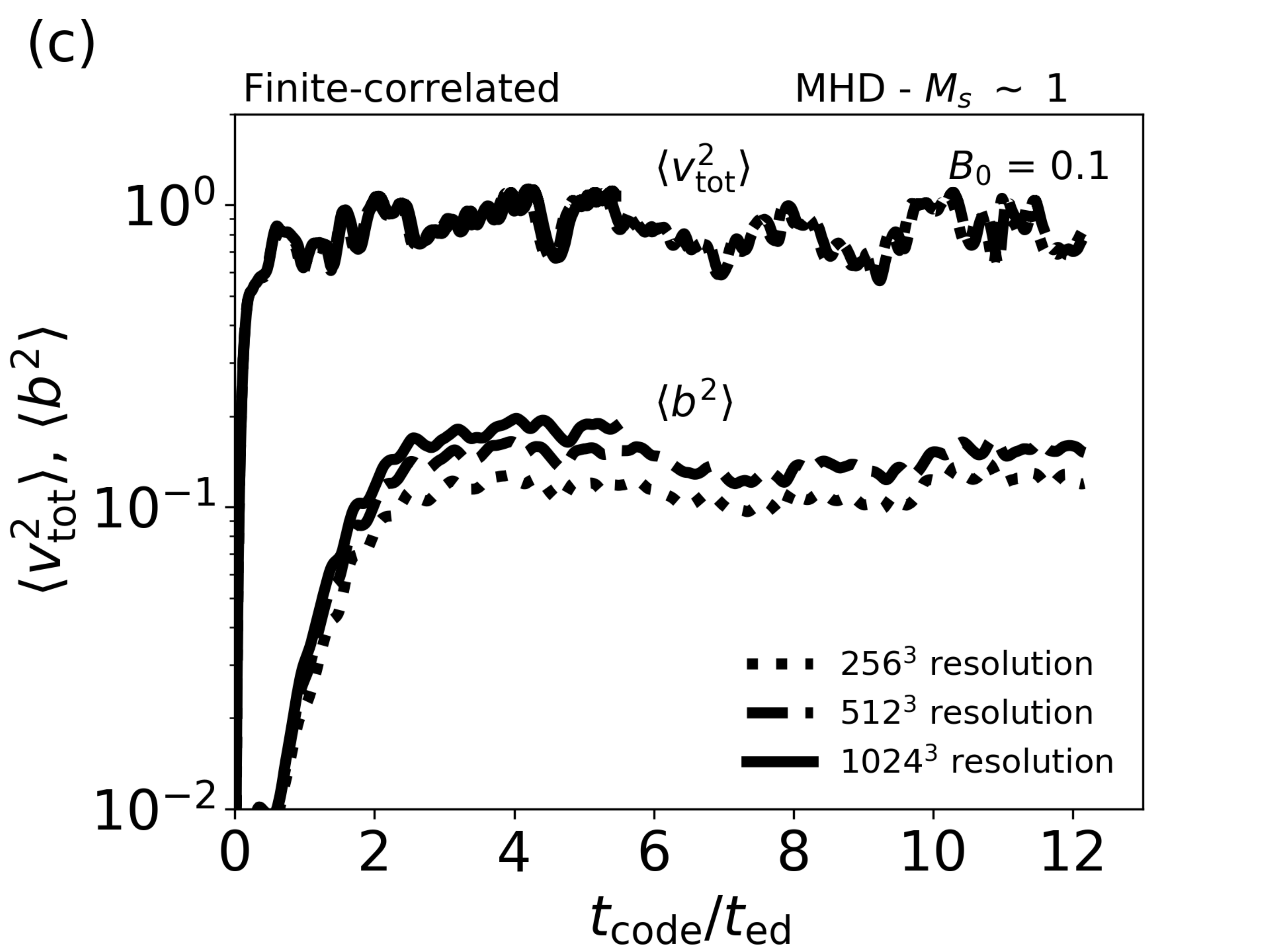} 
\caption{Time evolution of $\langle \vtotsq\rangle $ and $\langle \bsq\rangle$ for MHD turbulence simulations with high numerical resolutions and different values of $B_0$. (a) The same as Figure \ref{fig:fig13}(b), but for $512^3$ resolution. (b) Comparison for $M_s$ and numerical resolution effects in the case of $B_0$ = 0.001. Solid curves with blue, red, and green colors represent $M_s$ $\sim$ 0.5, $\sim$ 1, and $\sim$ 3 for $256^3$ resolution, respectively. Red dotted curves denote $512^3$ resolution simulation with $M_s$ $\sim$ 1. (c) Resolution study only for $\bzero$ = 0.1 and $M_s$ $\sim$ 1. Dotted, dashed, and solid curves indicate $256^3$, $512^3$, and $1024^3$ resolutions, respectively.
\label{fig:fig14}}
\end{figure*}

\begin{figure}
\centering
\includegraphics[scale=0.25]{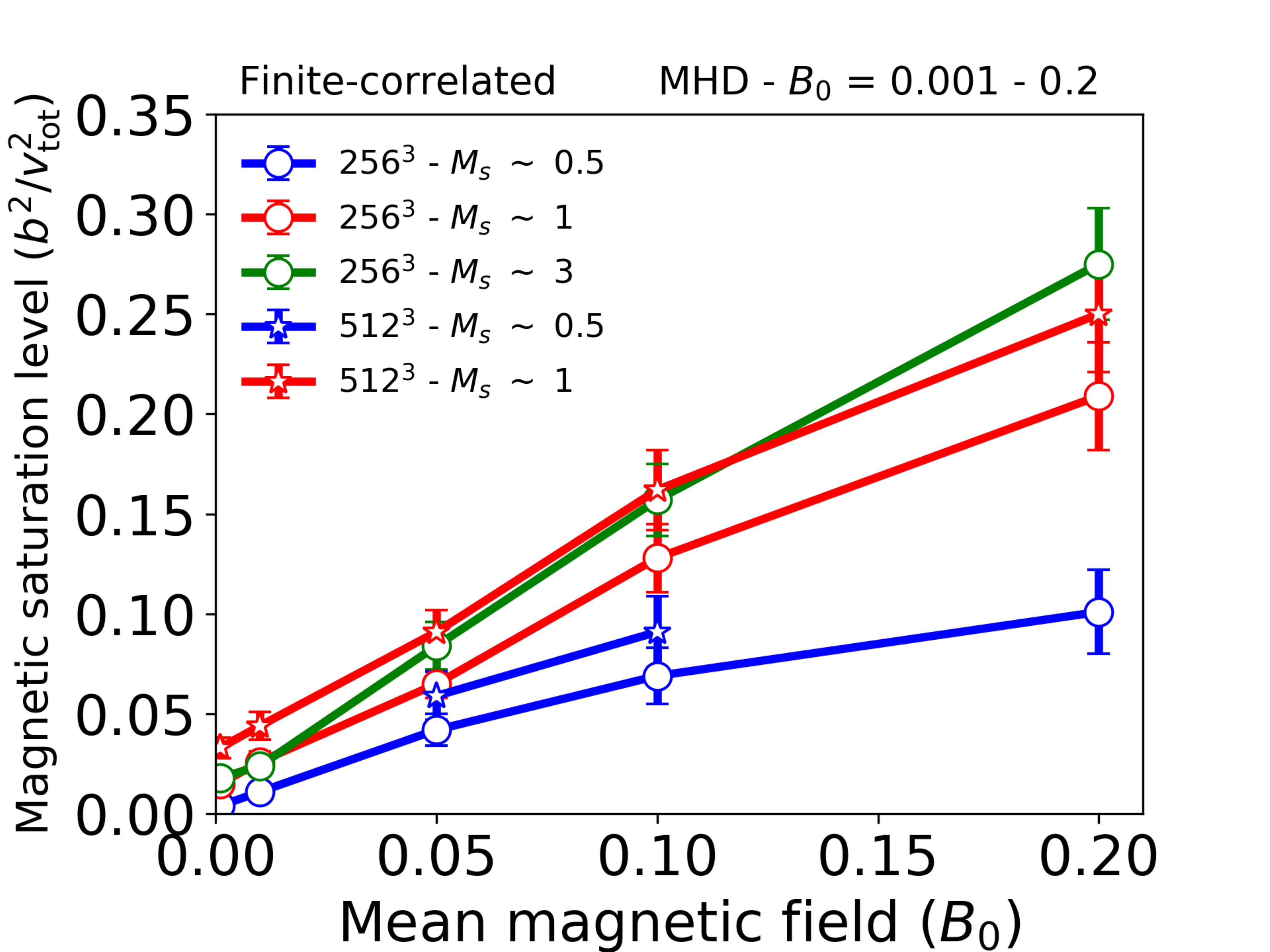}
\caption{Average values of the magnetic saturation level as a function of the mean magnetic field ($\bzero$) for MHD turbulence simulations with $\mach$ $\lesssim$ 3 and $B_0$ $\leq$ 0.2. Circles with blue, red, and green colors represent $256^3$ resolution simulations with $M_s$ $\sim$ 0.5, $\sim$ 1, and $\sim$ 3, respectively. Blue and red stars denote $512^3$ resolution simulations with $M_s$ $\sim$ 0.5 and $\sim$ 1, respectively. The error bars show standard deviations. In Table \ref{tab:tab1}, the average values and the time intervals for taking average are shown. 
\label{fig:fig15}}
\end{figure}


\begin{figure*}[ht!]
\centering
\includegraphics[scale=0.2]{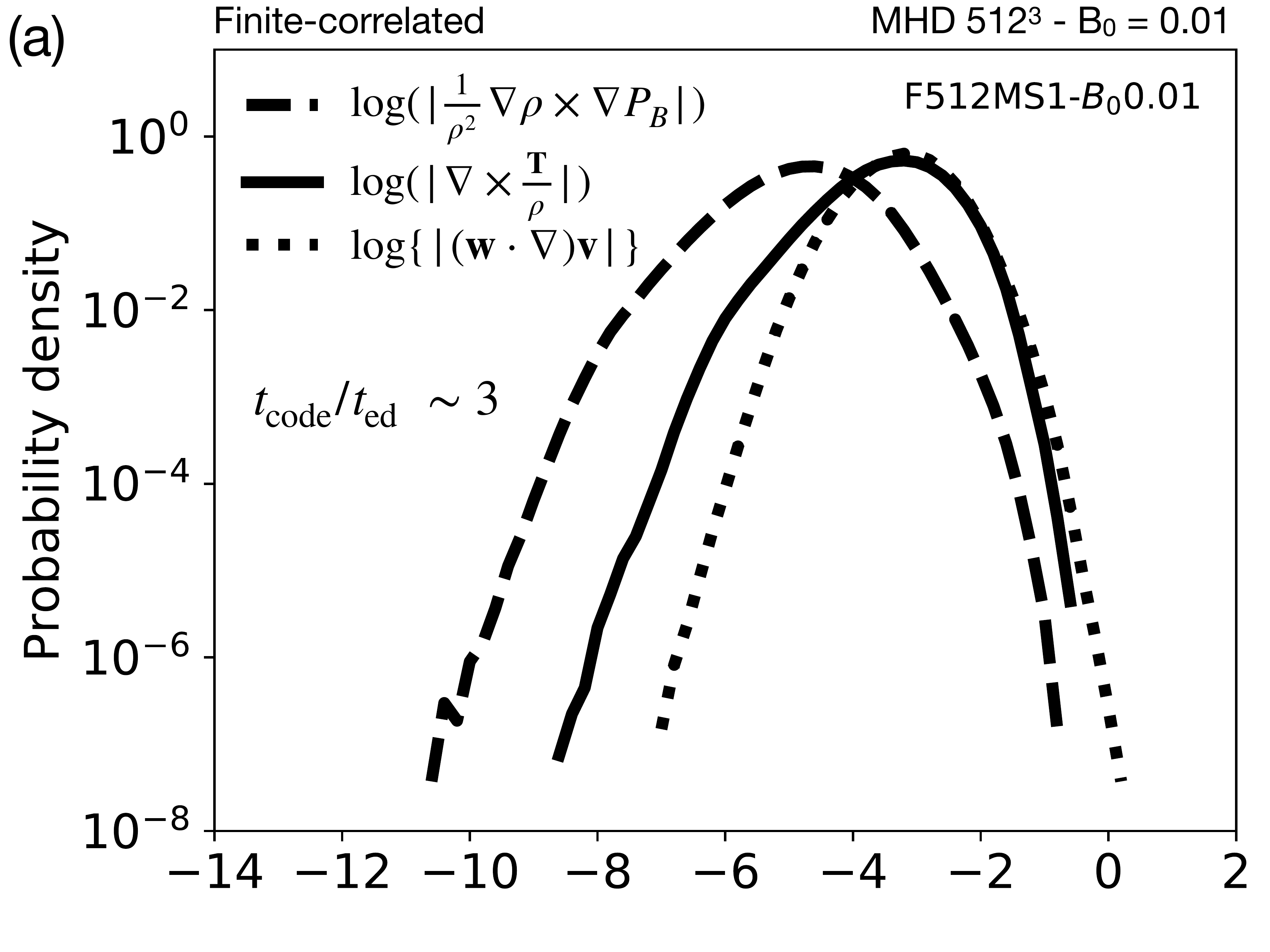}
\includegraphics[scale=0.2]{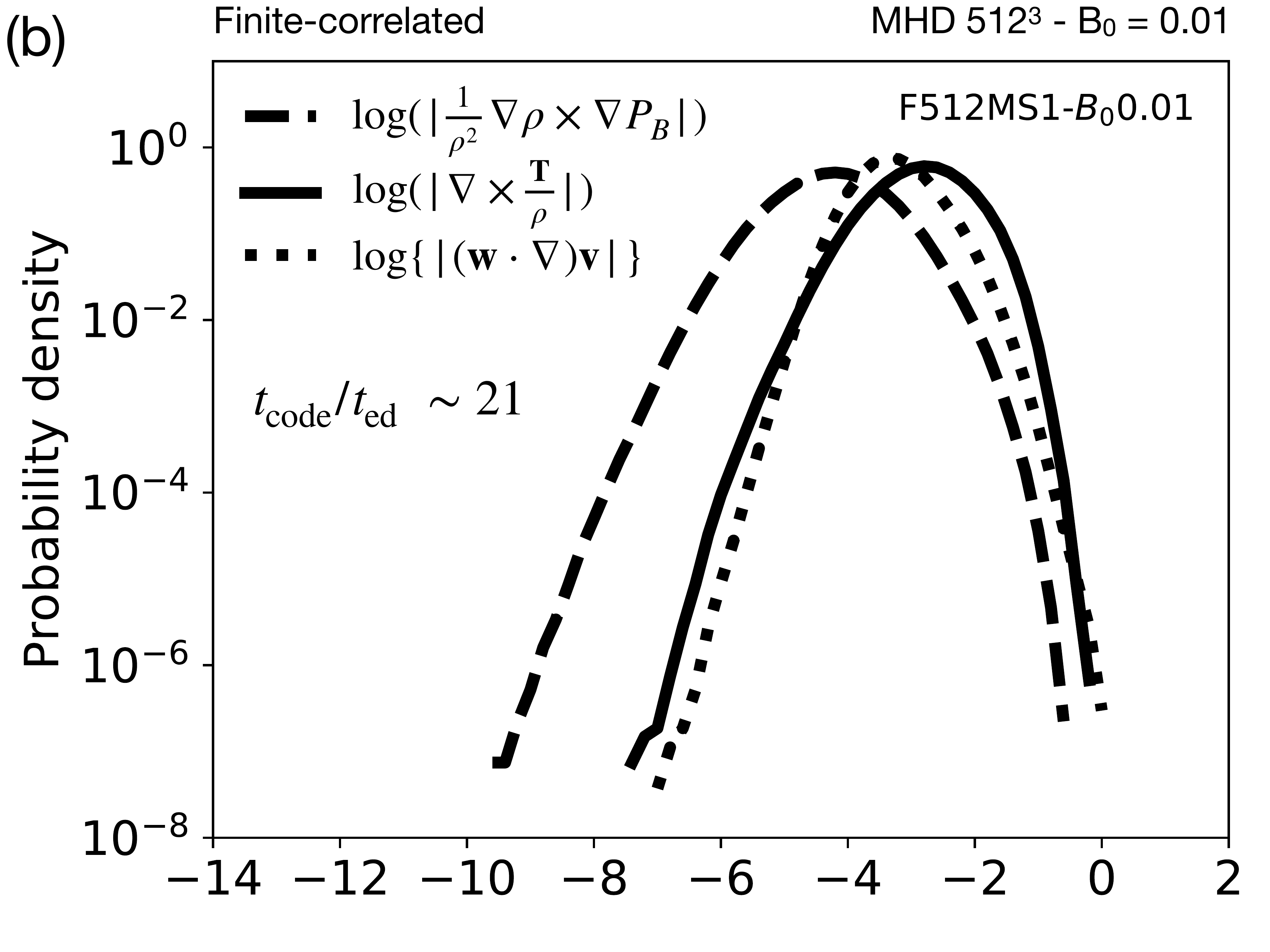} \\
\includegraphics[scale=0.2]{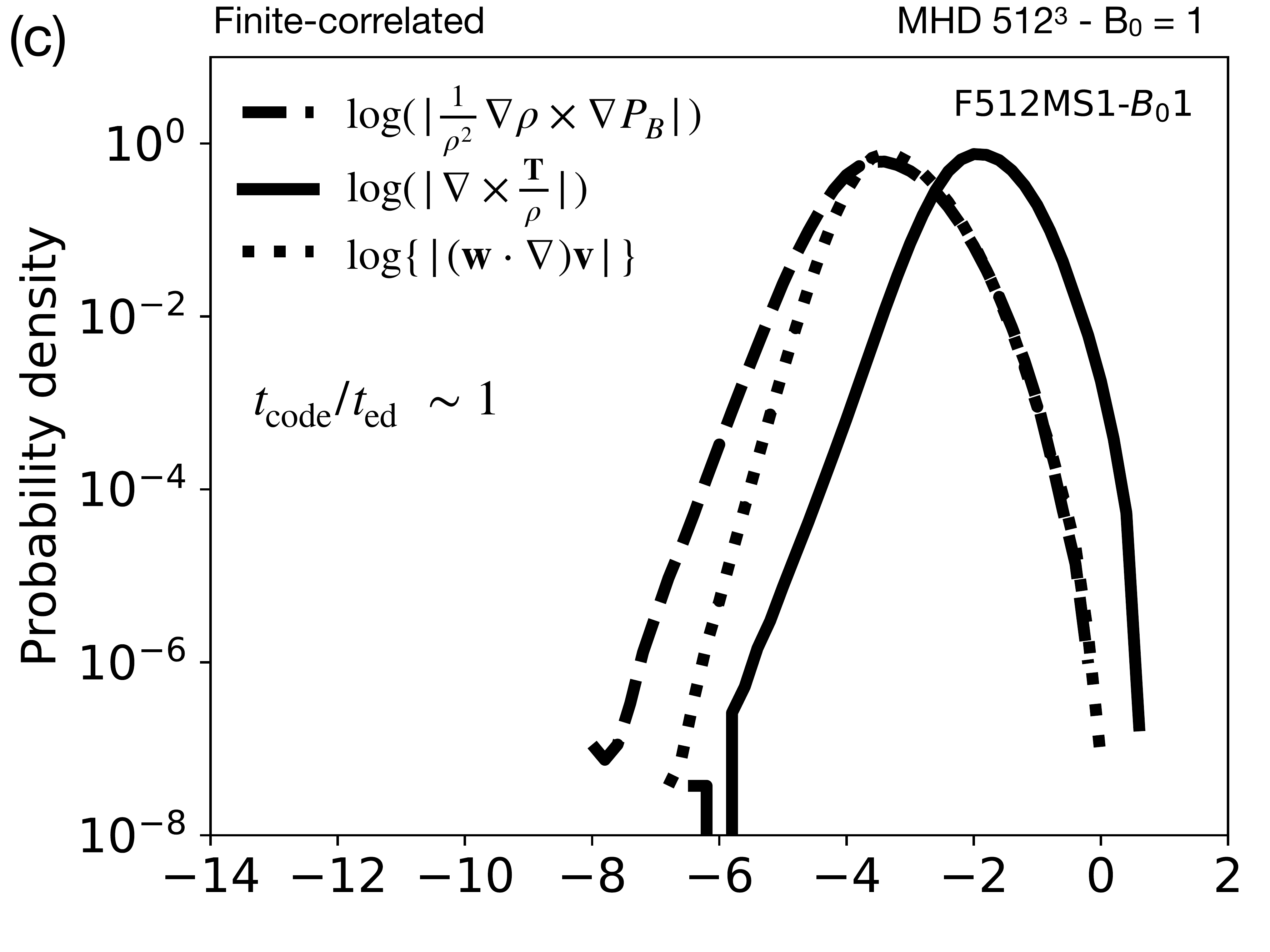} 
\includegraphics[scale=0.2]{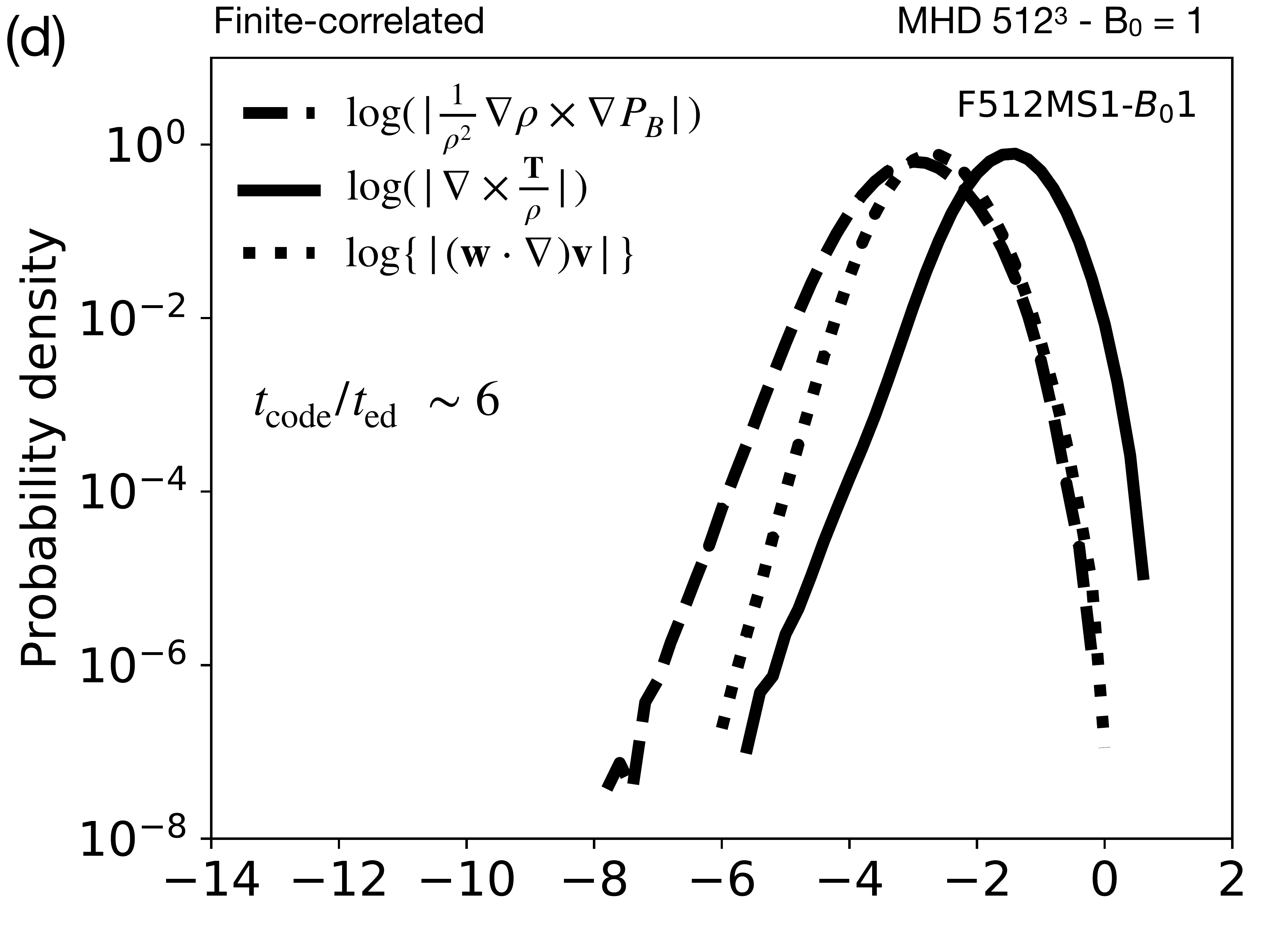}
\caption{PDFs of $\mathrm{log}(|\frac{1}{\rho^2}\nabla \rho \times \nabla P_B|)$ (dashed curves), $\mathrm{log}(|\nabla \times \frac{\mathbf{T}}{\rho}|)$ (solid curves), and $\mathrm{log}\{|(\mathbf{w}\cdot \nabla)\mathbf{u}|\}$ (dotted curves) for MHD simulations with $\mach$ $\sim$ 1 and $512^3$ resolution in the case of the finite-correlated compressive driving. Upper panels: $\bzero$ = 0.01. Lower panels: $\bzero$ = 1.0. The PDFs in the left panels are calculated before saturation, and those in the right panels after saturation. The time at which each PDF is calculated is shown in each panel. 
\label{fig:fig16}}
\end{figure*}


First of all, Figures \ref{fig:fig9}(a) and \ref{fig:fig9}(b) obviously reveal the effect of $\mach$ on the growth of $\langle \bsq \rangle$: although the level of $\langle \vtotsq\rangle $ at saturation is almost identical regardless of $\mach$, that of $\langle \bsq\rangle $ depends on $\mach$. When we compare $\langle b^2 \rangle$ for $M_s$ $\sim$ 0.5 (see the blue dotted curves) and $\sim$ 1 (see the red dotted curves) in those figures, the values for $M_s$ $\sim$ 1 are larger. However, when we compare $\langle b^2 \rangle$ for $M_s$ $\sim$ 1 and $\sim$ 3 (see the green dotted curves), the behaviors of the finite-correlated (left panel) and the delta-correlated (middle panel) drivings are different: the finite-correlated compressive driving yields similar level of $\langle \bsq\rangle $ between $\mach$ $\sim$ 1 and $\sim$ 3, while the delta-correlated compressive driving produces larger $\langle \bsq\rangle$ for $M_s$ $\sim$ 3.  Second, both Figures \ref{fig:fig9}(a) and \ref{fig:fig9}(b) have a common feature: $\langle \bsq \rangle$ grows fast when t $\lesssim$ 5 and it gradually levels off when t $\gtrsim$ 5. In some simulations, $\langle \bsq \rangle$ reaches saturation level relatively quickly, while in others, it does more or less slowly. Lastly, as opposed to the case of $\langle \vsolsq \rangle$, Figure \ref{fig:fig9}(c) clearly shows resolution dependence of $\langle \bsq\rangle $, with the level of $\langle \bsq\rangle$ at saturation being increasing with the numerical resolution for both driving schemes. Previous numerical studies reported such resolution effect on turbulence dynamo for solenoidally driven turbulence (for incompressible MHD turbulence, see, e.g., \citealt{Cho09}; for compressible MHD turbulence, see, e.g., \citealt{Ryu08}).

Figure \ref{fig:fig10} summarizes the results from Figure \ref{fig:fig9}. The figure shows the magnetic saturation level as a function of $\mach$. Black and magenta circles (squares) correspond to the finite-correlated and the delta-correlated compressive drivings at $256^3$ ($512^3$) resolution, respectively. The magnetic saturation level is clearly dependent on $\mach$, driving schemes, and the numerical resolution. First, for the range of $M_s$ presented in Figure \ref{fig:fig10}, the two driving schemes show different behaviors when we fix the numerical resolution: although it keeps increasing as $\mach$ increases for the delta-correlated compressive driving, it is not the case for the finite-correlated compressive driving. That is, the magnetic saturation level for the finite-correlated compressive driving peaks at $M_s$ $\sim$ 1 and slightly decreases for $M_s$ $\gtrsim$ 1. Second, for the same numerical resolution, we can clearly see that the finite-correlated compressive driving yields a larger magnetic saturation level than the delta-correlated compressive driving at a similar $\mach$. Third, when we compare the magnetic saturation level for $256^3$ and $512^3$ resolutions, it is larger for the latter resolution.

\subsubsection{Strong $B_0$ Cases ($B_0$ = 1)\label{sec:sec4.1.2}}
Figure \ref{fig:fig11} shows time evolution of $\langle \vtotsq\rangle$ and $\langle \bsq\rangle$ for $\bzero$ = 1. We consider the finite-correlated and the delta-correlated compressive drivings separately in Figures \ref{fig:fig11}(a) and \ref{fig:fig11}(b), respectively. In the figures, blue, red, cyan, and orange solid (dotted) curves represent $\langle \vtotsq\rangle$ $(\langle \bsq\rangle)$ for $\mach$ $\sim$ 0.5, $\sim$ 1, $\sim$ 3, and $\sim$ 10, respectively. The levels of $\langle \bsq\rangle$ at saturation in the figures are not very sensitive to $\mach$ for both driving schemes. Figure \ref{fig:fig11}(c) shows results of resolution study for the finite-correlated compressive driving in the case of $M_s$ $\sim$ 1. Dashed and solid curves denote $256^3$ and $512^3$ numerical resolutions, respectively. We can see from Figure \ref{fig:fig11}(c) that the resolution effect on $\langle \bsq\rangle$ is not very significant.
 
Figure \ref{fig:fig12} shows the magnetic saturation level as a function of $M_s$ for $\bzero$ = 1.0. Black and magenta circles (squares) correspond to the finite-correlated and the delta-correlated compressive drivings for $256^3$ ($512^3$) resolution, respectively. Although weak, the magnetic saturation level shows dependence on $\mach$: it goes up as $M_s$ increases when $M_s$ $\lesssim$ 1 and it decreases as $M_s$ increases when $M_s$ $\gtrsim$ 1 for both driving schemes. In addition, numerical resolution effect on the magnetic saturation level is hardly pronounced especially for the finite-correlated compressive driving.

%
\begin{figure*}[t!]
\centering 
\includegraphics[scale=0.15]{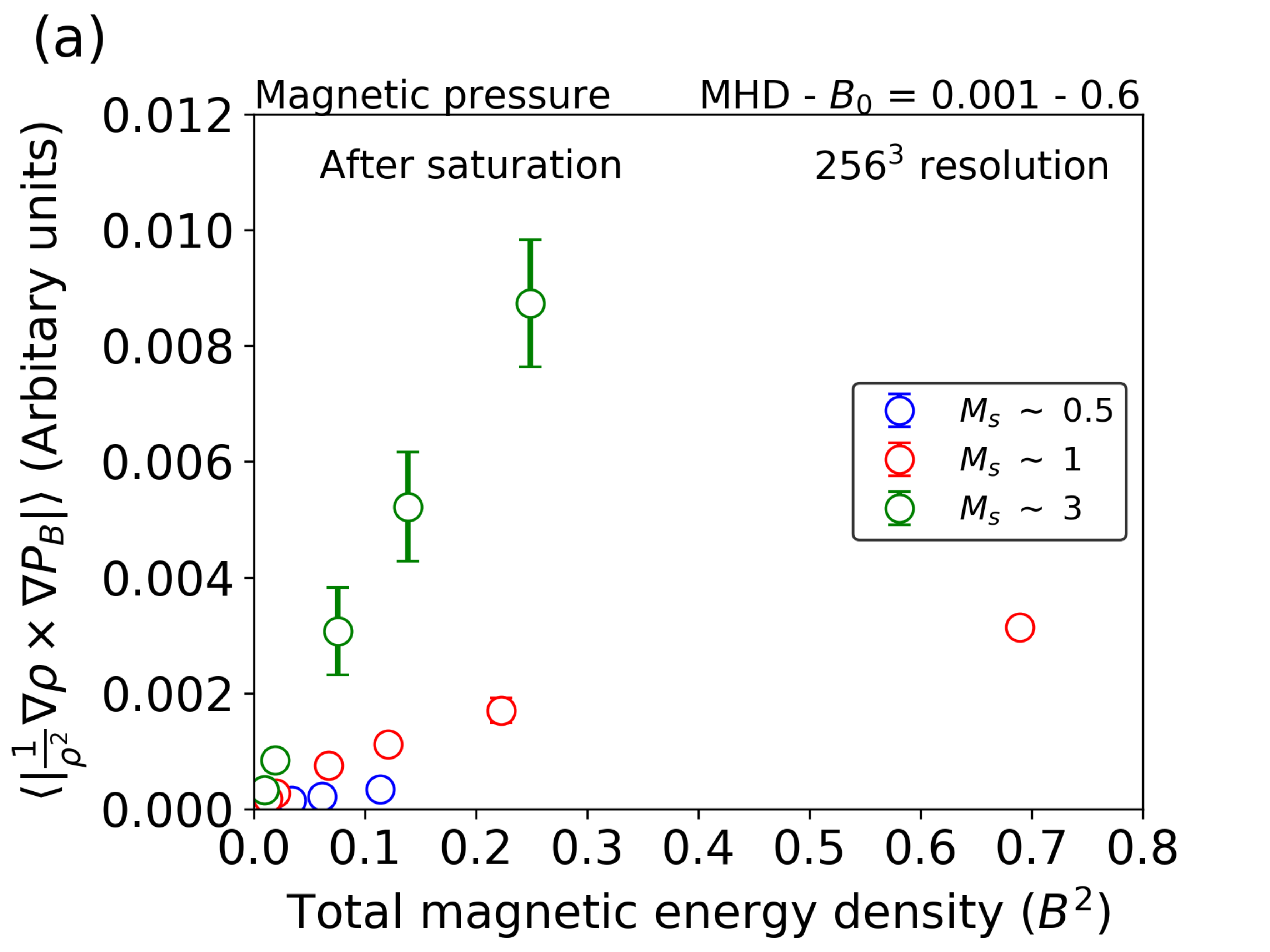}
\includegraphics[scale=0.15]{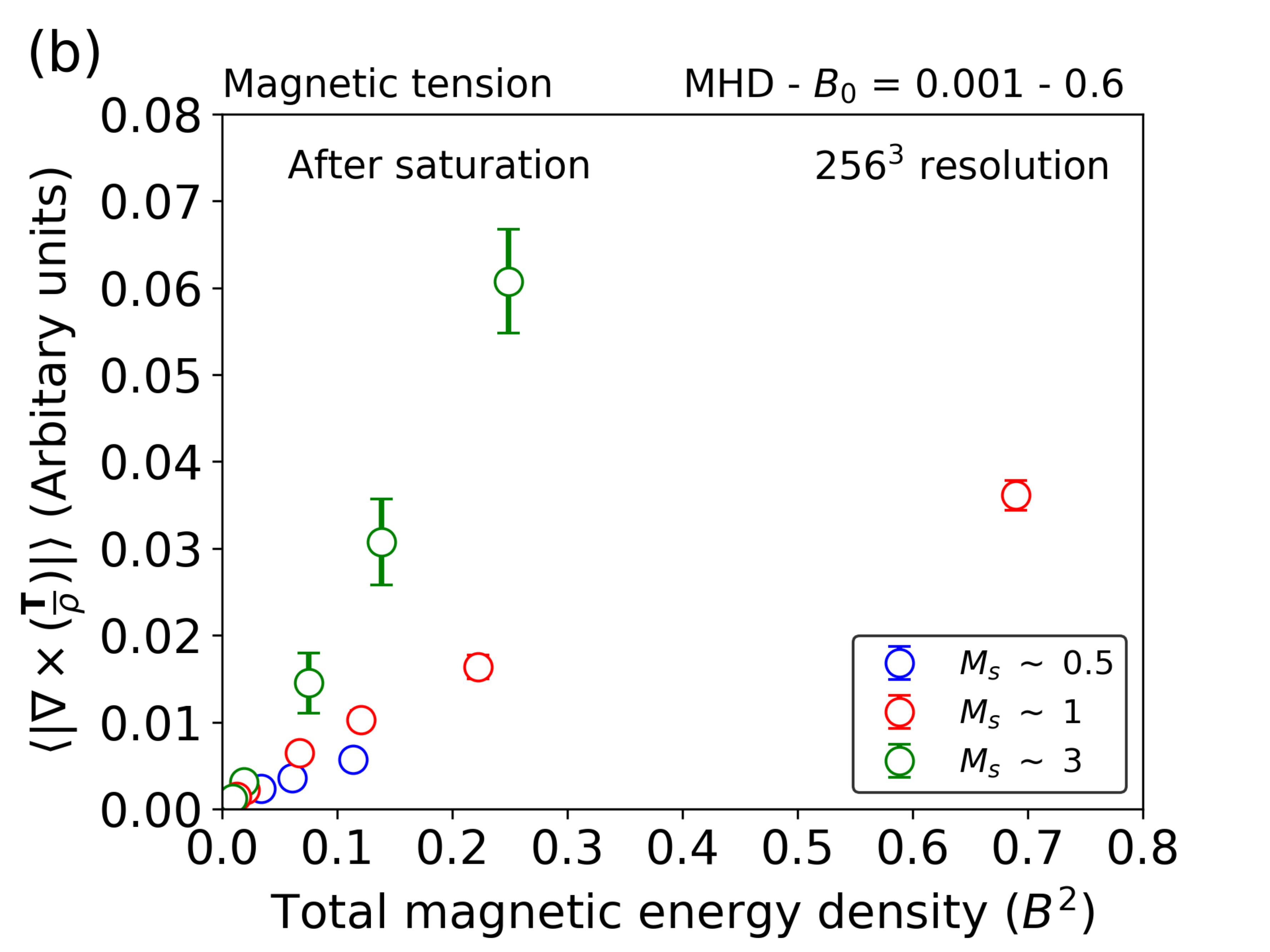} 
\includegraphics[scale=0.15]{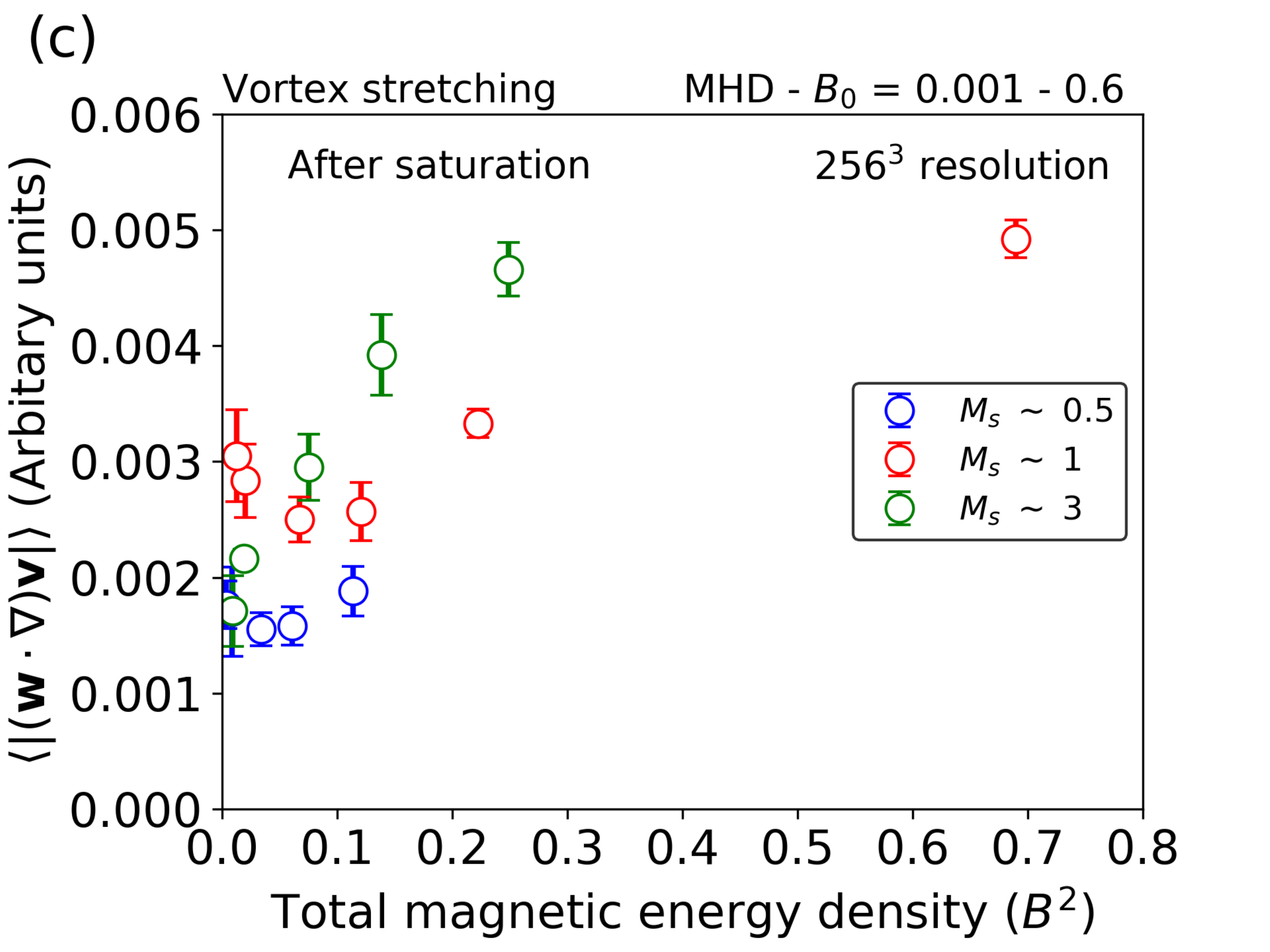} 
\caption{Average values of $\langle|\frac{1}{\rho^2}\nabla \rho \times \nabla P_B|\rangle$ (left panel), $\langle|\nabla \times \frac{\mathbf{T}}{\rho}|\rangle$ (middle panel), and $\langle|(\mathbf{w}\cdot \nabla)\mathbf{v}|\rangle$ (right panel) for MHD simulations with $256^3$ resolution and $B_0$ $\leq$ 0.6 as a function of total magnetic energy density ($B^2$). Blue, red, and green circles in each panel denote $M_s$ $\sim$ 0.5, $\sim$ 1, and $\sim$ 3, respectively. The error bars represent standard deviations. We take the average values after saturation.
\label{fig:fig17}}
\end{figure*}


\subsection{Effects of the Mean Magnetic Field\label{sec:sec4.2}}

In this subsection, we deal with the effect of the mean magnetic field $\bzero$ on small-scale dynamo in turbulence driven by compressive driving. Figure \ref{fig:fig13} shows time evolution of $\langle \vtotsq\rangle$ (solid curves) and $\langle \bsq\rangle$ (dotted curves) for different mean magnetic field strengths for the finite-correlated compressive driving. From left to right, Figures \ref{fig:fig13}(a)-(c) are for $M_s$ $\sim$ 0.5, $\sim$ 1, and $\sim$ 3, respectively. Cyan, brown, blue, orange, and black curves in each panel indicate $\bzero$ = 0.01, 0.05, 0.1, 0.2, and 0.6, respectively. Note that in each panel we present results for various values of $B_0$. The vertical axis is in logarithmic scale in all panels. As we can see from the figure, the level of $\langle \bsq\rangle$ at saturation increases with $B_0$ for all $M_s$'s. In general, when $B_0$ is weaker, the saturation level is lower and it takes more time to reach it. 

In Figure \ref{fig:fig14}, we present time evolution of $\langle \vtotsq\rangle$  and $\langle \bsq\rangle$ from additional MHD simulations with either higher numerical resolutions or different value of $B_0$. In Figure \ref{fig:fig14}(a), we present results for $512^3$ resolution. In Figure \ref{fig:fig14}(b), we present results for $B_0$ = 0.001. In Figure \ref{fig:fig14}(c), we compare effects of numerical resolution for fixed $B_0$ and $M_s$.

Figure \ref{fig:fig14}(a) is the same as Figure \ref{fig:fig13}(b), but for $512^3$ resolution: in this figure, cyan, brown, blue, orange, and black solid (dotted) curves denote time evolution of $\langle \vtotsq \rangle$ ($\langle \bsq \rangle$) for $B_0$ = 0.01, 0.05, 0.1, 0.2, and 0.6 at $M_s$ $\sim$ 1, respectively. We can clearly see almost the same evolutions of both $\langle \vtotsq \rangle$ and $\langle \bsq \rangle$ as those for $256^3$ resolution (compare Figure \ref{fig:fig14}(a) with Figure \ref{fig:fig13}(b)). However, $512^3$ resolution gives slightly higher saturation values of $\langle \bsq \rangle$ than $256^3$ resolution (see Figure \ref{fig:fig15} for details).

Figure \ref{fig:fig14}(b) simultaneously shows effects of both $M_s$ and numerical resolution on $\langle \bsq \rangle$ in the case of $B_0$ = 0.001. In this figure, blue solid, red solid, and green solid curves correspond to $M_s$ $\sim$ 0.5, $\sim$ 1, and $\sim$ 3 for $256^3$ resolution, respectively. Red dotted curves are for $512^3$ resolution in the case of $M_s$ $\sim$ 1. First of all, time evolution of $\langle \bsq \rangle$ for $M_s$ $\sim$ 0.5 clearly shows slower growth of $\langle \bsq \rangle$ compared to that of $M_s$ $\sim$ 1 and $\sim$ 3: $\langle \bsq \rangle$ for $M_s$ $\gtrsim$ 1 saturates roughly after 60$t_{\textrm{ed}}$ but that for $M_s$ $\sim$ 0.5 grows until 140$t_{\textrm{ed}}$. Furthermore, the level of $\langle \bsq \rangle$ at saturation is similar between $M_s$ $\sim$ 1 and $\sim$ 3 and that for $M_s$ $\sim$ 0.5 is lower. Second, when we compare time evolution of $\langle \bsq \rangle$ for $256^3$ and $512^3$ resolutions at $M_s$ $\sim$ 1 (see red solid and red dotted curves in Figure \ref{fig:fig14}(b)), it is obvious for $\langle \bsq \rangle$ of the latter resolution to grow faster and to exhibit a higher level of $\langle \bsq \rangle$ than that of the former resolution. 

Figure \ref{fig:fig14}(c) shows resolution study in the case of $B_0$ = 0.1 and $M_s$ $\sim$ 1. Dotted, dashed, and solid curves indicate $256^3$, $512^3$, and $1024^3$ resolutions, respectively. According to the figure, three different resolution simulations produce almost same time evolution of $\langle \vtotsq \rangle$. On the other hand, a simulation with a higher numerical resolution results in a higher level of $\langle \bsq \rangle$ at saturation. However, we can clearly see that the change of the level of $\langle b^2 \rangle$ at saturation with numerical resolution is not significant in compressively driven turbulence.

Figure \ref{fig:fig15} denotes the magnetic saturation level as a function of $\bzero$ for the simulations with $M_s$ $\lesssim$ 3. Blue, red, and green circles indicate $M_s$ $\sim$ 0.5, $\sim$ 1, and $\sim$ 3 for $256^3$ resolution, respectively. Blue and red stars correspond to $M_s$ $\sim$ 0.5 and $\sim$ 1 for $512^3$ resolution, respectively. The figure reveals approximately linear relation between $\bzero$ and the magnetic saturation level, 
\begin{equation}
\langle b^2 \rangle/\langle v_{\textrm{tot}}^2 \rangle = cB_0 + d, 	
\end{equation}
regardless of $M_s$ and numerical resolution. Here $\textit{c}$ and $\textit{d}$ are constants with proper dimensions. If we compare circles ($256^3$) and stars ($512^3$) at the same $M_s$ (i.e., the same color), we can clearly see that $512^3$ resolution gives higher saturation levels than $256^3$ resolution. The slopes of the linear relation are very similar for both resolutions. Note, however, that the increase of the magnetic saturation level with numerical resolution is not significant compared to that for solenoidally driven turbulence (see Section \ref{sec:sec6} for further discussion). If we compare results for different $M_s$'s, the slope becomes slightly steeper as $M_s$ increases at the same numerical resolution.

\begin{figure*}[t!]
\centering 
\includegraphics[scale=0.15]{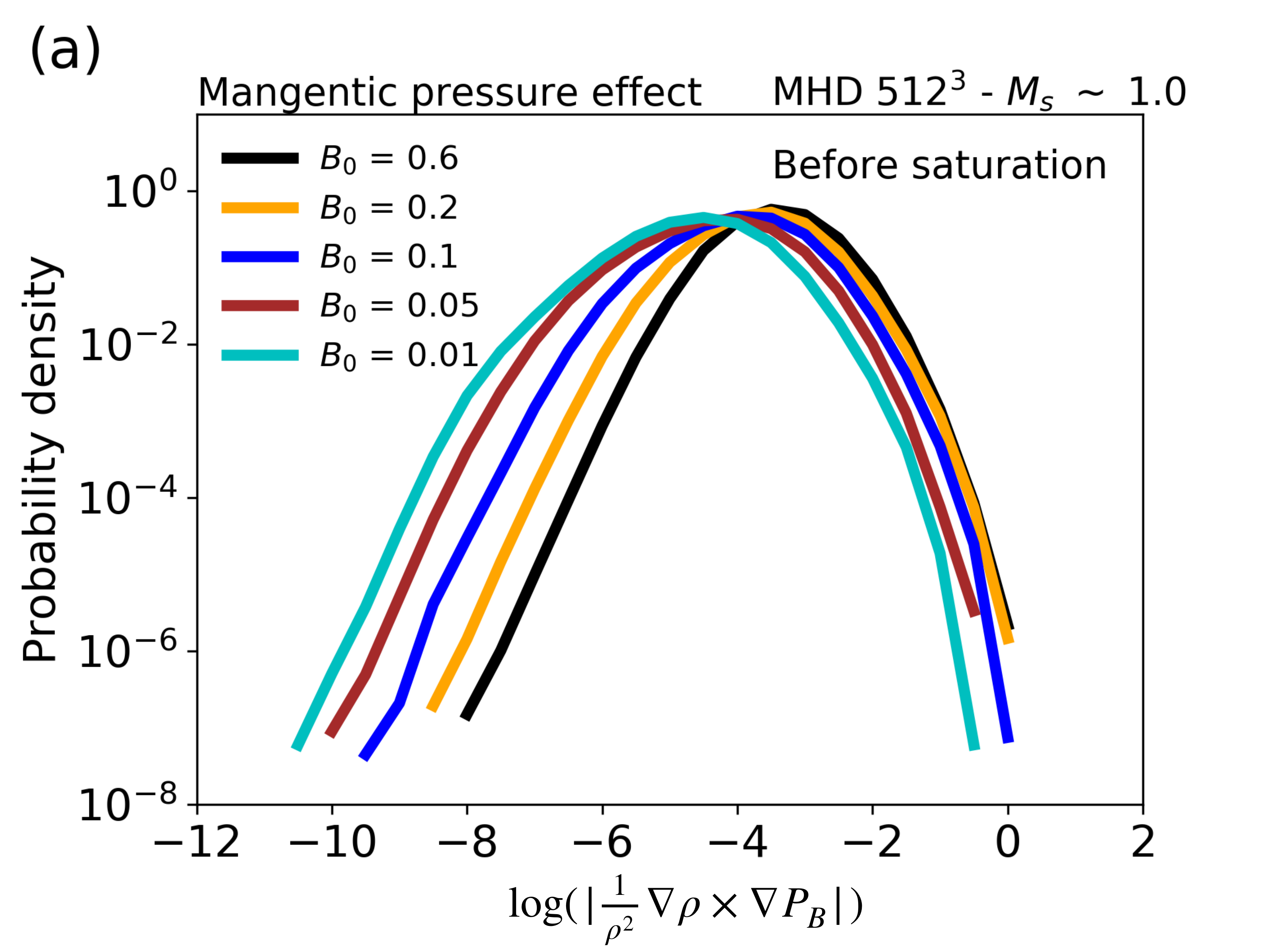}
\includegraphics[scale=0.15]{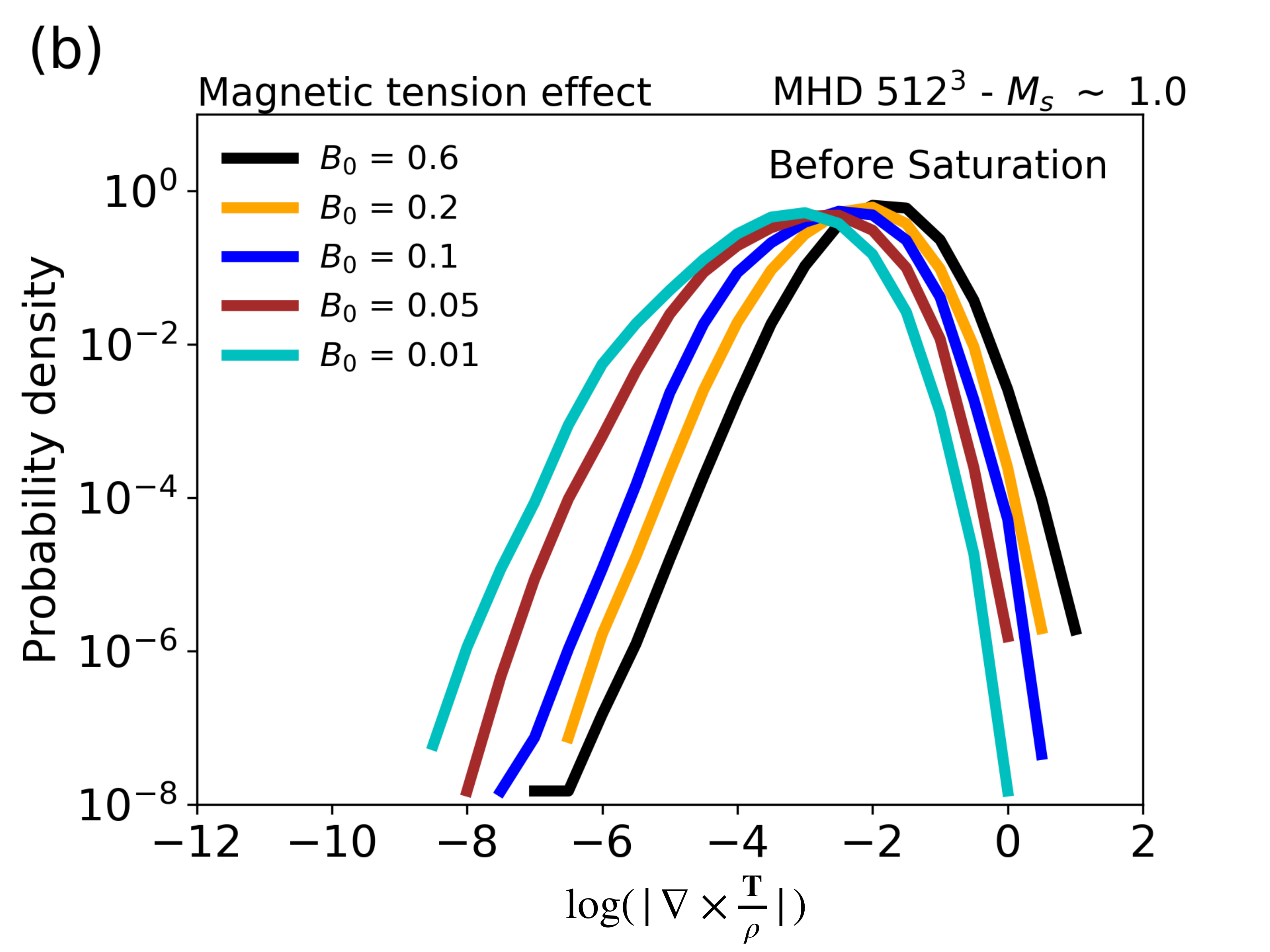} 
\includegraphics[scale=0.15]{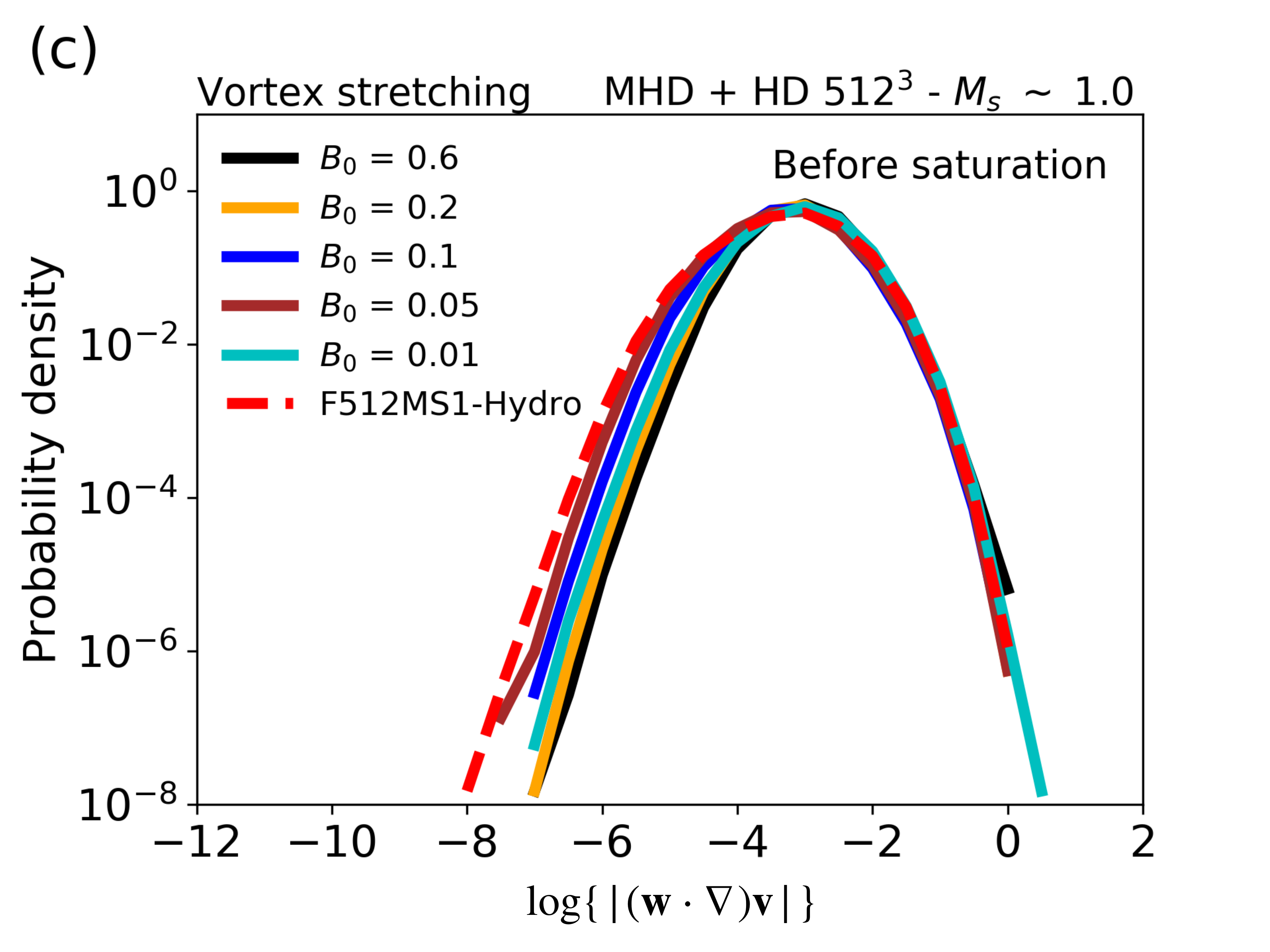} 
\caption{PDFs of $\mathrm{log}(|\frac{1}{\rho^2}\nabla \rho \times \nabla P_B|)$ (left panel), $\mathrm{log}(|\nabla \times \frac{\mathbf{T}}{\rho}|)$ (middle panel), and $\mathrm{log}\{|(\mathbf{w}\cdot \nabla)\mathbf{v}|\}$ (right panel) for MHD simulations with $M_s$ $\sim$ 1 and $512^3$ resolution. Cyan, brown, blue, orange, and black curves in each panel represent $\bzero$ = 0.01, 0.05, 0.1, 0.2, and 0.6, respectively. In the right panel, red dashed curve indicates HD simulation, F512MS1-Hydro. The PDFs are calculated before saturation.
\label{fig:fig18}}
\end{figure*}


\begin{figure*}[ht!]
\centering 
\includegraphics[scale=0.15]{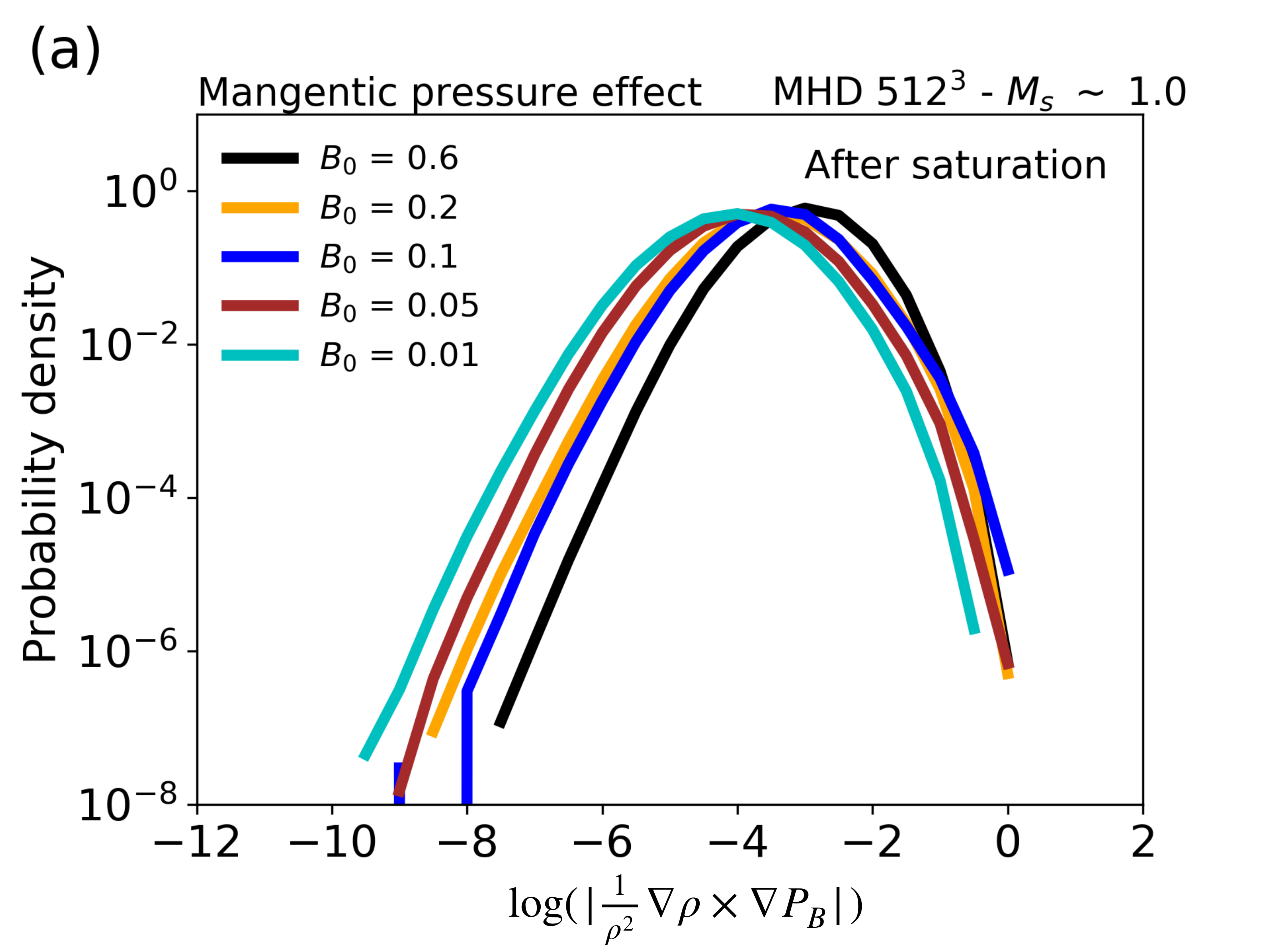}
\includegraphics[scale=0.15]{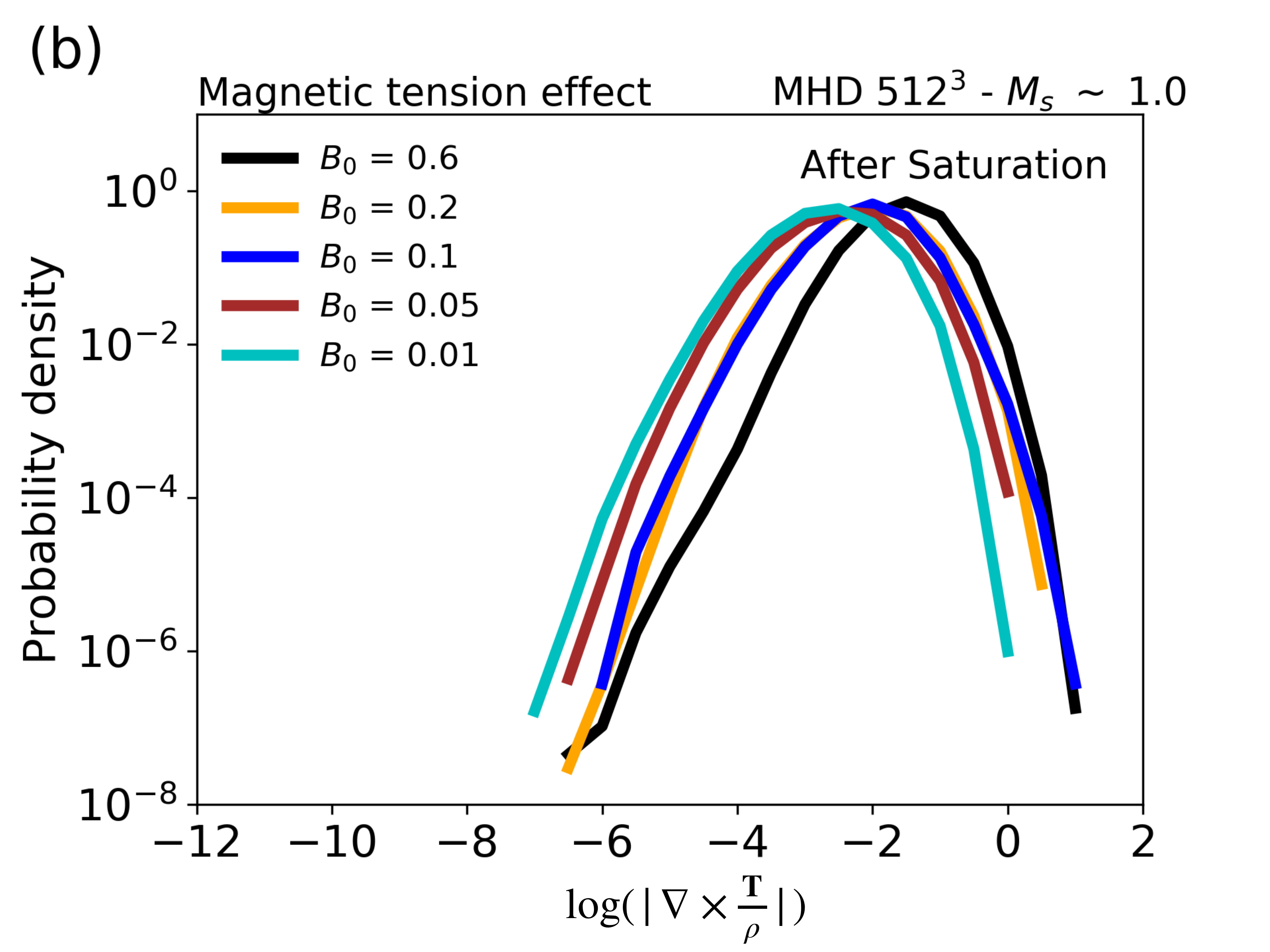} 
\includegraphics[scale=0.15]{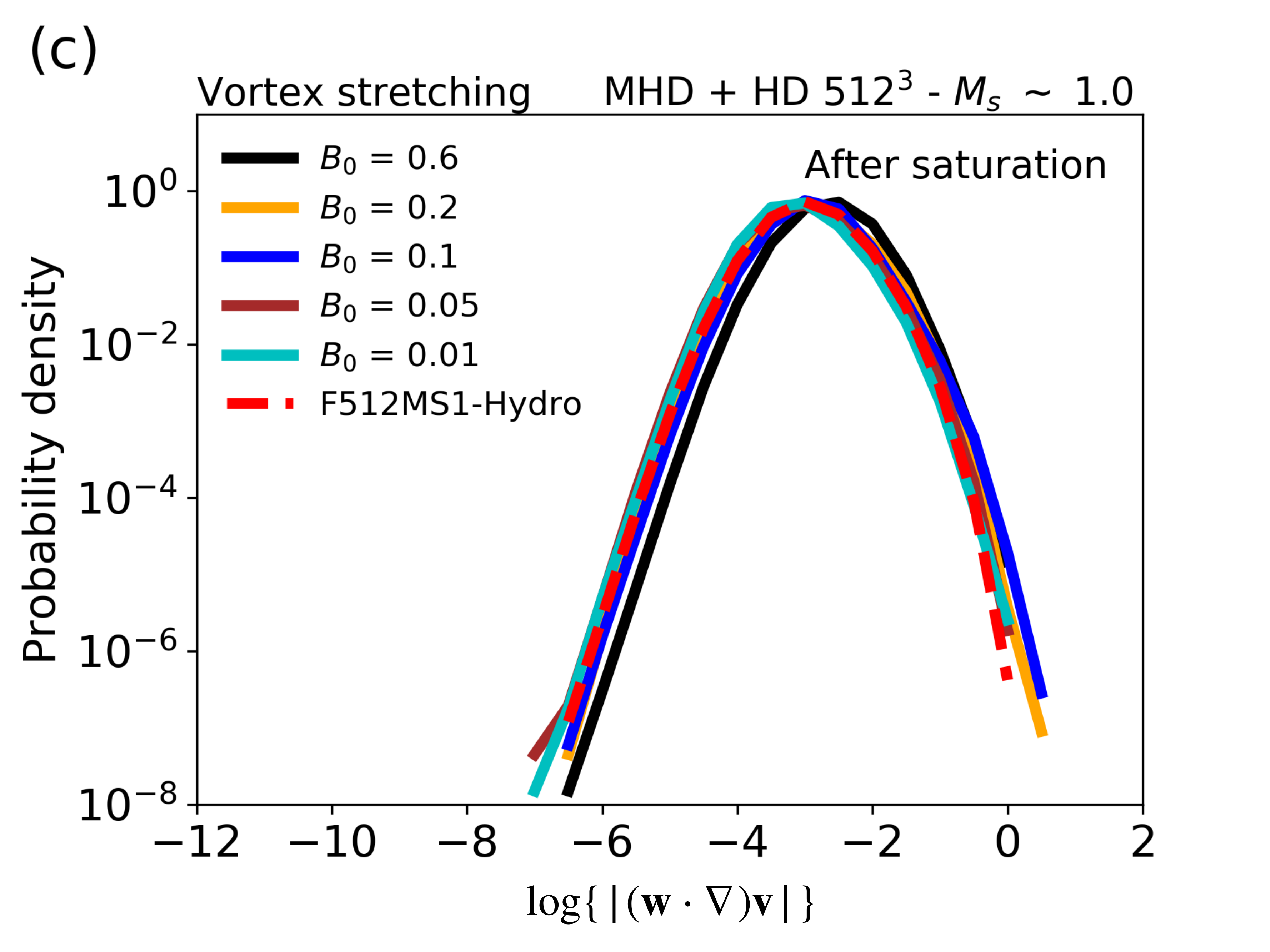} 
\caption{The same as Figure \ref{fig:fig18}, but for the PDFs calculated after saturation.
\label{fig:fig19}}
\end{figure*}


\begin{figure}
\centering
\includegraphics[scale=0.25]{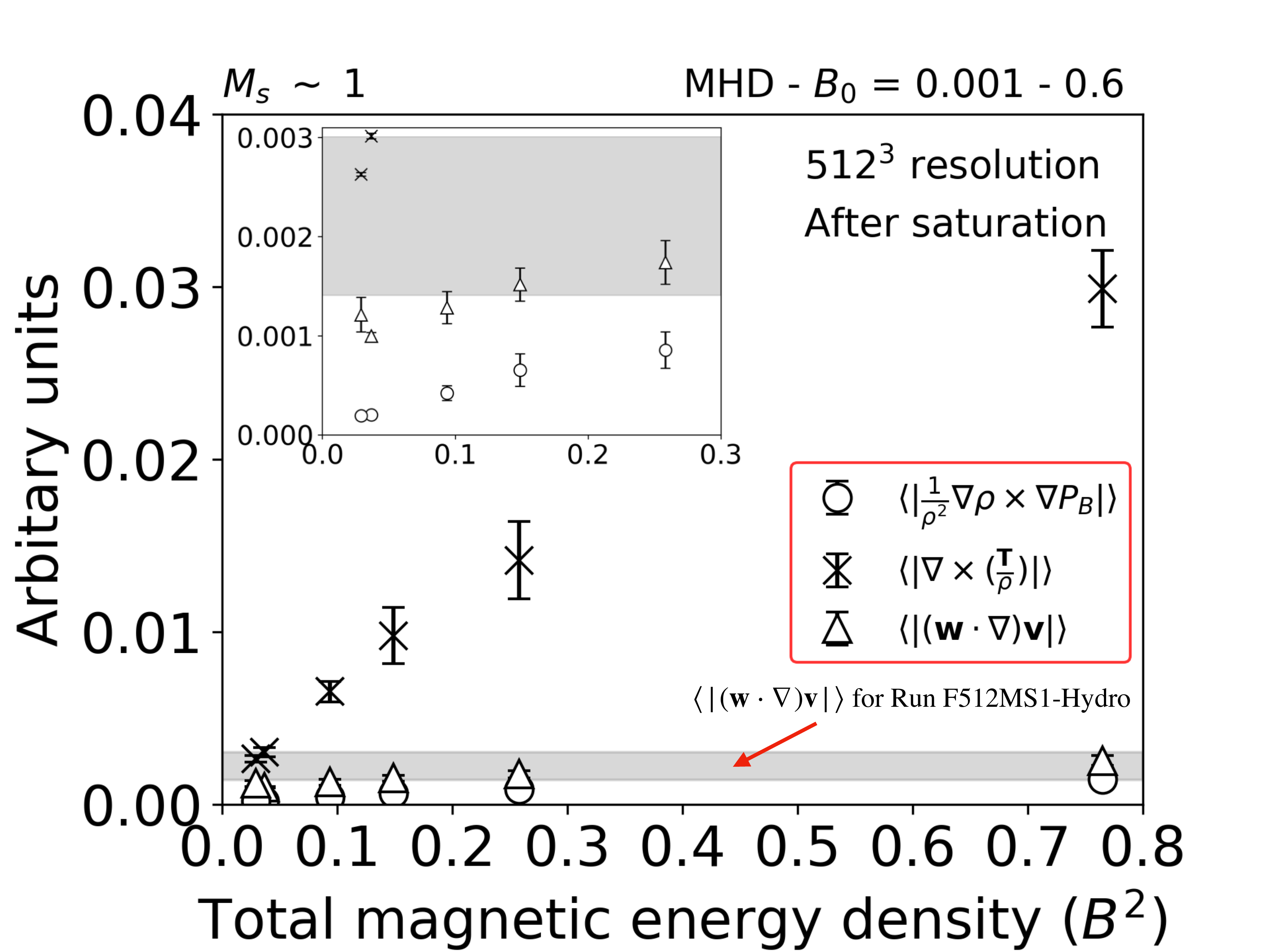}
\caption{Average values of $\langle|\frac{1}{\rho^2}\nabla \rho \times \nabla P_B|\rangle$ (circles), $\langle|\nabla \times \frac{\mathbf{T}}{\rho}|\rangle$ (``X" markers), and $\langle|(\mathbf{w}\cdot \nabla)\mathbf{v}|\rangle$ (triangles) at saturation as a function of total magnetic energy density ($B^2$). The grey shaded region represents 1$\sigma$-dispersion about the average value of $\langle|(\mathbf{w}\cdot \nabla)\mathbf{v}|\rangle$ for the HD simulation, Run F512MS1-Hydro. The error bars represent the standard deviations. The inset in the upper left corner shows a zoom to clearly illustrate the trend of $\langle|\frac{1}{\rho^2}\nabla \rho \times \nabla P_B|\rangle$ and $\langle|(\mathbf{w}\cdot \nabla)\mathbf{v}|\rangle$ with $B^2$. 
\label{fig:fig20}}
\end{figure}



\section{Discussion on Solenoidal Ratio\label{sec:sec5}}
In Section \ref{sec:sec3}, we have studied effects of both the sonic Mach number $(\mach)$ and the mean field strength $(\bzero)$ on the generation of solenoidal velocity component in compressively driven turbulence. We have found that the solenoidal component produced by compressive driving is dependent on $M_s$ and $\bzero$. In addition, when $M_s$ is similar and the numerical resolution is same, the finite-correlated compressive driving generates more solenoidal velocity component than the delta-correlated compressive driving. We discuss these findings and their implications in this section. 

\subsection{Dependence of Solenoidal Ratio on $\mach$ and $\bzero$\label{sec:sec5.1}}
Let us discuss how larger $M_s$ and $\bzero$ result in higher solenoidal ratio. To study the generation of solenoidal velocity component, it is helpful to write down the vorticity equation\footnote{As noted by \citet{Porter15}, solenoidal velocity component and the vorticity are not exactly same. While the former, $v_{\textrm{sol}} \propto l^{1/3}$, is dominant in energy injection scale, the latter, $w \propto l^{-2/3}$, concentrates near the dissipation scale, where $l$ is a scale in the inertial range and $v_{\textrm{sol}}$ and $w$ are solenoidal velocity and vorticity associated with the scale, respectively. However, since the vorticity is essentially solenoidal, and the generation of vorticity is accompanied by that of solenoidal velocity, we expect that the vorticity equation can trace the generation of the solenoidal velocity component.}, ``which is obtained from the curl of the Navier-Stokes equation with magnetic (Maxwell) stresses, $\mathbf{j} \times \mathbf{B}$, added" \citep{Porter15}: 
\begin{equation}
\label{eq:eq6}
 \begin{aligned}
\frac{\partial{\mathbf{w}}}{\partial{t}} = -\nabla\cdot \left(\mathbf{v}w\right)+\left(\mathbf{w} \cdot \nabla \right)\mathbf{v}+ \frac{1}{\rho^2}\nabla\rho \times \nabla P_T  \\ 
+ \nabla \times \left(\frac{\mathbf{T}}{\rho}\right) + \nu \left(\nabla^2w+\nabla \times \mathbf{G}\right)+\nabla \times \mathbf{f},
 \end{aligned}
\end{equation}
where $\mathbf{w}$ (= $\nabla \times \mathbf{v}$) is the vorticity, $P_{T}$ (= $p$ + $P_{B}$) is the sum of the gas pressure, $p$, and the magnetic pressure, $P_B$ = (1/2)$B^2$, $\mathbf{T}$ (= $\mathbf{B} \cdot \nabla \mathbf{B}$) is the magnetic tension, and $\nu$ is the kinematic viscosity. In the viscous term, $\mathbf{G}$ is defined by $\mathbf{G}$ = (1/$\rho$)$\nabla \rho \cdot \mathbf{S}$, where $\mathbf{S}$ is the standard traceless strain tensor (see, e.g., \citealt{Mee06}).

\begin{figure*}[ht!]
\centering
\includegraphics[scale=0.15]{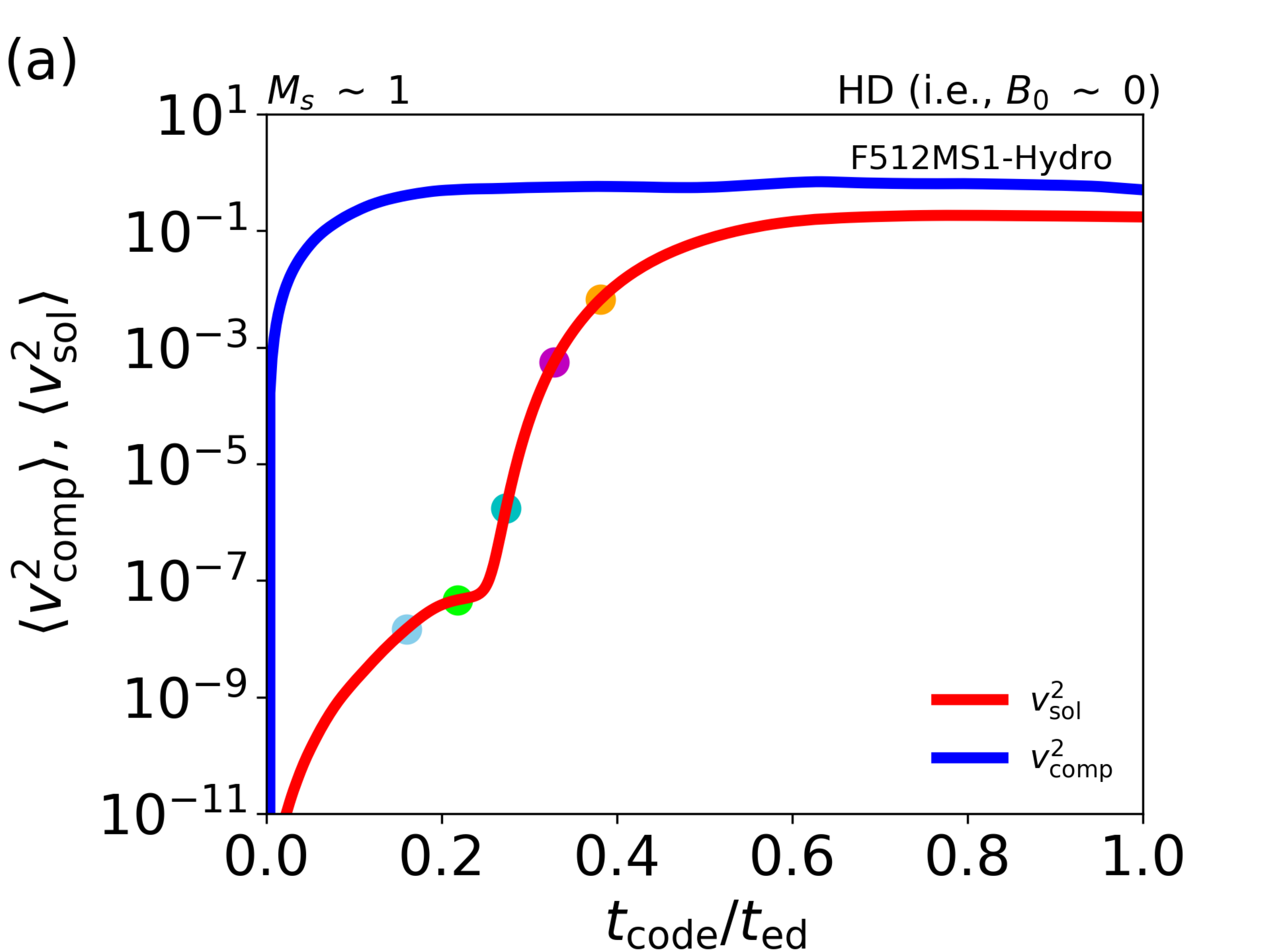}
\includegraphics[scale=0.15]{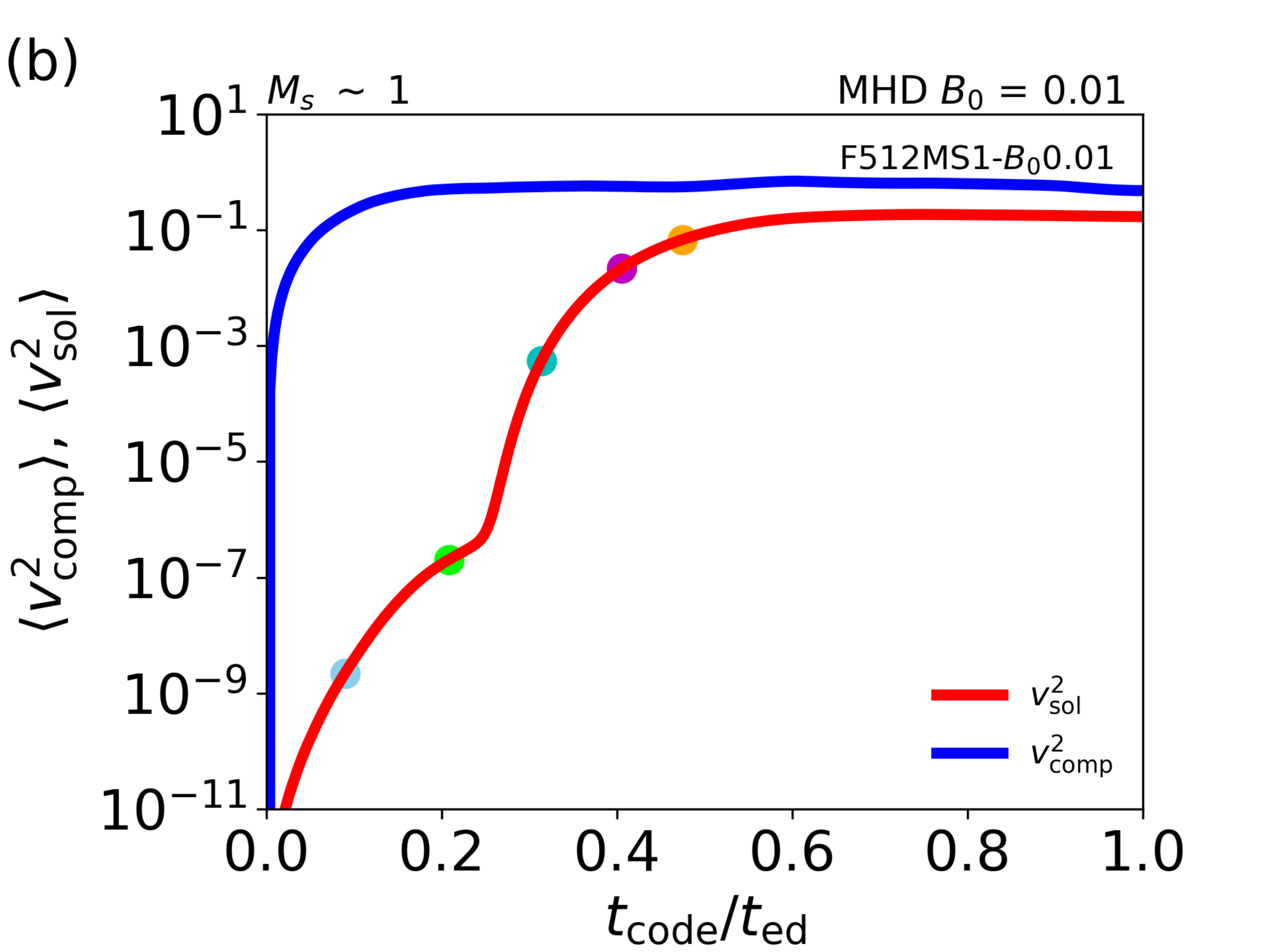} 
\includegraphics[scale=0.15]{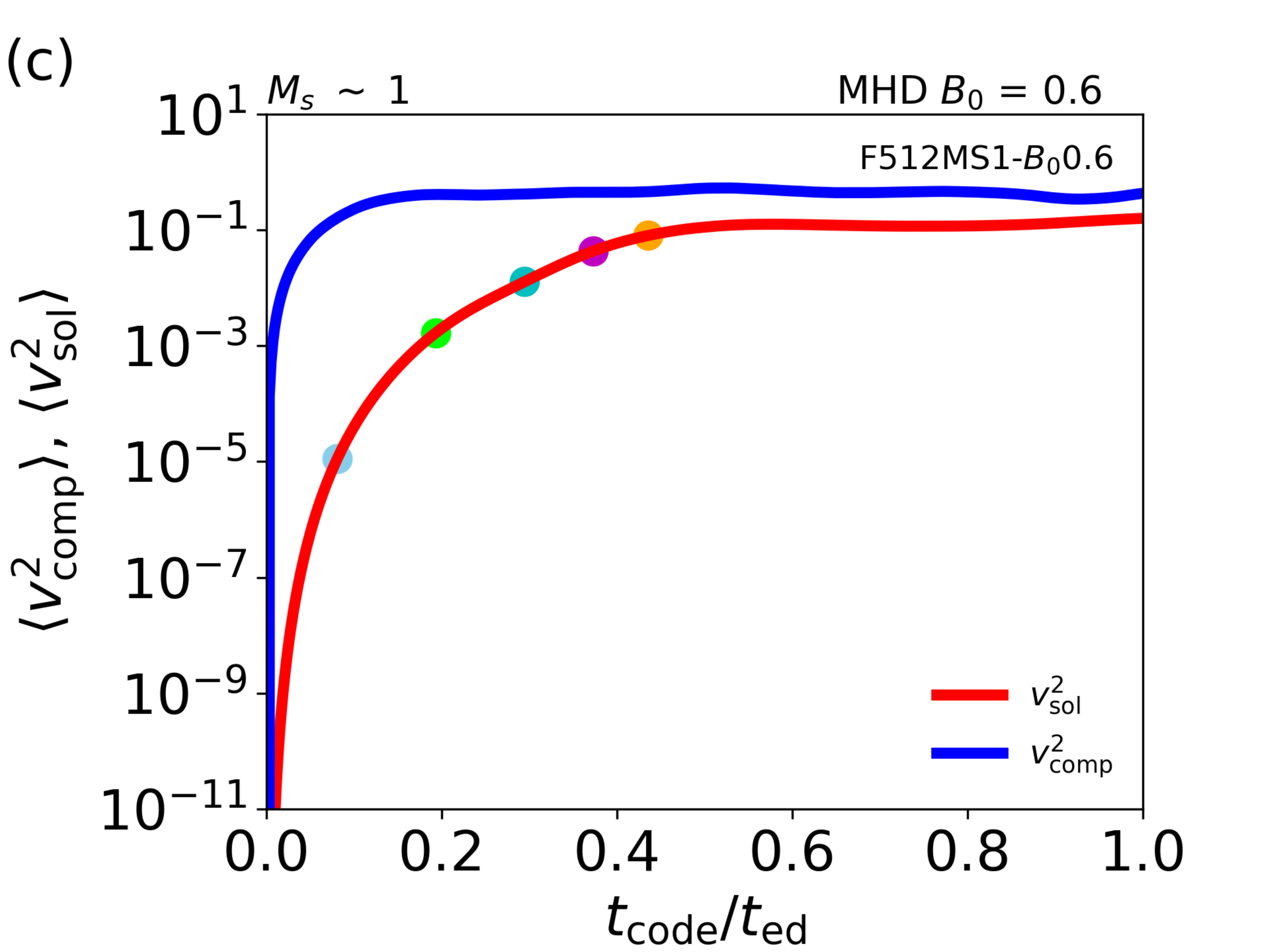} \\
\includegraphics[scale=0.15]{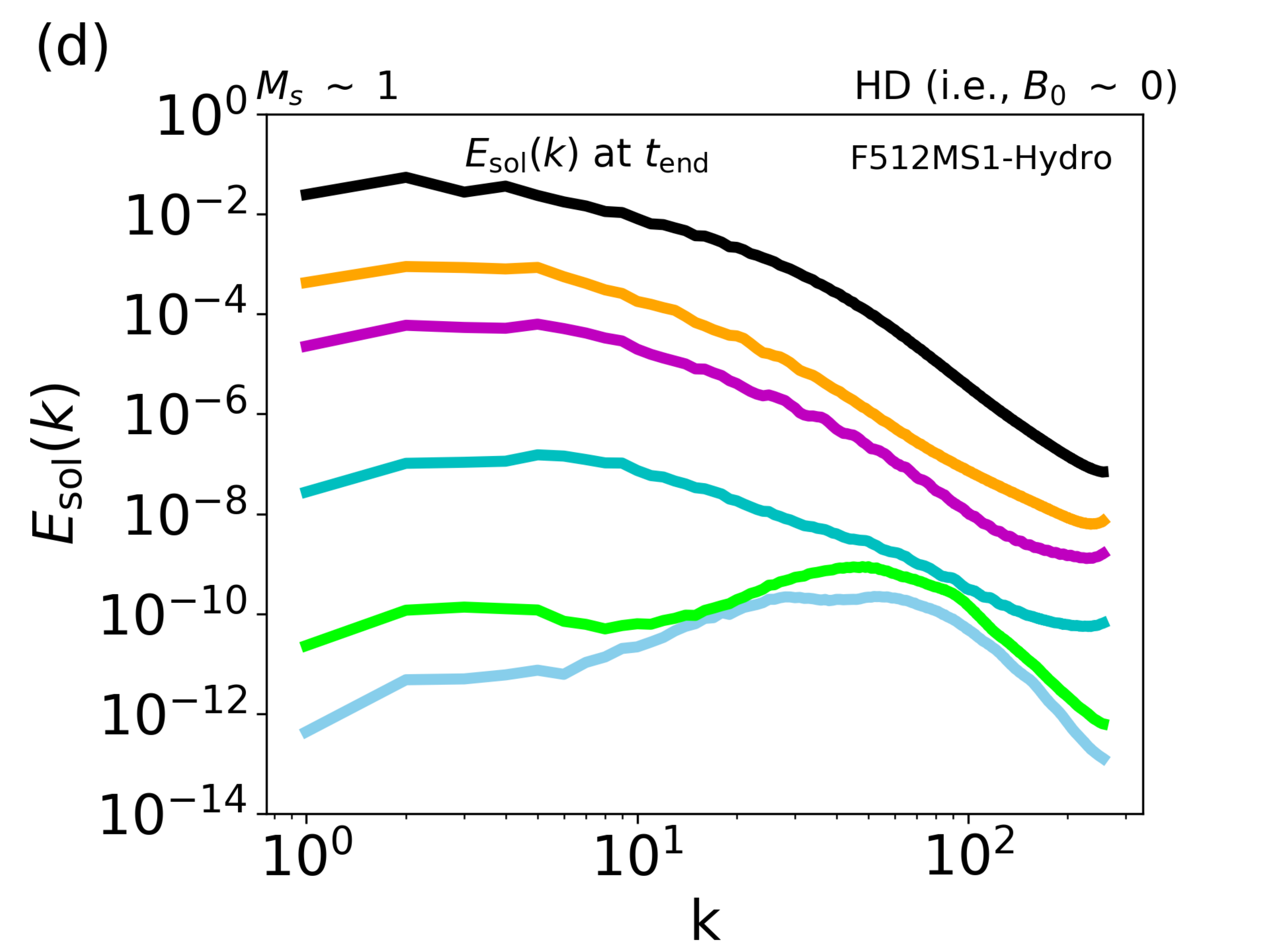}
\includegraphics[scale=0.15]{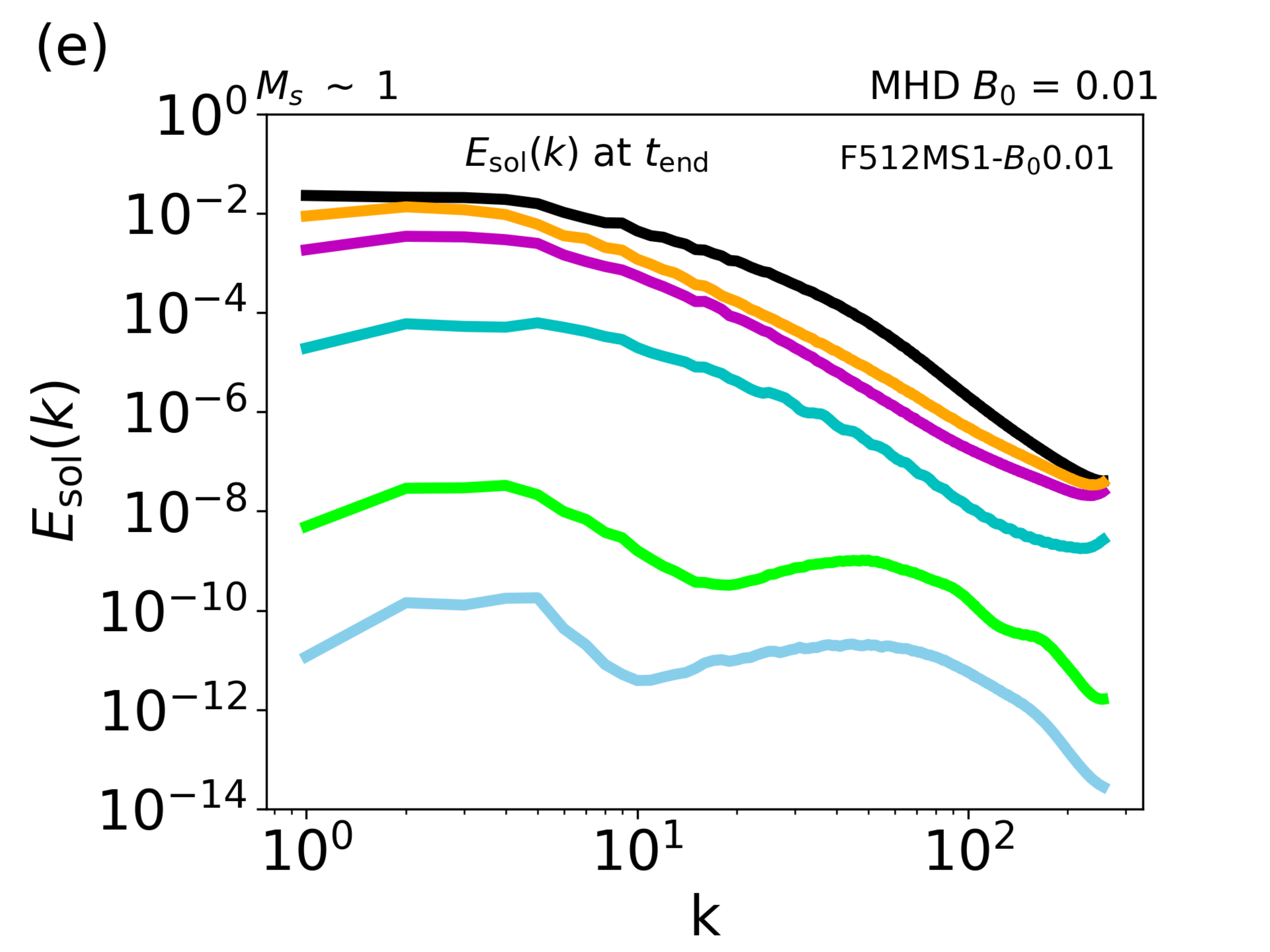} 
\includegraphics[scale=0.15]{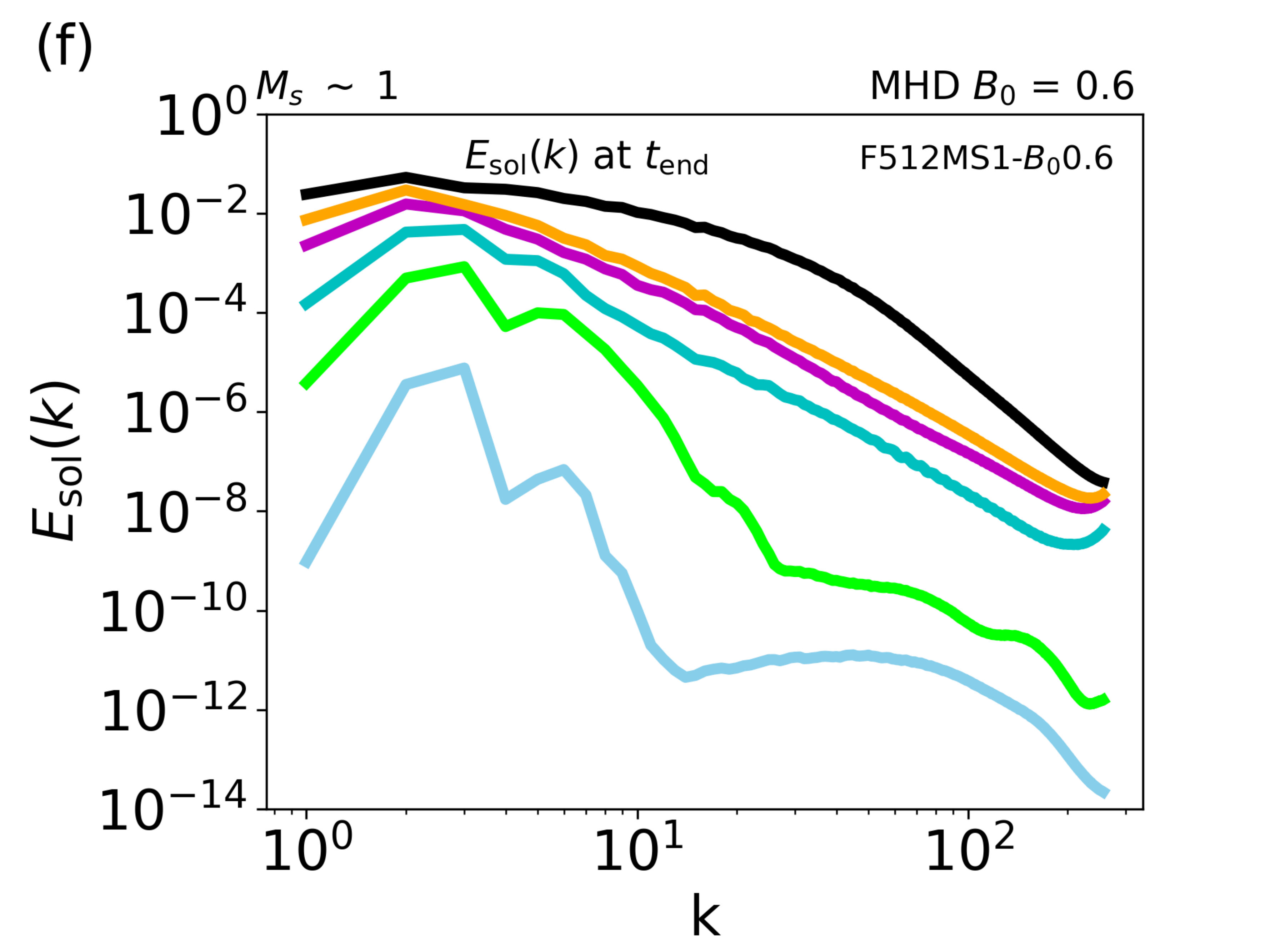}
\caption{Time evolution of $\langle v_{\textrm{comp}}^2\rangle $ and $\langle \vsolsq\rangle$ (upper panels) and the power spectra of $\mathbf{v}_{\textrm{sol}}$ (lower panels) at the beginning of simulations. Left panels: Run F512MS1-Hydro. Middle panels: Run F512MS1-$\bzero$0.01. Right panels: Run F512MS1-$\bzero$0.6. Blue and red curves in the upper panels denote $\langle v_{\textrm{comp}}^2\rangle $ and $\langle \vsolsq\rangle$, respectively. Each colored circle in the upper panels denotes the moment at which the power spectrum with matching color in the lower panels is drawn. Black curves in the lower panels show power spectra after saturation. 
\label{fig:fig21}}
\end{figure*}


 In our simulations, velocity and hence vorticity are zero at the beginning. Then, how is vorticity generated later on? We note that the last term on the right-hand side (rhs) in Equation (\ref{eq:eq6}) is zero in our simulations with compressive driving. In addition, since we assume the isothermal equation of state, $p$ $\propto$ $\rho$, the baroclinic source term by gas pressure, $(1/\rho^2)(\nabla \rho \times \nabla p)$, vanishes. Then, let us first assume that $B_0$ is zero or extremely small. In this case, all terms, except the viscous term (the second-last term in Equation (\ref{eq:eq6})), on the rhs are either exactly zero or almost zero when time is small. Our numerical simulations are for ideal MHD so that $\nu$ is zero formally. However, on the grid resolution-scale, numerical dissipation can resemble the viscosity, which makes it possible for the viscous term to act numerically. Therefore, the viscous term can seed the vorticity across density gradients even in the absence of initial vorticity in compressively driven turbulence \citep{Mee06,F11}. If $B_0$ is not very small, it is possible that the fourth term on the rhs generates vorticity even from the beginning\footnote{At t = 0, the magnetic tension ($\mathbf{T}$) is zero. Therefore, the fourth term on the rhs of Equation (\ref{eq:eq6}) is zero at t = 0. However, as soon as magnetic field lines are perturbed by turbulent motions, the tension term begins to work.}. Once seed vorticity is generated, the second term on the rhs can contribute to amplification of vorticity.

 Here we quantify the following source/amplification terms: vortex stretching term, $(\mathbf{w} \cdot \nabla)\mathbf{v}$, the magnetic pressure term, $(1/\rho^2)(\nabla \rho \times \nabla P_{B})$, and the magnetic tension term, $\nabla \times (\mathbf{T}/\rho)$. We consider only the finite-correlated compressive driving in this subsection.

Figure \ref{fig:fig16} shows PDFs of the source terms of vorticity for Run F512MS1-$\bzero$0.01 (upper panels) and Run F512MS1-$\bzero$1 (lower panels). Dotted, dashed, and solid curves in each panel denote logarithm of the vortex stretching, magnetic pressure, and magnetic tension terms, respectively. The left and the right panels correspond to the PDFs calculated before and after saturation, respectively. We show the time at which each PDF is calculated in the left side of each panel.

We can note from the figure that the magnetic pressure and tension effects are not dominant in the case of $\bzero$ = 0.01 (see dashed and solid curves in the upper panels); the former is weakest, and the latter is comparable to the stretching effect irrespective of whether turbulence saturates or not. On the other hand, in the case of $\bzero$ = 1, we can clearly see that the magnetic tension effect is strongest, and the other two effects are weaker than the tension term and comparable with each other. Hence, we can suggest that, when $B_0$ is very strong, the magnetic tension effect is highly effective in generating vorticity and solenoidal velocity component. In fact, we can explain the trend in Figure \ref{fig:fig8}(b) in terms of the magnetic tension term: as $B_0$ increases, the magnetic tension effect increases, which in turn produces more solenoidal velocity component.

Figure \ref{fig:fig17} illustrates effect of $M_s$ on the vorticity equation for $256^3$ resolution and $B_0$ $\leq$ 0.6. From left to right, Figures \ref{fig:fig17}(a)-(c) show average values of $\langle|\frac{1}{\rho^2}\nabla \rho \times \nabla P_B|\rangle$, $\langle|\nabla \times \frac{\mathbf{T}}{\rho}|\rangle$, and $\langle|(\mathbf{w}\cdot \nabla)\mathbf{v}|\rangle$ at saturation, respectively. In each panel, blue, red, and green circles correspond to $M_s$ $\sim$ 0.5, $\sim$ 1, and $\sim$ 3, respectively. The horizontal axis in each panel is total magnetic energy density ($B^2$).

First, we can clearly note that the magnetic pressure and tension terms are roughly proportional to $B^2$ for all $M_s$'s, with those for $M_s$ $\sim$ 3 being especially sensitive to $B^2$. This is understandable because the former contains $B^2$ and the latter $\mathbf{B} \cdot \nabla \mathbf{B}$. Second, according to Figure \ref{fig:fig17}(c), the vortex stretching term does not show strong dependence on $B^2$ for all $M_s$'s. More precisely, it stays nearly constants when $B^2$ $\lesssim$ 0.1 and deviates slightly from the constants when $B^2$ $>$ 0.1. Third, if we compare the scales of the vertical axes in those panels, the magnetic tension term is in general largest. Lastly, the middle panel of Figure \ref{fig:fig17} shows that magnetic tension term is larger when $M_s$ is larger, which means that compressively driven turbulence with a higher $M_s$ can generate vorticity more efficiently than that with a lower $M_s$ mainly by the magnetic tension effect. 

Figures \ref{fig:fig18} and \ref{fig:fig19} are for the examination of the effect of $\bzero$ in the case of $\mach$ $\sim$ 1 and $512^3$ resolution. In those figures, cyan, brown, blue, orange, and black curves correspond to $\bzero$ = 0.01, 0.05, 0.1, 0.2, and 0.6, respectively. The red dashed line in the right panels represent the HD simulation, Run F512MS1-Hydro. The PDFs before saturation are drawn in Figure \ref{fig:fig18} and those after saturation in Figure \ref{fig:fig19}. It is obvious from the figures that the PDFs for the magnetic pressure and the tension terms have larger mean values as $\bzero$ increases, and those for the stretching term remain almost same. Such a clear trend illustrates that the effect of magnetic fields on generation of solenoidal modes becomes stronger as $\bzero$ increases. 

\begin{figure}[t!]
\includegraphics[scale=0.25]{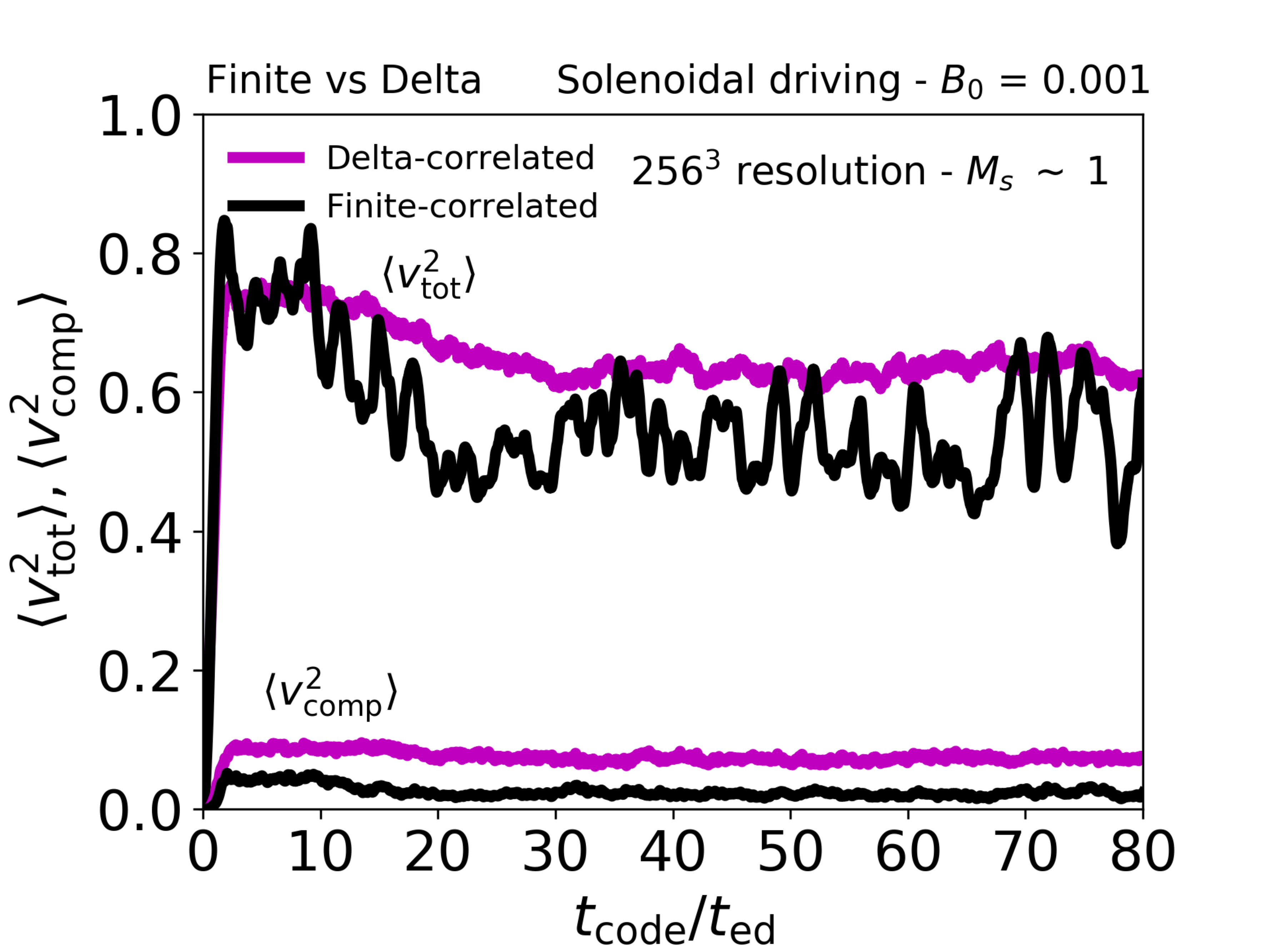}
\caption{Time evolution of $\langle \vtotsq \rangle$ and $\langle \vcompsq \rangle$ for $\textit{solenoidally}$ driven turbulence with $\bzero$ = 0.001 and $\mach$ $\sim$ 1. Magenta and black curves denote the delta-correlated and the finite-correlated solenoidal drivings, respectively. Note that we show $\langle \vcompsq \rangle$, not $\langle \vsolsq \rangle$ in this figure. 
\label{fig:fig22}}
\end{figure}

Figure \ref{fig:fig20} shows average values of $\langle|\frac{1}{\rho^2}\nabla \rho \times \nabla P_B|\rangle$, $\langle|\nabla \times \frac{\mathbf{T}}{\rho}|\rangle$, and $\langle|(\mathbf{w}\cdot \nabla)\mathbf{v}|\rangle$ after saturation, which are denoted by circles, ``X" markers, and triangles, respectively, as a function of $B^2$. The grey shaded region shows 1$\sigma$-dispersion about the average value of $\langle|(\mathbf{w}\cdot \nabla)\mathbf{v}|\rangle$ for the HD simulation, Run F512MS1-Hydro. The inset in the upper left corner is a zoom to clearly show the trend of $\langle|\frac{1}{\rho^2}\nabla \rho \times \nabla P_B|\rangle$ and $\langle|(\mathbf{w}\cdot \nabla)\mathbf{v}|\rangle$ with $B^2$.

The figure shows that the magnetic tension term is largest and the inset shows that the magnetic pressure term is smallest for all $B^2$'s. As in Figure \ref{fig:fig17}, the magnetic tension term is roughly proportional to $B^2$. Note that its dependence on $B^2$ is very steep and it becomes much larger than others when $B^2$ $\gtrsim$ 0.1. This implies that as $B^2$ increases, tension of magnetic field lines becomes more important and Alfvenization of compressively driven turbulence happens. As a consequence, when $B_0$ is large (i.e., $B^2$ is large), compressively driven turbulence yields a significant amount of solenoidal velocity component by the efficient role of the magnetic tension.

So far, we have discussed why large $B_0$ results in substantial solenoidal modes in turbulence driven by compressive driving. We have found that the effect of magnetic tension is most important. To clearly demonstrate this, we present in Figure \ref{fig:fig21} power spectra of $\mathbf{v}_{\textrm{sol}}$ for $t_{\textrm{code}}/t_{\textrm{ed}}$ $\leq$ 1. The upper panels of Figure \ref{fig:fig21} show the time evolution of both $\langle \vsolsq\rangle $ (red curves) and $\langle v_{\textrm{comp}}^2\rangle $ (blue curves), and the lower panels show the power spectra of $\mathbf{v}_{\textrm{sol}}$. The left, the middle, and the right panels correspond to Run F512MS1-Hydro, Run F512MS1-$\bzero$0.01, and Run F512MS1-$\bzero$0.6, respectively. The circles with different colors in the upper panels denote moments at which power spectra in the lower panels are calculated. The black curves in the lower panels are the power spectra after saturation. 

We can immediately note the difference in the power spectra. If we compare sky-blue curves in the lower panels, the power spectra peak at different wavenumbers. The peak wavenumber for the HD simulation (left panel) is k $\approx$ 80, which is closer to the dissipation scale, and moves to smaller wavenumbers as times go on. For $\bzero$ = 0.6 (right panel), the power spectrum (in sky-blue) peaks at k $\approx$ 2.5, corresponding to the driving scale, and then goes up without changing the peak wavenumber. The weak magnetic field case (middle panel) is intermediate between the two extreme cases. 

We interpret the result of Figure \ref{fig:fig21} as follows. In the absence of magnetic field, as in the case of Run F512MS1-Hydro (left panel of Figure \ref{fig:fig21}), we can reduce Equation (\ref{eq:eq6}) to 
\begin{equation}
\label{eq:eq7}
 \begin{aligned}
\frac{\partial{\mathbf{w}}}{\partial{t}} \sim -\nabla\cdot(\mathbf{v}w)+(\mathbf{w} \cdot \nabla)\mathbf{v} + \nu(\nabla^2w+\nabla \times \mathbf{G}).
 \end{aligned}
\end{equation}
Since the first term on the rhs is conservative advection of the vorticity, the only contributing terms to the vorticity generation are the stretching (the second on the rhs) and the viscous dissipation (the third on the rhs) terms. In our simulations, vorticity is zero at the beginning and the compressive driving does not produce any vorticity later. Therefore, the vorticity in compressively driven HD turbulence is initially generated by the viscous dissipation term, which is responsible for the peak near the dissipation scale, and then amplified by the stretching effect (see also \citealt{Mee06,F11}). As a consequence of the stretching effect, the peak position moves to smaller k's as time goes on. On the other hand, when the mean magnetic field is initially strong, as in the case of Run F512MS1-$\bzero$0.6 (right panel of Figure \ref{fig:fig21}), magnetic tension effect directly contributes to generating solenoidal velocity component from the beginning, thus the power spectrum can have a peak near the driving scale. When the mean magnetic field is initially weak, as in the case of Run F512MS1-$\bzero$0.01 (middle panel of Figure \ref{fig:fig21}), the process seems more complicated, and we suggest that the stretching and magnetic field effects play roles in conjunction because their strengths are comparable (see Figures \ref{fig:fig16}(a) and \ref{fig:fig16}(b)). 

\begin{figure}
\includegraphics[scale=0.25]{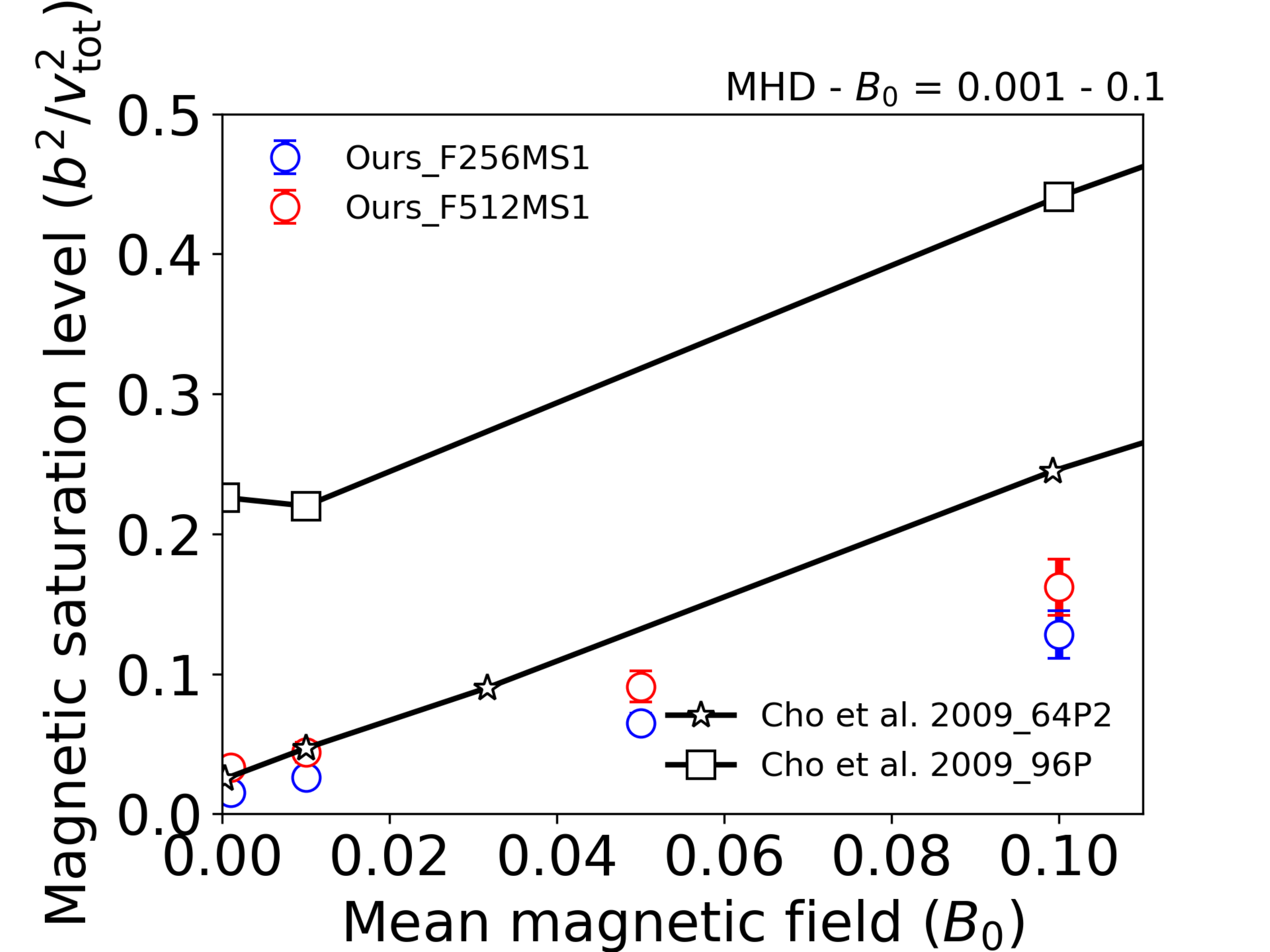}
\caption{Comparison of the magnetic saturation level from our simulations for $M_s$ $\sim$ 1 with that from simulations of \citet{Cho09}, who used solenoidal driving and considered incompressible turbulence. Blue circles and red circles correspond to the our result from $256^3$ and $512^3$ resolution simulations, respectively. Black stars and black squares denote 64P2 and 96P simulation results from \citet{Cho09}, respectively. According to their notation, 64 or 96 refers to the number of grid points in each spatial direction, and P2 or P refers to physical viscosity.
\label{fig:fig23}}
\end{figure}

\subsection{Effects of Correlation Timescale of Forcing Vectors\label{sec:sec5.2}}
In Sections \ref{sec:sec3} and \ref{sec:sec4}, we have found that the delta-correlated compressive driving results in lower solenoidal ratio and magnetic saturation level than the finite-correlated compressive driving with a similar $M_s$ and the same numerical resolution. If a similar (in terms of direction and magnitude) driving is applied for a sufficiently long time, as in the case of the finite-correlated compressive driving, coherent generation of solenoidal component occurs and the solenoidal velocity component can have a sufficiently long time to maintain its vortical motions once generated. We can have more clear picture about this in terms of vorticity: if a coherent driving is applied for a long time, continued stretching of vorticity occurs until the forcing vector changes significantly, and as a result, stretching can be efficient. However, if a driving  changes its amplitude and direction in a very short time interval, as in the case of the delta-correlated compressive driving, stretching of vorticity can be inefficient due to the frequent and abrupt change of the forcing vector. Therefore, it is understandable that the finite-correlated compressive driving generates more solenoidal velocity component. This argument is applicable to both HD and MHD turbulence.

 In addition, the finite-correlated driving can generate stronger small-scale magnetic field, which in turn helps turbulence driven by the finite-correlated driving to produce more solenoidal component in MHD turbulence. Similar to stretching of vorticity, stretching of magnetic field lines can be more efficient in turbulence driven by the finite-correlated driving. Thus, the finite-correlated compressive driving can generate a stronger random magnetic field, which implies that the effect of magnetic field on the generation of solenoidal velocity component is more significant in the finite-correlated compressive driving. Therefore, we can conclude that the finite-correlated compressive driving generates more solenoidal velocity component.
 
Timescale of forcing vectors can also affect generation of compressive modes in $\textit{solenoidally}$ driven turbulence. When compressive modes are generated in solenoidally driven turbulence, they will produce density fluctuations. If timescale of forcing is sufficiently long, as in the finite-correlated solenoidal driving, there can be time for gas pressure to (partially) counteract generation of the density fluctuations, which will (partially) suppress generation of compressive modes. There will not be such a suppression in the delta-correlated solenoidal driving case. Therefore, we expect that the latter driving scheme produces more compressive modes. Figure \ref{fig:fig22} clearly demonstrates that this argument is valid for solenoidally driven turbulence. In the figure, magenta and black curves correspond to the delta-correlated and the finite-correlated solenoidal drivings, respectively. We only consider $\mach$ $\sim$ 1 and $256^3$ resolution. Unlike other figures presented in Section \ref{sec:sec3}, we present the time evolution of $\langle v_{\textrm{comp}}^2\rangle$. As we can see, the delta-correlated solenoidal driving produces a higher level of $\langle v_{\textrm{comp}}^2\rangle$ at saturation. This argument can also explain why the delta-correlated solenoidal driving has wider density PDFs \citep{YHS16}.

\section{Discussion on Small-Scale Turbulence Dynamo\label{sec:sec6}}
As mentioned in Section \ref{sec:sec4.2}, small-scale turbulence dynamo in compressively driven turbulence is not significantly sensitive to numerical resolution. To compare this with dynamo in solenoidally driven turbulence, we refer to \citet{Cho09}, who studied small-scale turbulence dynamo in incompressible turbulence driven by a solenoidal driving. 

Figure \ref{fig:fig23} shows the comparison. Blue and red circles indicate the magnetic saturation levels for $256^3$ and $512^3$ resolutions from our simulations, respectively. Black stars and black squares represent the magnetic saturation levels of the simulations with $64^3$ and $96^3$ resolutions from \citet{Cho09}, respectively.

First, we can see from Figure \ref{fig:fig23} that both solenoidally and compressively driven turbulence show linear relation between $B_0$ and the magnetic saturation level. However, we note that the slopes are steeper for incompressible turbulence driven by the solenoidal driving. Second, the figure obviously shows inefficient turbulence dynamo induced by the compressive driving: when $B_0$ $\lesssim$ 0.01, the magnetic saturation levels from our $256^3$ (blue circles) and $512^3$ (red circles) resolution simulations are comparable to those from $64^3$ resolution  simulations of solenoidally driven incompressible turbulence (black stars). When $B_0$ $>$ 0.01, even $64^3$ resolution simulations of incompressible turbulence driven by the solenoidal driving show higher magnetic saturation levels than compressible turbulence with $512^3$ resolution driven by the compressive driving. Third, for solenoidally driven turbulence, magnetic saturation level is very sensitive to numerical resolution (compare black squares with black stars), while it is not for compressively driven turbulence (compare red circles with blue ones). Therefore, the comparison suggests that a compressive driving would not effectively amplify small-scale magnetic fields even though numerical resolution becomes very high. 

Due to inefficiency of turbulence dynamo in compressively driven turbulence, it is very difficult to estimate the saturation level of magnetic energy density in the limit of a very large numerical resolution. However, we may conjecture at least the upper limit for the saturation level. In solenoidally driven turbulence with unit magnetic Prandtl number, the magnetic saturation level is slightly less than unity. Since solenoidal motions are mainly responsible for magnetic field growth, it is not plausible for magnetic energy to be larger than solenoidal energy. If this is true, we expect that the magnetic saturation level in the limit of a very high numerical resolution is less than 0.25 for $M_s$ $\sim$ 1, which is the solenoidal ratio for runs with no or a very small mean magnetic field. We will address this issue elsewhere.

\section{Summary\label{sec:sec7}}
In this paper, we have studied generation of solenoidal velocity component and small-scale magnetic field in compressively driven turbulence. In this regard, we have quantified the effects of the sonic Mach number $(\mach)$ and the mean magnetic field $(\bzero)$. Moreover, we have considered two different driving schemes in terms of different correlation timescale of forcing vectors, a finite-correlated driving and a delta-correlated driving. Our main findings are as follows:

\begin{enumerate}
	\item \textit{The effect of the sonic Mach number ($\mach$) on the generation of solenoidal velocity component.} We have shown that the higher $\mach$ is, the more solenoidal velocity component is generated in compressively driven turbulence in both strong and weak mean magnetic field cases. 
	\item \textit{The effect of mean magnetic field ($\bzero$) on the generation of solenoidal velocity component.} We have found that when $\bzero$ is small, compressive driving yields solenoidal velocity component as similar as hydrodynamic turbulence. However, when $\bzero$ exceeds a certain value, it produces more solenoidal velocity component than hydrodynamic turbulence.		
	\item \textit{The effect of $\mach$ on the generation of small-scale magnetic field component.} We have examined that, when $\bzero$ is small, the saturation level of the small-scale magnetic field component peaks at $\mach$ $\sim$ 1 for the finite-correlated compressive driving and shows monotonic increase when $\mach$ $\lesssim$ 3 for the delta-correlated compressive driving. When $\bzero$ is very large, both driving schemes show similar behaviors: they produce maximum saturation level at $\mach$ $\sim$ 1, and the level gradually decreases as $M_s$ increases.
	\item \textit{The effect of $\bzero$ on the generation of small-scale magnetic field component.} We have revealed that as $\bzero$ increases, more small-scale magnetic field components are generated in compressively driven turbulence. Moreover, saturation level of the magnetic field follows approximately a linear relation with $B_0$ as in solenoidally driven turbulence.
	\item \textit{The effect of numerical resolution.} We have shown that generation of solenoidal velocity component is virtually independent of numerical resolution and that of small-scale magnetic field is mildly sensitive to numerical resolution in compressively driven turbulence.
	\item \textit{The effect of a driving scheme}. When $M_s$ is similar and the numerical resolution is same, we have found that the finite-correlated driving always generates more solenoidal velocity component than the delta-correlated driving. The trend is also observed in the case of solenoidally driven turbulence. 
\end{enumerate}
We have analyzed the vorticity equation to examine the effects of $\mach$ and $\bzero$ on generation of solenoidal velocity component provided by compressive driving:
\begin{enumerate}
	\item For hydrodynamic turbulence, viscous dissipation initially generates vorticity, and then vortex stretching amplifies it.
	\item For MHD turbulence with strong mean magnetic fields, magnetic tension is in effect; it directly produces solenoidal modes from the beginning, which is responsible for large amounts of solenoidal velocity component in cases of strong mean magnetic field. For weak mean magnetic field cases, vortex stretching and magnetic field play roles simultaneously.
\end{enumerate}

In addition, we have discussed small-scale dynamo in compressively driven turbulence. We have compared small-scale turbulence dynamo by compressive driving with that by solenoidal driving from \citet{Cho09}. We have found that the magnetic saturation levels from our $256^3$ and $512^3$ resolution simulations are comparable to those from their $64^3$ resolution simulations in weak mean magnetic field regime, which implies inefficient dynamo action in compressively driven turbulence. We have obtained that the solenoidal ratio is $\sim$ 0.25 for compressively driven turbulence with no or a very weak mean magnetic field. Since it is not plausible for magnetic energy to be larger than solenoidal energy, we may conjecture that the magnetic saturation level at an arbitrarily high numerical resolution is less than 0.25 for $M_s$ $\sim$ 1. \\

This paper has been expanded from a chapter of Jeonghoon Lim’s Master thesis. This work is supported by the National R $\&$ D Program through the National Research Foundation of Korea Grants funded by the Korean Government (NRF-2016R1A5A1013277 and NRF-2016R1D1A1B02015014).
\bibliographystyle{apj}
\bibliography{ref}

\begin{thebibliography}{}
\expandafter\ifx\csname natexlab\endcsname\relax\def\natexlab#1{#1}\fi
\providecommand{\url}[1]{\href{#1}{#1}}
\providecommand{\dodoi}[1]{doi:~\href{http://doi.org/#1}{\nolinkurl{#1}}}
\providecommand{\doeprint}[1]{\href{http://ascl.net/#1}{\nolinkurl{http://ascl.net/#1}}}
\providecommand{\doarXiv}[1]{\href{https://arxiv.org/abs/#1}{\nolinkurl{https://arxiv.org/abs/#1}}}

\bibitem[{{Batchelor}(1950)}]{1950RSPSA.201..405B}
{Batchelor}, G.~K. 1950, Proceedings of the Royal Society of London Series A,
  201, 405

\bibitem[{{Bertoglio} {et~al.}(2001){Bertoglio}, {Bataille}, \&
  {Marion}}]{Berto2001}
{Bertoglio}, J.-P., {Bataille}, F., \& {Marion}, J.-D. 2001, Physics of Fluids,
  13, 290

\bibitem[{{Boldyrev} {et~al.}(2002){Boldyrev}, {Nordlund}, \&
  {Padoan}}]{Boldrev2002}
{Boldyrev}, S., {Nordlund}, {\r{A}}., \& {Padoan}, P. 2002, \apj, 573, 678.
\newblock \doarXiv{astro-ph/0111345}

\bibitem[{{Brandenburg} \& {Subramanian}(2005)}]{2005PhR...417....1B}
{Brandenburg}, A., \& {Subramanian}, K. 2005, \physrep, 417, 1.
\newblock \doarXiv{astro-ph/0405052}

\bibitem[{{Brunetti} \& {Jones}(2014)}]{BJ14}
{Brunetti}, G., \& {Jones}, T.~W. 2014, International Journal of Modern Physics
  D, 23, 1430007.
\newblock \doarXiv{1401.7519}

\bibitem[{{Carilli} \& {Taylor}(2002)}]{CT02}
{Carilli}, C.~L., \& {Taylor}, G.~B. 2002, \araa, 40, 319.
\newblock \doarXiv{astro-ph/0110655}

\bibitem[{{Cho}(2014)}]{Cho14}
{Cho}, J. 2014, \apj, 797, 133.
\newblock \doarXiv{1410.1893}

\bibitem[{{Cho} \& {Lazarian}(2002)}]{CL02}
{Cho}, J., \& {Lazarian}, A. 2002, \prl, 88, 245001.
\newblock \doarXiv{astro-ph/0205282}

\bibitem[{{Cho} \& {Lazarian}(2003)}]{Cho03}
---. 2003, \mnras, 345, 325.
\newblock \doarXiv{astro-ph/0301062}

\bibitem[{{Cho} \& {Vishniac}(2000)}]{CV00}
{Cho}, J., \& {Vishniac}, E.~T. 2000, \apj, 538, 217.
\newblock \doarXiv{astro-ph/0003404}

\bibitem[{{Cho} {et~al.}(2009){Cho}, {Vishniac}, {Beresnyak}, {Lazarian}, \&
  {Ryu}}]{Cho09}
{Cho}, J., {Vishniac}, E.~T., {Beresnyak}, A., {Lazarian}, A., \& {Ryu}, D.
  2009, \apj, 693, 1449.
\newblock \doarXiv{0812.0817}

\bibitem[{{Cho} \& {Yoo}(2012)}]{Cho12}
{Cho}, J., \& {Yoo}, H. 2012, \apj, 759, 91.
\newblock \doarXiv{1209.6130}

\bibitem[{{Crutcher}(2012)}]{Cr2012}
{Crutcher}, R.~M. 2012, \araa, 50, 29

\bibitem[{{Elmegreen} \& {Scalo}(2004)}]{ES04}
{Elmegreen}, B.~G., \& {Scalo}, J. 2004, \araa, 42, 211.
\newblock \doarXiv{astro-ph/0404451}

\bibitem[{{Federrath}(2013)}]{F13}
{Federrath}, C. 2013, \mnras, 436, 1245.
\newblock \doarXiv{1306.3989}

\bibitem[{{Federrath} {et~al.}(2011){Federrath}, {Chabrier}, {Schober},
  {Banerjee}, {Klessen}, \& {Schleicher}}]{F11}
{Federrath}, C., {Chabrier}, G., {Schober}, J., {et~al.} 2011, \prl, 107,
  114504.
\newblock \doarXiv{1109.1760}

\bibitem[{{Federrath} {et~al.}(2008){Federrath}, {Klessen}, \& {Schmidt}}]{F08}
{Federrath}, C., {Klessen}, R.~S., \& {Schmidt}, W. 2008, \apjl, 688, L79.
\newblock \doarXiv{0808.0605}

\bibitem[{{Federrath} {et~al.}(2009){Federrath}, {Klessen}, \&
  {Schmidt}}]{Fe09}
---. 2009, \apj, 692, 364.
\newblock \doarXiv{0710.1359}

\bibitem[{{Federrath} {et~al.}(2010){Federrath}, {Roman-Duval}, {Klessen},
  {Schmidt}, \& {Mac Low}}]{F10}
{Federrath}, C., {Roman-Duval}, J., {Klessen}, R.~S., {Schmidt}, W., \& {Mac
  Low}, M.~M. 2010, \aap, 512, A81.
\newblock \doarXiv{0905.1060}

\bibitem[{{Govoni} \& {Feretti}(2004)}]{GF04}
{Govoni}, F., \& {Feretti}, L. 2004, International Journal of Modern Physics D,
  13, 1549.
\newblock \doarXiv{astro-ph/0410182}

\bibitem[{{Haugen} {et~al.}(2003){Haugen}, {Brandenburg}, \&
  {Dobler}}]{Haugen03}
{Haugen}, N. E.~L., {Brandenburg}, A., \& {Dobler}, W. 2003, \apjl, 597, L141.
\newblock \doarXiv{astro-ph/0303372}

\bibitem[{Haugen {et~al.}(2004)Haugen, Brandenburg, \& Dobler}]{Haugen04}
Haugen, N. E.~L., Brandenburg, A., \& Dobler, W. 2004, Phys. Rev. E, 70,
  016308.
\newblock \url{https://link.aps.org/doi/10.1103/PhysRevE.70.016308}

\bibitem[{{Hennebelle} \& {Inutsuka}(2019)}]{HI19}
{Hennebelle}, P., \& {Inutsuka}, S.-i. 2019, Frontiers in Astronomy and Space
  Sciences, 6, 5.
\newblock \doarXiv{1902.00798}

\bibitem[{{Hitomi Collaboration} {et~al.}(2016){Hitomi Collaboration},
  {Aharonian}, {Akamatsu}, {Akimoto}, {Allen}, {Anabuki}, {Angelini}, {Arnaud},
  {Audard}, {Awaki}, {Axelsson}, {Bamba}, {Bautz}, {Blandford}, {Brenneman},
  {Brown}, {Bulbul}, {Cackett}, {Chernyakova}, {Chiao}, {Coppi}, {Costantini},
  {de Plaa}, {den Herder}, {Done}, {Dotani}, {Ebisawa}, {Eckart}, {Enoto},
  {Ezoe}, {Fabian}, {Ferrigno}, {Foster}, {Fujimoto}, {Fukazawa}, {Furuzawa},
  {Galeazzi}, {Gallo}, {Gandhi}, {Giustini}, {Goldwurm}, {Gu}, {Guainazzi},
  {Haba}, {Hagino}, {Hamaguchi}, {Harrus}, {Hatsukade}, {Hayashi}, {Hayashi},
  {Hayashida}, {Hiraga}, {Hornschemeier}, {Hoshino}, {Hughes}, {Iizuka},
  {Inoue}, {Inoue}, {Ishibashi}, {Ishida}, {Ishikawa}, {Ishisaki}, {Itoh},
  {Iyomoto}, {Kaastra}, {Kallman}, {Kamae}, {Kara}, {Kataoka}, {Katsuda},
  {Katsuta}, {Kawaharada}, {Kawai}, {Kelley}, {Khangulyan}, {Kilbourne},
  {King}, {Kitaguchi}, {Kitamoto}, {Kitayama}, {Kohmura}, {Kokubun}, {Koyama},
  {Koyama}, {Kretschmar}, {Krimm}, {Kubota}, {Kunieda}, {Laurent}, {Lebrun},
  {Lee}, {Leutenegger}, {Limousin}, {Loewenstein}, {Long}, {Lumb}, {Madejski},
  {Maeda}, {Maier}, {Makishima}, {Markevitch}, {Matsumoto}, {Matsushita},
  {McCammon}, {McNamara}, {Mehdipour}, {Miller}, {Miller}, {Mineshige},
  {Mitsuda}, {Mitsuishi}, {Miyazawa}, {Mizuno}, {Mori}, {Mori}, {Moseley},
  {Mukai}, {Murakami}, {Murakami}, {Mushotzky}, {Nagino}, {Nakagawa},
  {Nakajima}, {Nakamori}, {Nakano}, {Nakashima}, {Nakazawa}, {Nobukawa},
  {Noda}, {Nomachi}, {O'Dell}, {Odaka}, {Ohashi}, {Ohno}, {Okajima}, {Ota},
  {Ozaki}, {Paerels}, {Paltani}, {Parmar}, {Petre}, {Pinto}, {Pohl}, {Porter},
  {Pottschmidt}, {Ramsey}, {Reynolds}, {Russell}, {Safi-Harb}, {Saito},
  {Sakai}, {Sameshima}, {Sato}, {Sato}, {Sato}, {Sawada}, {Schartel},
  {Serlemitsos}, {Seta}, {Shidatsu}, {Simionescu}, {Smith}, {Soong}, {Stawarz},
  {Sugawara}, {Sugita}, {Szymkowiak}, {Tajima}, {Takahashi}, {Takahashi},
  {Takeda}, {Takei}, {Tamagawa}, {Tamura}, {Tamura}, {Tanaka}, {Tanaka},
  {Tanaka}, {Tashiro}, {Tawara}, {Terada}, {Terashima}, {Tombesi}, {Tomida},
  {Tsuboi}, {Tsujimoto}, {Tsunemi}, {Tsuru}, {Uchida}, {Uchiyama}, {Uchiyama},
  {Ueda}, {Ueda}, {Ueno}, {Uno}, {Urry}, {Ursino}, {de Vries}, {Watanabe},
  {Werner}, {Wik}, {Wilkins}, {Williams}, {Yamada}, {Yamaguchi}, {Yamaoka},
  {Yamasaki}, {Yamauchi}, {Yamauchi}, {Yaqoob}, {Yatsu}, {Yonetoku}, {Yoshida},
  {Yuasa}, {Zhuravleva}, \& {Zoghbi}}]{Hitomi16}
{Hitomi Collaboration}, {Aharonian}, F., {Akamatsu}, H., {et~al.} 2016, \nat,
  535, 117.
\newblock \doarXiv{1607.04487}

\bibitem[{{Kritsuk} {et~al.}(2007){Kritsuk}, {Norman}, {Padoan}, \&
  {Wagner}}]{Kritsuk07}
{Kritsuk}, A.~G., {Norman}, M.~L., {Padoan}, P., \& {Wagner}, R. 2007, \apj,
  665, 416.
\newblock \doarXiv{0704.3851}

\bibitem[{{Kritsuk} {et~al.}(2010){Kritsuk}, {Ustyugov}, {Norman}, \&
  {Padoan}}]{Kritsuk10}
{Kritsuk}, A.~G., {Ustyugov}, S.~D., {Norman}, M.~L., \& {Padoan}, P. 2010, in
  Astronomical Society of the Pacific Conference Series, Vol. 429, Numerical
  Modeling of Space Plasma Flows, Astronum-2009, ed. N.~V. {Pogorelov},
  E.~{Audit}, \& G.~P. {Zank}, 15.
\newblock \doarXiv{0912.0546}

\bibitem[{{Kulsrud} {et~al.}(1997){Kulsrud}, {Cen}, {Ostriker}, \&
  {Ryu}}]{Kulsrud97}
{Kulsrud}, R.~M., {Cen}, R., {Ostriker}, J.~P., \& {Ryu}, D. 1997, \apj, 480,
  481.
\newblock \doarXiv{astro-ph/9607141}

\bibitem[{{Larson}(1981)}]{L1981}
{Larson}, R.~B. 1981, \mnras, 194, 809

\bibitem[{{Mac Low} \& {Klessen}(2004)}]{MK04}
{Mac Low}, M.-M., \& {Klessen}, R.~S. 2004, Reviews of Modern Physics, 76, 125.
\newblock \doarXiv{astro-ph/0301093}

\bibitem[{{McKee} \& {Ostriker}(2007)}]{MO07}
{McKee}, C.~F., \& {Ostriker}, E.~C. 2007, \araa, 45, 565.
\newblock \doarXiv{0707.3514}

\bibitem[{{Mee} \& {Brandenburg}(2006)}]{Mee06}
{Mee}, A.~J., \& {Brandenburg}, A. 2006, \mnras, 370, 415.
\newblock \doarXiv{astro-ph/0602057}

\bibitem[{{Porter} {et~al.}(2015){Porter}, {Jones}, \& {Ryu}}]{Porter15}
{Porter}, D.~H., {Jones}, T.~W., \& {Ryu}, D. 2015, \apj, 810, 93.
\newblock \doarXiv{1507.08737}

\bibitem[{{Ryu} {et~al.}(2008){Ryu}, {Kang}, {Cho}, \& {Das}}]{Ryu08}
{Ryu}, D., {Kang}, H., {Cho}, J., \& {Das}, S. 2008, Science, 320, 909.
\newblock \doarXiv{0805.2466}

\bibitem[{{Ryu} {et~al.}(2012){Ryu}, {Schleicher}, {Treumann}, {Tsagas}, \&
  {Widrow}}]{Ryu12}
{Ryu}, D., {Schleicher}, D.~R.~G., {Treumann}, R.~A., {Tsagas}, C.~G., \&
  {Widrow}, L.~M. 2012, \ssr, 166, 1.
\newblock \doarXiv{1109.4055}

\bibitem[{{Schekochihin} {et~al.}(2004){Schekochihin}, {Cowley}, {Taylor},
  {Maron}, \& {McWilliams}}]{Schekochihin04}
{Schekochihin}, A.~A., {Cowley}, S.~C., {Taylor}, S.~F., {Maron}, J.~L., \&
  {McWilliams}, J.~C. 2004, \apj, 612, 276.
\newblock \doarXiv{astro-ph/0312046}

\bibitem[{{Schekochihin} {et~al.}(2007){Schekochihin}, {Iskakov}, {Cowley},
  {McWilliams}, {Proctor}, \& {Yousef}}]{Schekochihin07}
{Schekochihin}, A.~A., {Iskakov}, A.~B., {Cowley}, S.~C., {et~al.} 2007, New
  Journal of Physics, 9, 300.
\newblock \doarXiv{0704.2002}

\bibitem[{{Schuecker} {et~al.}(2004){Schuecker}, {Finoguenov}, {Miniati},
  {B{\"o}hringer}, \& {Briel}}]{Sc04}
{Schuecker}, P., {Finoguenov}, A., {Miniati}, F., {B{\"o}hringer}, H., \&
  {Briel}, U.~G. 2004, \aap, 426, 387.
\newblock \doarXiv{astro-ph/0404132}

\bibitem[{{Vazza} {et~al.}(2017){Vazza}, {Jones}, {Br{\"u}ggen}, {Brunetti},
  {Gheller}, {Porter}, \& {Ryu}}]{Vazza17}
{Vazza}, F., {Jones}, T.~W., {Br{\"u}ggen}, M., {et~al.} 2017, \mnras, 464,
  210.
\newblock \doarXiv{1609.03558}

\bibitem[{{Yoon} \& {Cho}(2019)}]{YHS19}
{Yoon}, H., \& {Cho}, J. 2019, \apj, 880, 137

\bibitem[{{Yoon} {et~al.}(2016){Yoon}, {Cho}, \& {Kim}}]{YHS16}
{Yoon}, H., {Cho}, J., \& {Kim}, J. 2016, \apj, 831, 85

\end{thebibliography}
\end{document}